\documentclass[twocolumn,twocolappendix,trackchanges]{aastex631}
\newcommand{\pfrac}{$P_{\text{frac}}$}
\newcommand{\sigpfrac}{$\sigma_{P_{\text{frac}}}$}

\newcommand{\ceoto}{\mbox{C$^{18}$O(2--1)}}
\newcommand{\ttcooz}{\mbox{$^{13}$CO(1--0)}}
\newcommand{\ttcoto}{\mbox{$^{13}$CO(2--1)}}
\newcommand{\Ngas}{$N_{\text{gas}}$}
\newcommand{\smallngas}{$n_{\text{gas}}$}
\newcommand{\NHt}{$N_{\text{H$_2$}}$}

\newcommand{\kms}     {km\,s$^{-1}$}
\newcommand{\tco}{\mbox{$^{13}$CO(1--0)}}

\newcommand{\nht}{\mbox{NH$_3$(1,1)}}
\defcitealias{Coude2025}{C25}
\defcitealias{Stephens2022}{S22}
\defcitealias{Zucker2015}{Z15}
\defcitealias{Zucker2018b}{Z18}

\shortauthors{Stephens et al.}
\begin{document}

\title{Magnetic Fields in the Bones of the Milky Way}

\author{Ian W. Stephens}
\affiliation{Department of Earth, Environment, and Physics, Worcester State University, Worcester, MA 01602, USA \href{mailto:istephens@worcester.edu}{istephens@worcester.edu}}
\affiliation{Center for Astrophysics $|$ Harvard \& Smithsonian, 60 Garden Street, Cambridge, MA 02138, USA} 

\author{Simon Coud\'e}
\affiliation{Department of Earth, Environment, and Physics, Worcester State University, Worcester, MA 01602, USA \href{mailto:istephens@worcester.edu}{istephens@worcester.edu}}
\affiliation{Center for Astrophysics $|$ Harvard \& Smithsonian, 60 Garden Street, Cambridge, MA 02138, USA} 

\author{Philip C. Myers}
\affiliation{Center for Astrophysics $|$ Harvard \& Smithsonian, 60 Garden Street, Cambridge, MA 02138, USA} 

\author{Catherine Zucker}
\affiliation{Center for Astrophysics $|$ Harvard \& Smithsonian, 60 Garden Street, Cambridge, MA 02138, USA} 

\author{James M. Jackson}
\affiliation{Green Bank Observatory,155 Observatory Road, Green Bank, WV 24944, USA}
%
%
\author{B-G Andersson}
\affiliation{McDonald Observatory, University of Texas at Austin, 2515 Speedway Boulevard, Austin, TX 78712, USA}
\author{Rowan Smith}
\affiliation{SUPA, School of Physics and Astronomy, University of St Andrews, North Haugh, St Andrews, KY16 9SS, UK}
\author{Archana Soam}
\affiliation{Indian Institute of Astrophysics, II Block, Koramangala, Bengaluru 560034, India}
%
%
%
\author{Patricio Sanhueza}
\affiliation{Department of Astronomy, School of Science, The University of Tokyo, 7-3-1 Hongo, Bunkyo, Tokyo 113-0033, Japan}
%
%
\author{Taylor Hogge}
\affiliation{Institute for Astrophysical Research, Boston University, 725 Commonwealth Avenue, Boston MA 02215, USA}
\author{Howard A. Smith}
\affiliation{Center for Astrophysics $|$ Harvard \& Smithsonian, 60 Garden Street, Cambridge, MA 02138, USA} 
\author{Giles Novak}
\affiliation{Center for Interdisciplinary Exploration and Research in Astrophysics (CIERA), and Department of Physics \& Astronomy, Northwestern University, 2145 Sheridan Rd., Evanston, IL 60208, USA}
%
%
\author{Sarah Sadavoy}
\affiliation{Department of Physics, Engineering and Astronomy, Queen's University, 64 Bader Lane, Kingston, ON, K7L 3N6, Canada}
\author{Thushara Pillai}
\affiliation{MIT Haystack Observatory, 99 MIllstone Road, Westford, MA, USA, 01826}
\author{Zhi-Yun Li}
\affiliation{Astronomy Department, University of Virginia, Charlottesville, VA 22904, USA}
\author{Leslie W. Looney}
\affiliation{Department of Astronomy, University of Illinois, 1002 West Green Street, Urbana, IL 61801, USA}
\author{Koji Sugitani}
\affiliation{Graduate School of Science, Nagoya City University, Mizuho-ku, Nagoya, Aichi 467-8501, Japan}
\author{Andr\'es E. Guzm\'an}
\affiliation{National Astronomical Observatory of Japan, National Institute of Natural Sciences, 2-21-1 Osawa, Mitaka, Tokyo 181-8588, Japan}
\author{Alyssa Goodman}
\affiliation{Center for Astrophysics $|$ Harvard \& Smithsonian, 60 Garden Street, Cambridge, MA 02138, USA} 
\author{Takayoshi Kusune}
\affiliation{Graduate School of Science, Nagoya University, Chikusa-ku, Nagoya 464-8602, Japan}
%
%
%
%
%
\author{Miaomiao Zhang}
\affiliation{Purple Mountain Observatory, and Key Laboratory for Radio Astronomy, Chinese Academy of Sciences, 210008 Nanjing, PR China} 
%
%
%
%
%
%
\author{Nicole Karnath}
\affiliation{Space Science Institute, 4765 Walnut St., Ste. B, Boulder, CO 80301, USA}
\affiliation{Center for Astrophysics $|$ Harvard \& Smithsonian, 60 Garden Street, Cambridge, MA 02138, USA}

\author{Jessy Marin}
\affiliation{Jodrell Bank Centre for Astrophysics, Department of Physics and Astronomy, University of Manchester, Oxford Road, Manchester, M13 9PL, UK}
\affiliation{Center for Astrophysics $|$ Harvard \& Smithsonian, 60 Garden Street, Cambridge, MA 02138, USA} 


\begin{abstract}
Stars primarily form in galactic spiral arms within dense, filamentary molecular clouds. The largest and most elongated of these molecular clouds are referred to as ``bones," which are massive, velocity-coherent filaments (lengths $\sim$20 to $>$100 pc, widths $\sim$1--2 pc) that run approximately parallel and in close proximity to the Galactic plane. While these bones have been generally well characterized, the importance and structure of their magnetic fields (B-fields) remain largely unconstrained. Through the SOFIA Legacy program FIELDMAPS, we mapped the B-fields of 10 bones in the Milky Way. We found that their B-fields are varied, with no single preferred alignment along the entire spine of the bones. At higher column densities, the spines of the bones are more likely to align perpendicularly to the B-fields, although this is not ubiquitous, and the alignment shows no strong correlation with the locations of identified young stellar objects. We estimated the B-field strengths across the bones and found them to be $\sim$30--150\,$\mu$G at pc scales. Despite the generally low virial parameters, the B-fields are strong compared to the local gravity, suggesting that B-fields play a significant role in resisting global collapse. Moreover, the B-fields may slow and guide gas flow during dissipation. Recent star formation within the bones may be due to high-density pockets at smaller scales, which could have formed before or simultaneously with the bones.
\end{abstract}

\section{Introduction} \label{sec:intro}
Most stars form within the molecular clouds of a galaxy's spiral arms, and parts of these star-forming clouds are filamentary with widths of $\sim$0.1\,pc \citep[e.g.,][]{Andre2010,Hacar2023}. Infrared dark clouds (IRDCs) within the Milky Way have been identified using a variety of telescopes, including the \emph{Infrared Space Observatory}, the \emph{Midcourse Space Experiment}, and the \emph{Spitzer Space Telescope} \citep[e.g.,][]{Perault1996, Price2001, Benjamin2003, Simon2006, Carey2009}. Among these, some IRDCs were found to be exceptionally elongated, such as Nessie \citep{Jackson2010,Goodman2014}, with large widths of approximately $\sim$1\,pc and lengths ranging from tens to over 100\,pc. These structures have been referred to as the ``bones of the Milky Way" since they are the densest and largest-scale coherent star-forming structures in the Milky Way \citep{Goodman2014}. \citet[][henceforth, Z15, Z18]{Zucker2015,Zucker2018b} identified 18 such bones, using the criteria that these bones lie mostly parallel and close to the Galactic plane, they are continuous and kinematically coherent, and they have aspect ratios exceeding 20:1. Since stars form in filamentary molecular clouds within spiral arms, studying star formation in these elongated bones provides a detailed view of how mass collects in the magnetized spiral potential to form stars. With the extensive catalog of ancillary data available for these bones, many of their properties such as length, width, temperature, and density have been well constrained. However, the magnetic fields (henceforth B-fields) within these bones, which could potentially guide the flow of mass and/or slow the formation of stars, remain poorly constrained.

The most common method for probing B-fields in star-forming clouds is through polarimetric observations of thermal dust emission. Dust grains align with their short axes along the B-field direction, resulting in their long axes being perpendicular to the B-fields \citep[e.g.,][]{Andersson2015}. Therefore, the thermal emission from these dust grains is polarized in a direction perpendicular to the B-fields. By mapping the polarization direction across an entire region, the plane-of-sky direction of the B-fields can be inferred by rotating all the polarization angles by 90$^\circ$. However, the B-field direction inferred from this method provides only a projected angle on the plane of the sky. By convention, we refer to these angles as B-field ``vectors," even though they are not true vectors.


At large scales in the interstellar medium toward Gould Belt clouds, polarimetric observations from \textit{Planck} have shown that the inferred B-fields tend to be more parallel to low-density elongated structures and more perpendicular to high-density elongated structures \citep{Planck35}. Simulations \citep{Soler2013,Planck35,SolerHennebelle2017} suggest that these observations are consistent with Alfv\'enic or sub-Alfv\'enic turbulence, indicating that B-fields are likely to be dynamically important. However, the resolution of these observations was coarse (10$\arcmin$ or 0.4\,pc for the nearest clouds) and did not resolve the filamentary structure within these clouds. A follow-up higher resolution observational analysis of a single Gould Belt cloud, L1688, showed that the transition from parallel to perpendicular alignment with increasing column density is consistent with \textit{Planck} \citep{LeeDennis2021}. Galactic bones, which are located much farther away than the Gould Belt clouds, have widths of only a few arcminutes and are therefore not resolved by \textit{Planck}. 

The Stratospheric Observatory for Infrared Astronomy (SOFIA), a Boeing 747SP aircraft equipped with a 2.7\,m telescope, was ideal for probing B-fields in bones. Specifically, its polarimeter, the High-resolution Airborne Wideband Camera Plus (HAWC+) \citep{Dowell2010,Harper2018}, was well-suited for tracing their B-fields because (1) it observed in the far-infrared where thermal dust emission from cold clouds is the brightest, (2) it could resolve the distant bone clouds with large maps, and (3) its sensitivity was sufficient to detect diffuse and dense dust emission.  Moreover, as a SOFIA instrument, HAWC+ could target all bones, as they lie in both the Northern and Southern skies. We utilized SOFIA/HAWC+ for an observational SOFIA Legacy program called FIlaments Extremely Long and Dark: a MAgnetic Polarization Survey (FIELDMAPS), which mapped B-fields for 10 of the 18 bones identified in the \citetalias{Zucker2018b} catalog. These 10 bones were selected because they had the most contrast with the background emission from the Galactic plane, allowing for them to be observed in a reasonable amount of time \citep[for details, see][henceforth, C25]{Coude2025}. The first results were published for the G47 bone in \citet[][henceforth, S22]{Stephens2022}. \citetalias{Stephens2022} found that B-fields are not always perpendicular to the G47 bone but tend to be perpendicular in the highest density regions where stars are forming. They also found that B-fields at pc scales tend to be 20–100\,$\mu$G, which is strong enough to inhibit gravitational collapse in most of the bone (i.e., subcritical), except possibly in the highest density regions where stars are forming. A separate analysis of G47 FIELDMAPS data by \citet{Jadhav2025} found similar results in that the energy density of the B-fields dominates over that of turbulence and gravity. \citet{Ngoc2023} found strong fields (subcritical) toward the Snake bone (also referred to as G11 or G11.11\,--\,0.12) using SOFIA data as well, which will be re-analyzed here with an improved data reduction. Analysis of the B-field strengths for all 10 bones will be investigated in this paper.

The data release for the FIELDMAPS survey is given in \citetalias{Coude2025}. \citetalias{Coude2025} used both FIELDMAPS and $Planck$ data to analyze the general alignment of fields with the Galactic plane and the large-scale, linear orientation of the bones. The $Planck$ data, which cannot resolve the widths of the bones given the resolution of 5$\arcmin$ ($\sim$2--8\,pc spatial resolution given the varying distances of the bones), showed that B-fields are primarily parallel to the Galactic plane, which is consistent with B-fields following the spiral arms found in both Galactic and extragalactic studies \citep[e.g.,][]{Clemens2020,Borlaff2021}. The SOFIA data, on the other hand, showed B-fields that were closer to random, with a very slight overall preference for fields to be perpendicular to the large-scale orientation of the bones, in stark contrast to that found with $Planck$. This contrast is consistent with the fact that the direction of B-fields in individual star-forming regions show no dependence on their location within the Galaxy \citep{Stephens2011}. Since bones exhibit significant curvature and deviate from the linear geometries adopted for simplicity in \citetalias{Coude2025}, a detailed analysis of the morphological alignment between B-fields and the bone structure is warranted, which will be investigated in detail in this paper.

This paper extends the analysis of \citetalias{Stephens2022} to all 10 bones. In Section~\ref{sec:obs}, we discuss the data used in the paper. In Section~\ref{sec:orientationbones}, we discuss the alignment between B-fields and the spines of the bones and discuss whether this alignment is related to where young stellar objects (YSOs) form. In Section~\ref{sec:sliding}, we analyze the effects of gravity and B-fields along the spine of the bone via a sliding box analysis. We discuss the findings in Section~\ref{sec:discussion} and summarize the findings in Section~\ref{sec:summary}.

\begin{figure*}
\begin{center}
 \includegraphics[width=0.85\textwidth]{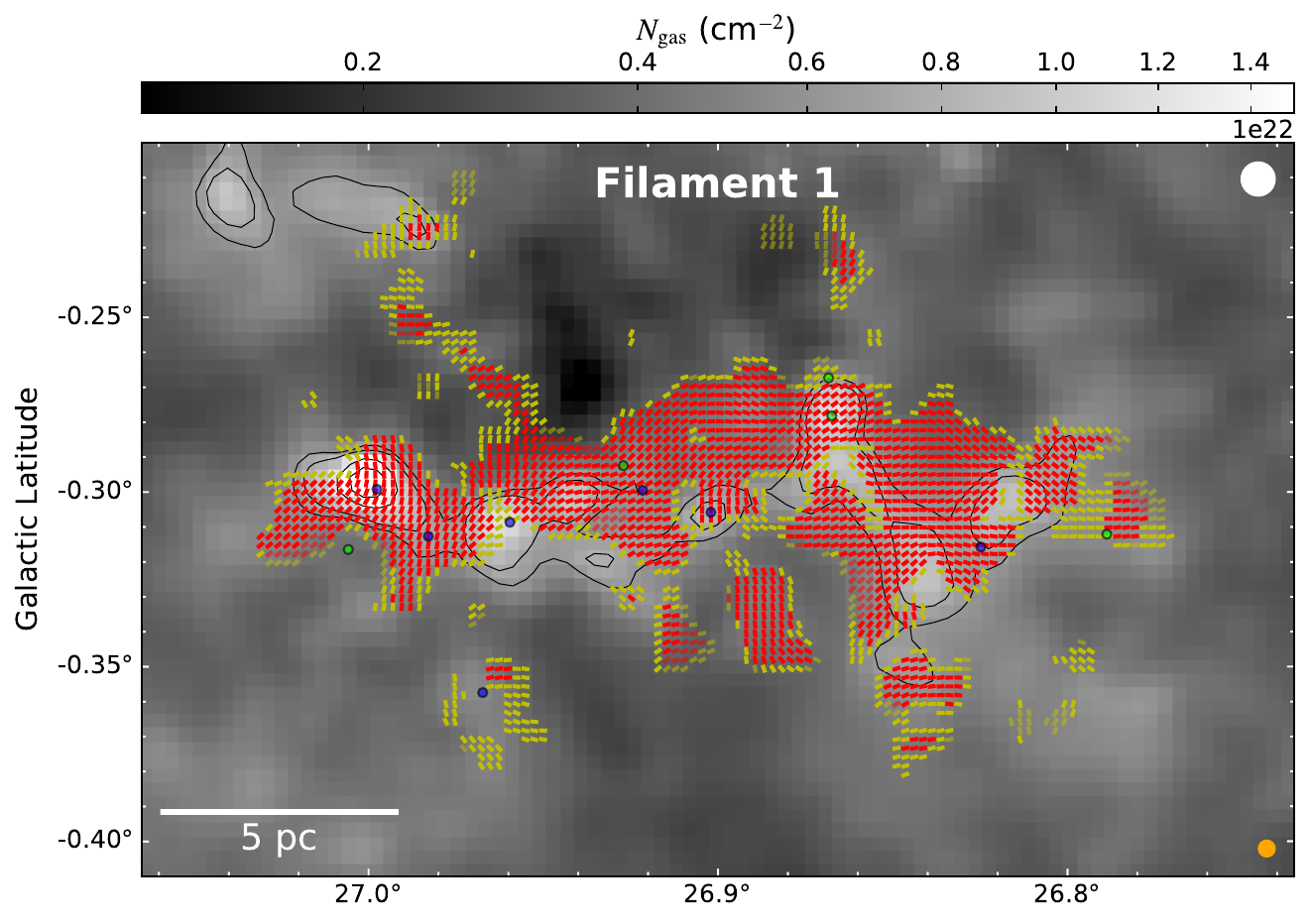}
 \includegraphics[width=0.85\textwidth]{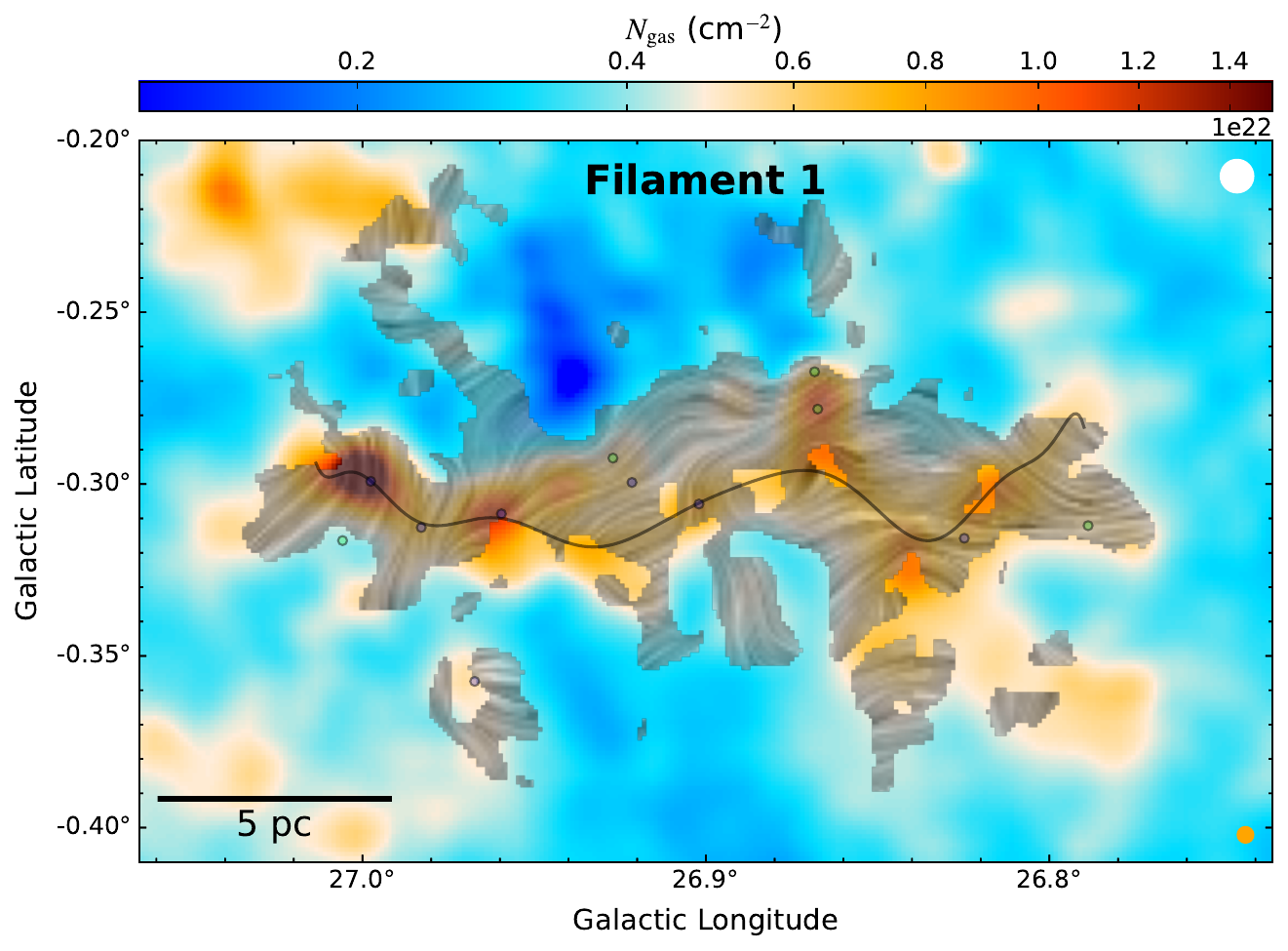}
 \end{center} \vspace{-16pt}
 \caption{
Top Panel: Filament~1 inferred B-field vectors overlaid on the Herschel-derived column density map, with contours for this figure and in Appendix~\ref{appendix:vectormaps} at [0.63, 0.75, 1.3, 1.8, 2.5, 5.0, 7.5]\,$\times$\,10$^{22}$\,cm$^{-2}$. Bright red and yellow vectors have measurements of \pfrac/\sigpfrac\,$>$\,3 and 2\,$\leq$\,\pfrac/\sigpfrac $\leq$\,3, respectively; faded vectors indicate where \pfrac\,$>$\,0.15. Blue and green circles mark known Class~I and~II YSOs, respectively (Section~\ref{sec:YSOs}).  Vectors are spaced by 2~HAWC+ pixels (9$\farcs$1), near Nyquist sampling. Bottom Panel: LIC map overlaid on column density, where the wavy pattern shows the B-field direction for \pfrac/\sigpfrac\,$>$\,1.5 and $I/\sigma_I$\,$>$\,10. The spine's polynomial fit is the solid line. The white beam (top right;  36$\farcs$4) shows the resolution for the column density map and the orange beam (bottom right; 18$\farcs$7) shows the resolution for SOFIA/HAWC+. \label{fig:fil1vectors}} 
 \end{figure*}

%

\section{Observations and Ancillary Data} \label{sec:obs}
\begin{deluxetable}{l@{}c@{}c@{}c@{}c@{}}
\tablecolumns{5}
\tabletypesize{\footnotesize}
\tablewidth{0pt}
\tablecaption{Bone Parameters and Polynomial fit Order \label{tab:fits}}
\tablehead{\colhead{Bone} & \colhead{Distance\tablenotemark{a}} & \colhead{FWHM size\tablenotemark{b}} & \colhead{Radius\tablenotemark{c}} & \colhead{Polynomial}\tablenotemark{d}  \vspace{-8pt} \\
\colhead{Name} & \colhead{(kpc)} & \colhead{(pc)} & \colhead{(pc)} & \colhead{Fit Order} 
}
\startdata
Filament 1 & 4.2\,$\pm$\,0.3 & 1.5 & 1.1 & 19 \\
Filament 2 & 3.8\,$\pm$\,0.4 & 1.1 & 0.84 & 22 \\
Filament 4 & 4.3\,$\pm$\,0.5 & 1.8 & 1.3 & 14 \\
Filament 5 & 4.0\,$\pm$\,0.4 & 1.3 & 0.95 & 11 \\
Snake\tablenotemark{e} & 4.1\,$\pm$\,0.2 & 1.4 & 1.0 & 35 \\
Filament 8 & 1.3\,$\pm$\,0.1 & 0.7 & 0.46 & 20 \\
Filament 10 & 2.9\,$\pm$\,0.2 & 1.2 & 0.84 & 45 \\
G24 & 5.3\,$\pm$\,0.6 & 1.7 & 1.3 & 12 \\
G47 & 4.2\,$\pm$\,0.7 & 1.1 & 0.83 & 24 \\
G49 & 5.21$^{+0.26}_{-0.23}$ & 1.4 & 1.0 & 33 
\enddata

\tablenotetext{a}{Distances are from \citetalias{Coude2025}. See \citetalias{Coude2025} for bone coordinates. The distances to the Snake and G49 are derived from water maser parallax measurements \citep{Wu2014,LiJJ2022}, while the remaining sources use kinematic distances.}
\tablenotetext{b}{Median, deconvolved FWHM widths of the filamentary bones from \citetalias{Zucker2018b}}
\tablenotetext{c}{Calculated from the FWHM width using the correction factor from \citet{vanHoof2000}.} 
\tablenotetext{d}{Polynomial fit order number used to fit the spine of each bone (see Section~\ref{sec:spines}).}
\tablenotetext{e}{Also called Filament~6 in \citetalias{Zucker2015,Zucker2018b}.}
\end{deluxetable}

We analyze SOFIA/HAWC+ polarization data from the FIELDMAPS survey in conjunction with ancillary data and parameters estimated in other studies. In the subsections below, we will describe details of the SOFIA observations, column density maps, the bone's spine orientation, molecular line data, and the YSO catalog. Extended details about the FIELDMAPS survey are given in the data release paper \citepalias{Coude2025}.

Table~\ref{tab:fits} lists the distances to the bones, which were updated in \citetalias{Coude2025} given new Galactic models \citep{Reid2019} and maser parallax measurements \citep{Wu2014,LiJJ2022}. Notably, the Snake, Filament~8, and G47 show substantial changes in their distances compared to \citetalias{Zucker2018b}. The table also lists the deconvolved Full Width at Half Maximum (FWHM) sizes, as discussed in detail in \citetalias{Zucker2018b}. The deconvolved FWHM sizes are based on the assumption of a cylindrical geometry for the bones, where Gaussians are fitted to their radial column density profiles, and the median FWHM value is assigned to each bone. The analysis in this paper needs the radius of the bone, which can be estimated from the FWHM using a correction factor based on an assumed underlying structure and the resolution of the fitted observations (43$\arcsec$ in \citetalias{Zucker2018b}). We use the correction factor from \citet{vanHoof2000} assuming a disk-like geometry, as fits across the width of `cylindrical' filaments are disk-like. The estimated radii are given in the fourth column of Table~\ref{tab:fits}. The radius for G47 listed here of 0.83\,pc differs from the 1.6\,pc used in \citetalias{Stephens2022}, which was based on the \citetalias{Zucker2018b} flattening radius from a Plummer fit to the bone, and assumed G47 was at a larger distance.

\subsection{SOFIA/HAWC+ Observations}\label{sec:sofia_obs}
This paper uses Band~E (214\,$\mu$m) HAWC+ polarization data from the SOFIA telescope, where we mapped polarization across the 10 bones listed in Table~\ref{tab:fits}. The data primarily come from the FIELDMAPS SOFIA Legacy Program (project code 08\_0186; PI: I. Stephens), although data for the Snake are from a combination of programs 05\_0206 (PI: T. Pillai) and 06\_0027 (PI: I. Stephens). Data presented in \citet{Ngoc2023} are only from the 06\_0027 program. All bones required multiple pointings. Data from 08\_0186 used scan mode (on-the-fly mapping), while the data for the Snake used the chop-nod technique. HAWC+'s pipeline uses the CRUSH package \citep{Kovacs2008a,Kovacs2008b}, and the imaging modes and calibration are discussed in detail in \citet{Harper2018}. The Level~0 data were reprocessed through the SOFIA data reduction pipeline to create Level~4 products. Details of the FIELDMAPS data reduction process is discussed in-depth in \citetalias{Coude2025}, including a comparison between data reduction of chop-nod and scan mode mapping toward a section of Filament~5. The comparison of the mapping modes showed consistency between angles, suggesting that the foreground/background subtraction does not change the polarization morphology. However, when comparing $Herschel$ and HAWC+ data, \citetalias{Coude2025} showed that there is considerable spatial filtering of Stokes~I data. The analysis of SOFIA data in this paper primarily focuses on polarization angles, which are considered robustly determined and are unaffected by how much we spatially filter Stokes~I. The native angular resolution of SOFIA/HAWC+ Band~E is 18$\farcs$2, but the pipeline applies a smoothing of 4$\farcs$274, resulting in a final resolution of 18$\farcs$7. Given that the distances to the bones vary from 1.3 to 5.21\,kpc, the spatial resolution ranges from 0.12 to 0.47\,pc, which is sufficient to resolve the radius of each bone.

In general, the HAWC+ maps cover the majority of the dense material in the bones. However, there are a few exceptions where large portions (about half or more) were not completely mapped. Filament~5 was not fully mapped because the eastern part would have been time prohibitive due to the sensitivity required for that region. Filament~8 was not fully mapped as the SOFIA telescope was decommissioned before the observations could be completed. Additionally, some bones, such as G47, G49, and Filament~10, have their tail ends unmapped, as the required integration times were long relative to the minimal science gain. Nevertheless, the B-fields were mapped across the vast majority of the solid angle covered by these bones.

The HAWC+ instrument uses a rotating half-wave plate to measure the Stokes parameters $I$, $Q$, and $U$. Stokes~$I$ represents the total intensity (both polarized and unpolarized emission), while Stokes~$Q$ and~$U$ parameterize the linear components of the polarization. Typical uncertainties in $I$, $Q$, and $U$ varied significantly from bone to bone and within even a bone itself. The polarized intensity, $P_I = \sqrt{Q^2+U^2}$, has a bias toward positive values since it cannot be negative \citep{Vaillancourt2006}. This paper uses the de-biased polarization fraction following \citetalias{Coude2025}: 
\begin{equation}
P_{\text{frac}} = \frac{\sqrt{P_I - \sigma_{P_I}}}{I} ,
\end{equation}
where $\sigma_{P_I}$ is the error on the polarized intensity. The error in the polarization fraction is calculated via propagation of error:
\begin{equation}
\sigma_{P_{\text{frac}}} = \frac{P_I}{I} \sqrt{\left( \frac{\sigma_{P_I}}{P_I} \right)^2 + \left( \frac{\sigma_I}{I} \right)^2},
\end{equation}
where $\sigma_I$ is the error on Stokes~$I$.

The polarization angle in the plane of the sky is calculated via

\begin{equation}
\chi = \frac{1}{2} \text{arctan}\left( \frac{U}{Q} \right).
\end{equation}
For the calculation of $\chi$, we used the \texttt{arctan2} function in Python, which selects the correct quadrant based on the signs of $Q$ and $U$. In this paper, we show the inferred B-field direction, which is $\chi$ rotated by 90$^\circ$, i.e., $\theta_B = \chi + 90^\circ$. $\theta_B$ is a position angle measured counterclockwise from Galactic North. We use data for \pfrac/\sigpfrac\,$>$\,2 for the majority of our analysis. We choose this threshold rather than \pfrac/\sigpfrac\,$>$\,3 because our analysis does not depend on individual measurements but rather on an ensemble of polarization data, and statistically, the vast majority of the 2\,$<$\,\pfrac/\sigpfrac\,$<$\,3 measurements are real detections. As given in \citetalias{Coude2025}, the median flux uncertainties for strongly detected polarization vectors (\pfrac/$\sigma_{P_{\text{frac}}}$\,$>$\,3, $I$/$\sigma_I$\,$>$\,10, and \pfrac\,$<$\,30\%) across each bone varied from 0.04 to 0.24\,MJy\,arcsec$^{-2}$ for Stokes~$I$ and 0.055 to 0.30\,MJy\,arcsec$^{-2}$ for Stokes~$Q$ and~$U$. 

As done in \citetalias{Stephens2022}, we rotated the equatorial data products into Galactic coordinates using the \texttt{reproject} Python package \citep{Robitaille2020} and recalculated the magnetic field position angles following \citetalias{Coude2025}. The pixel size of the maps is 4$\farcs$55, which is one quarter of the native 18$\farcs$2 resolution. The pixel size differs from the 3$\farcs$7 pixel size used in \citetalias{Stephens2022}. Toward the later SOFIA observation cycles, 4$\farcs$55 became the default Band~E size in the pipeline and archive, so we opted to use this size for the entire bone sample.

In this work, we assume that B-fields align the grains, and the inferred B-field is perpendicular to the polarization. Radiative alignment torque (RAT) theory, which is commonly assumed to be the alignment mechanism of grains with the B-field, requires significant radiation to align grains along the B-field \citep[e.g.,][]{Andersson2015}. Most of the bones have high extinctions even in the infrared due to high column densities, which suggests that grains are aligning only on the surface of the clouds unless there is an internal illuminating source. Determining whether and where the B-field is aligned in clouds requires a careful analysis of the \pfrac\ and column density, and such relationships are shown in \citetalias{Coude2025}. Nevertheless, such an analysis needs to be nuanced because, for example, B-fields at the highest column densities often exhibit fine substructure which may be unresolved within a SOFIA beam and would result in a lower value for \pfrac. A deeper analysis of grain alignment in the bones would require modeling that is beyond the scope of this paper. Given these uncertainties about which grains are being aligned along the line of sight, our assumed inferred B-fields are considered as the ``most likely" direction of the B-field for each SOFIA beam. Since we are analyzing these data primarily in a statistical manner (i.e., many thousands of Nyquist-sampled measurements for each bone; \citetalias{Coude2025}), anomalies at particular locations in the bones are not likely to affect the overall results.

An example of the polarimetric map for Filament~1 is shown in the top panel of Figure~\ref{fig:fil1vectors}.  The bottom panel of this figure shows the line integral convolution (LIC) maps \citep{Cabral1993} for each bone, along with the polynomial fit to the spine (see Section~\ref{sec:spines}). The wavy pattern in the LIC map shows the direction of the field morphology, which helps visualize how the field varies along the bone. The polarimetric and LIC maps are shown for the rest of the bones in Appendix~\ref{appendix:vectormaps} (Figures~\ref{fig:fil2vectors}\,--\,\ref{fig:g49vectors}). The polarimetric images in this paper are presented slightly differently than in \citetalias{Coude2025}, which also used Stokes~$I$ contours to highlight the HAWC+ data. Here, the emphasis is on the relationship between B-field vectors, column density, and the bone's spine. Fine details can be seen by zooming in on the figures.




\subsection{Column Density Maps}\label{sec:column}
We used column density maps derived from modified blackbody fits of \textit{Herschel} data from the Hi-GAL program \citep{Molinari2010}. Rather than using the Hi-GAL pipeline for deriving column density maps for the bones, as used in \citetalias{Stephens2022} (following \citetalias{Zucker2018b}), here we used the technique described in \citet{Guzman2015}, which uses a single-temperature modified blackbody dust emission model with an improved background estimation technique that more effectively removes diffuse foreground and background emission. The \citet{Guzman2015} model is specifically used to study more accurately these higher-density structures that are often cooler and faint at 70 microns. More details on the column density maps are provided in \citetalias{Coude2025}. In this paper, masses $M$ will be calculated from column density $N$ via $M$ = $\mu m_H N$, where $\mu$ is the mean molecular weight and $m_H$ is the mass of a hydrogen atom. We opt to use column density maps of \Ngas\ (uses mean molecular weight per particle, $\mu$\,=\,2.37) rather than \NHt\ (uses mean molecular weight per hydrogen molecule, $\mu$ = 2.8) \citep{Kauffmann2008} since \Ngas\ is technically the more direct value used for mass calculations. The \Ngas\ maps are in Galactic coordinates and have a resolution of 36$\farcs$4. For the sliding box analysis used in Section~\ref{sec:sliding}, we regrid all the column density maps to have the same pixel size as the HAWC+ maps using \texttt{reproject\_exact} from the \texttt{reproject} Python package \citep{Robitaille2020}. Uncertainties in column densities are discussed in \citet{Guzman2015}, but the assumed gas-to-dust mass ratio of 100 is likely the dominant source of uncertainty.

\subsection{Spines and Orientation of Bones}\label{sec:spines}
In this paper, we will infer the general orientation of the B-fields and each bone. We follow the method used for G47 in \citetalias{Stephens2022}. In this approach, the orientation of each bone is determined by fitting a polynomial to its so-called ``spine," which is defined in \citetalias{Zucker2018b}. This spine traces the center of the elongation of the bone of the $Herschel$ column density maps. The spine of the bone was selected using the medial axis skeletonization technique via FilFinder and RadFil python packages \citep{Koch2015,Zucker2018c}, which identifies the central part of the bone based on column density contours. The goal in \citetalias{Zucker2018b} for these spines was primarily to measure the bone lengths. However, for Filament~5 and Filament~10, this approach did not work for large sections of these bones, as there are places where the bones split into two different filaments. We have since modified the spines for these two bones via RadFil to more closely follow the higher-density filament where the bones split, rather than the center of the two combined filaments. The algorithm for the spine skeletonization technique generates a final product of a pixelated boolean map of the spine's location, with the spine having a width of a single pixel (11$\farcs$5\,$\times$\,11$\farcs$5 for \citetalias{Zucker2018b}). 

From the pixelated spine maps, we then fit the spine with a polynomial, as described in \citetalias{Stephens2022}, so that the orientation of the bone can be determined and compared to the B-field orientation. The order of the fitted polynomials is given in Table~\ref{tab:fits}. The orientation of the bone along the spine is defined by determining the instantaneous slope (i.e., the derivative) of the polynomial at each pixel. In this paper, these orientations along the spine will be referred to as $\theta_{\text{fil}}$, and are measured counterclockwise from Galactic North. The spine maps and their fits will be presented in Section~\ref{sec:orientations}.


\subsection{Molecular Line Data}\label{sec:molecularlines}
To calculate the B-field strength and virial parameters along the spine of a bone (Section~\ref{sec:sliding} below), we require knowledge of the velocity dispersion. All the bones lie within the boundaries of several molecular line surveys. Although it is unclear precisely where the polarized dust emission originates (i.e., cloud surface versus the highest density regions; see short discussion on grain alignment in Section~\ref{sec:sofia_obs}), the polarized emission almost certainly comes from the bones. As such, for estimating the velocity dispersion, we used spectral lines that are commonly used as high-density tracers, i.e., spectral lines that have high critical densities. In this paper, we use the spectral lines \nht, \tco, \ttcoto, \ceoto, which have similar critical densities (on the order of 10$^3$\,cm$^{-3}$), which is well-suited for tracing the entire bones since their densities are $\sim$10$^3$ to 10$^4$\,cm$^{-3}$. Spectral lines with higher critical densities will mainly just trace the density peaks. As discussed in \citet{Friesen2017}, NH$_3$ is an excellent high-density tracer for bones since this gas does not suffer from depletion onto dust grains and is much less likely to be optically thick compared to more abundant molecules like CO.

For six of the more northern bones, we used the high-density tracer ammonia NH$_3$(1,1) from the RAMPS survey \citep{Hogge2018}, which utilized the 100\,m Green Bank Telescope (GBT), providing NH$_3$(1,1) cubes at 32$\arcsec$ resolution. Data for the full RAMPS survey is available via the Green Bank Observatory's website\footnote{\url{https://greenbankobservatory.org/portal/gbt/gbt-legacy-archive/ramps-data/}}. Using the same spectral setup as RAMPS, we also mapped NH$_3$(1,1) for G49 under project code GBT23A-288 with the GBT. Observations were taken over 5 days in 2023 May, June, and November. For these observations, we created four overlapping tiles that were scanned using the K-band Focal Plane Array (KFPA), and these were sufficient for observing G49 in its entirety. Three of the tiles were 9$\arcmin$\,$\times$\,9$\arcmin$ and one was 6$\farcm$4\,$\times$\,6$\farcm$4 (i.e., about half the area of the bigger tiles), and each tile includes at least one track that was scanned along longitude and latitude. Regardless of the tile size, each tile was scanned for an hour. Before and after each hour scan, we did a pointing and focus with the GBT. The final tile maps were a bit larger than the sizes above, as the KFPA footprint provided noisy edges. We used the GBTIDL pipeline for data reduction and followed \cite{Hogge2018} for baseline subtraction. This program involved increased integration time across the bone, allowing for slightly higher NH$_3$(1,1) sensitivity for G49 compared to the RAMPS survey. The sensitivity of the final maps varied spatially since observations took place at different nights and some tiles were repeated more than others. The rms noise for the main beam temperature was $\sim$0.3\,--\,0.5\,K per 0.2\,\kms channel. 

While most of the northern bones have almost all of their spine mapped well with \nht, Filament~4 and G24 do not; a RAMPS track covering a large part of Filament~4 had exceptionally poor sensitivity, while G24 only had a very small section covered. The lack of high-density tracers for these two bones will be noted throughout the paper.

High-resolution NH$_3$(1,1) data were not available for Filament~8, and Filament~10. For Filament~8 and Filament~10, we used C$^{18}$O(2--1) spectral line data from the SEDIGISM survey \citep{Schuller2021} as the high-density tracer. While C$^{18}$O(2--1) is more likely to suffer from depletion, it is typically not optically thick compared to the more abundant istopologues of CO. These observations were obtained with the APEX telescope at 30$\arcsec$ resolution and had a channel width of 0.25\,\kms.

Emission from high-density tracers was not detected along the entire length of each bone, primarily due to the limited sensitivity of the surveys. In regions where high-density tracers were not detected, the temperatures are typically warmer and the densities lower. In these lower-density areas, CO is expected to become a suitable alternative for measuring velocity dispersions, as it is less depleted and has lower optical depths as compared to higher-density areas. Accordingly, for these sections we use the lower-density tracer $^{13}$CO, which was available at sub-arcminute resolution for all bones. The spectral line transition of $^{13}$CO was either $J\,=\,1-0$ \citep{Jackson2006} or $J\,=\,2-1$ \citep{Schuller2021}, depending on the available survey. For seven of the bones, we used Galactic Ring Survey \citep{Jackson2006} FCRAO $^{13}$CO(1--0) data at 46$\arcsec$ resolution with a channel width of 0.21\,\kms. For the Snake, Filament~8, and Filament~10, we used SEDIGISM $^{13}$CO(2--1) data at 30$\arcsec$ resolution and a channel width of 0.25\,\kms \citep{Schuller2021}. For the latter two bones, Filament~8, and Filament~10, $^{13}$CO(2--1) is largely unused in this paper due to the detection of C$^{18}$O(2--1) throughout the polarimetric maps. However, we compare the linewidths of the two in Appendix~\ref{appendix:linewidths}. The spectral lines and surveys used for each bone for both low- and high-density tracers are also listed in Table~\ref{tab:lines}.

The spectral lines were fitted for both the high- and low-density tracers as described for G47 in \citetalias{Stephens2022} and references therein. In many cases, multiple velocity components were present for a given pixel, which may or may not be associated with the specific bone. We first discarded any velocity component that did not match the bone's systemic velocity given in \citetalias{Zucker2018b} since their velocity is either too high or low to be associated with the bone. If multiple velocity components remained associated with the bone, we selected the velocity component with the highest amplitude as a proxy for the higher density part of each bone. 

For the spectral line fits, the velocity dispersion is crucial for estimating B-field strengths using the DCF technique described in Section~\ref{sec:sliding}. We create a single velocity dispersion map from both low-density and high-density tracers, which will be extensively used in Section~\ref{sec:sliding}. In this map, we first regrid all the original velocity dispersion maps to have the same pixel size as the HAWC+ maps with the \texttt{reproject} Python package \citep{Robitaille2020}, using the function \texttt{reproject\_adaptive} with the parameter \texttt{bad\_value\_mode = `ignore'.} Then in this map, we use the high-density tracer velocity dispersion when available for a given pixel; otherwise, we use $^{13}$CO. Although it would generally be preferable to use a single high-density tracer throughout, as discussed above high-density tracers are often not detected in parts of the bone with lower column densities. In these regions the high-density tracer lines are not effectively excited by collisions, making them harder to detect. Nevertheless, as will be discussed in Section~\ref{sec:sliding}, when estimating a velocity dispersion in a particular area of a bone, we heavily favor those velocity dispersions measured from the high-density tracer. 

A point by point comparison of linewidths along the spine of the bone are presented in Appendix~\ref{appendix:linewidths}. The appendix shows that the measured velocity dispersions for low-density tracers are usually higher than high-density tracers, typically by a factor of 2--3. As such, in this paper we do not analyze in detail the parameters that are measured with low-density tracers (i.e., B-fields, critical ratios, virial parameters, and equilibrium parameters in Section~\ref{sec:sliding}).



\begin{deluxetable}{l@{}c@{}c@{}c@{}}
\tablecolumns{4}
\tabletypesize{\footnotesize}
\tablewidth{0pt}
\tablecaption{Molecular Lines used to Calculate Velocity Dispersions \label{tab:lines}}
\tablehead{\colhead{Bone} & \colhead{Low-Density} & \colhead{High-Density} \vspace{-8pt} \\
\colhead{Name} & \colhead{Tracer} & \colhead{Tracer} & \colhead{Surveys}
}
\startdata
Filament 1 & \tco & \nht & GRS, RAMPS\\
Filament 2 & \tco & \nht & GRS, RAMPS\\
Filament 4 & \tco & \nht & GRS, RAMPS\\
Filament 5 & \tco & \nht & GRS, RAMPS\\
Snake & \ttcoto & \nht & SEDIGISM, RAMPS\\
Filament 8 & \ttcoto & \ceoto & SEDIGISM\\
Filament 10 & \ttcoto & \ceoto & SEDIGISM\\
G24 & \tco & \nht & GRS, RAMPS\\
G47 & \tco & \nht & GRS, RAMPS\\
G49 & \tco & \nht & GRS, This study
\enddata
\tablecomments{GRS: Galactic Ring Survey \citep[46$\arcsec$ resolution;][]{Jackson2006}; RAMPS: Radio Ammonia Mid-plane Survey \citep[32$\arcsec$ resolution][]{Hogge2018}; SEDIGISM: Structure, Excitation and Dynamics of the Inner Galactic Interstellar Medium \citep[30$\arcsec$ resolution;][]{Schuller2021}}
\end{deluxetable}

\subsection{YSOs}\label{sec:YSOs}
To assess the star formation population across the bones, we use previously identified YSOs from \citet{ZhangMM2019}. This survey identified Class~I, Flat, and Class~II YSOs based on their photometric spectral energy distributions from various Galactic surveys, including data from \emph{Spitzer} and \emph{Herschel}. Class~0 YSOs were all combined with Class~I YSOs since it is difficult to accurately separate the two based on infrared imaging alone. As was done in \citet{ZhangMM2019}, we combine Class~I and Flat protostars since this combination represents the embedded protostellar population. Due to the shorter integration times and limited resolution of the \emph{Spitzer} and \emph{Herschel} surveys along with the highly embedded nature of these objects, the catalogs of the youngest YSO populations in the bones are incomplete. Nevertheless, these catalogs provide valuable insights into recent star formation locations within the bones.

The YSOs in the Snake were not originally included in \citet{ZhangMM2019}. These YSOs were identified using a method similar to that of \citet{ZhangMM2019}, but with the \citetalias{Zucker2018b} mask to define the boundary of the Snake.


\begin{figure*}
\begin{center}
 \includegraphics[width=\textwidth]{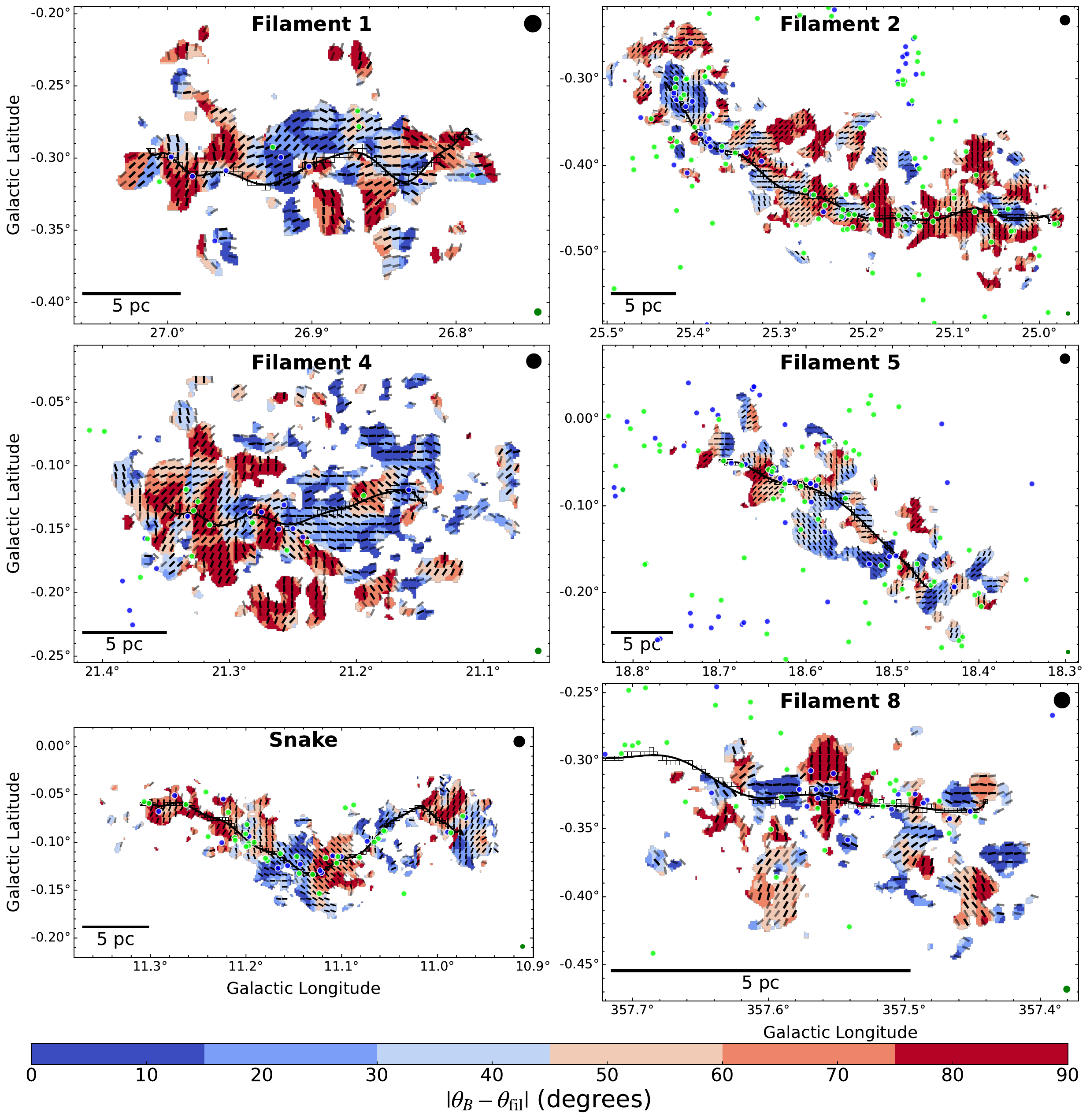}
 \end{center} \vspace{-16pt}
 \caption{B-field and spine angle difference, $|\theta_B -\theta_{\text{fil}}|$, shown in color-scale for $P_{\text{frac}}$/$\sigma_{P_{\text{frac}}} > 2$ for six of the bones (the other 4 are in Figure~\ref{fig:diffs2}). Squares show the spine pixels of the bones, and the gray line shows the polynomial fit to these pixels (Section~\ref{sec:spines}).  Black vectors have measurements of \pfrac/\sigpfrac\,$>$\,3, while faded black (gray) vectors are have 2\,$\leq$\,\pfrac/\sigpfrac $\leq$\,3. Blue and green circles show the location of known Class~I and~II YSOs, respectively (Section~\ref{sec:YSOs}).   The black beam on the top right is the column density resolution (36$\farcs$4) while the green beam on the bottom right is the SOFIA resolution (18$\farcs$7). Vectors are shown are separated by 5~HAWC+ pixels (22$\farcs$75 or $\sim$1.25 beams) to help better visualize the B-field morphology.  \label{fig:diffs1}}
 \end{figure*}
 
 \begin{figure*}
\begin{center}
 \includegraphics[width=\textwidth]{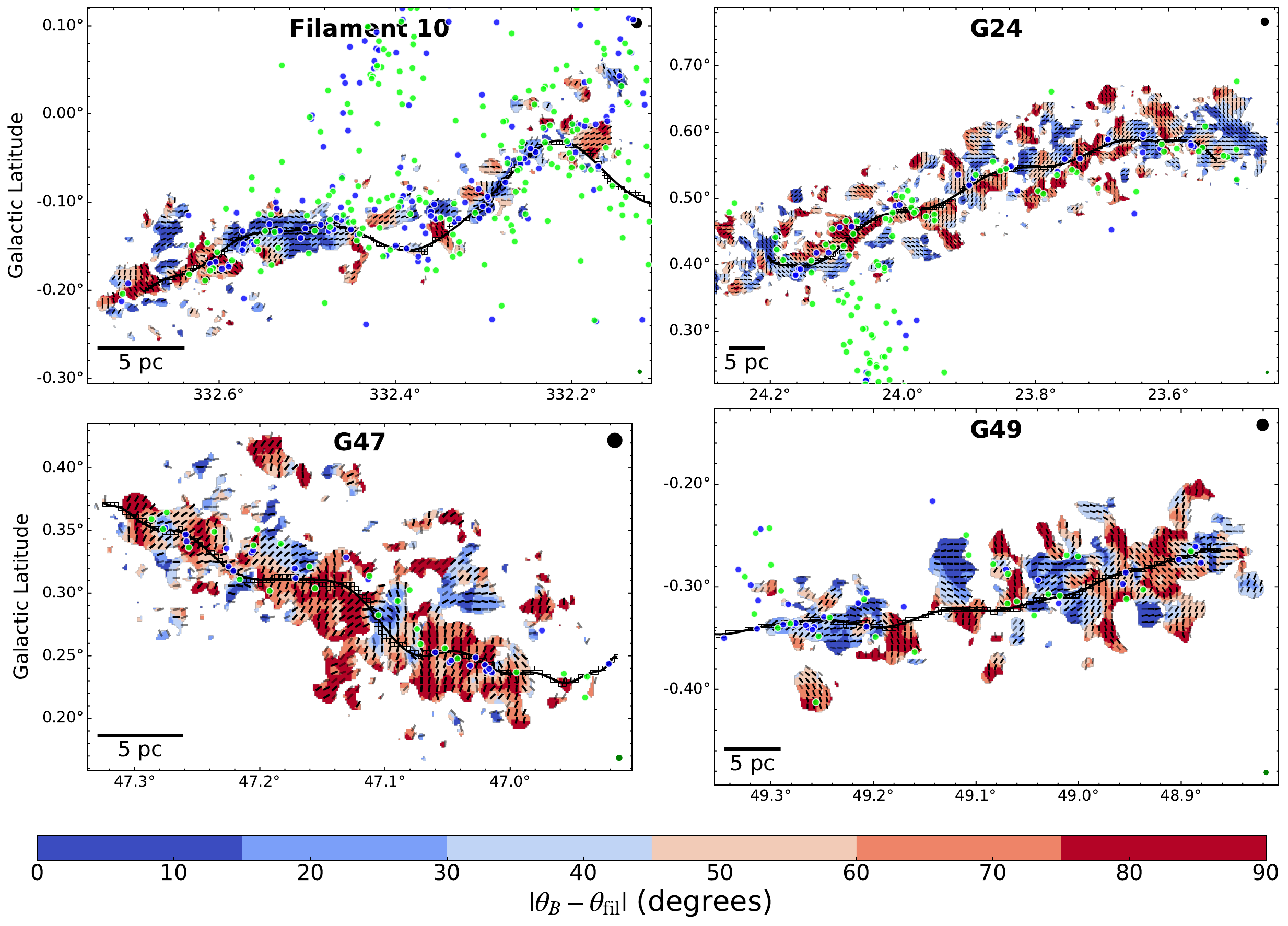}
 \end{center} \vspace{-16pt}
 \caption{B-field and spine angle difference, $|\theta_B -\theta_{\text{fil}}|$, for four of the bones. Caption is the same as Figure~\ref{fig:diffs1}, which shows the other six of the 10 bones.   \label{fig:diffs2}}
 \end{figure*}

\begin{figure*}[ht!]
\begin{center}
\includegraphics[width=2\columnwidth]{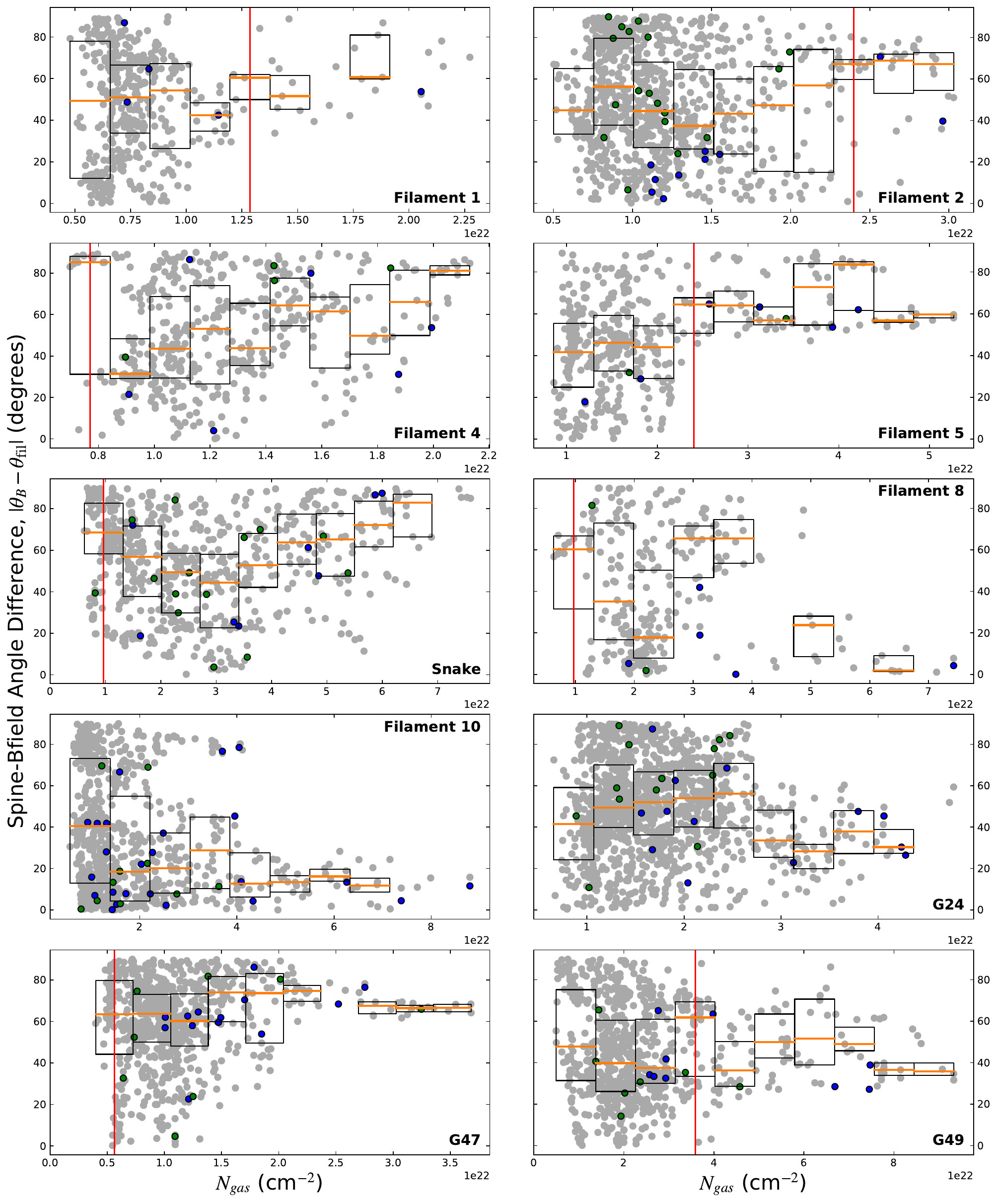}
\end{center}
\caption{\Ngas\ versus $|\theta_B -\theta_{\text{fil}}|$ for each bone, shown for every other HAWC+ pixel (approximately Nyquist sampled). Only data within 23$\arcsec$ of the bone (two spine pixels) are considered. The data is binned into 10 equal \Ngas\ ranges, and box plots are shown if there are at least 4 points in the bin. The box plots show the first and third quartiles in black and the median in orange. A red vertical line is shown for the first \Ngas\ box plot bin where $|\theta_B -\theta_{\text{fil}}|$\,$>$\,60$^\circ$, which indicates the first sign of perpendicularity with \Ngas. Blue and green circles show the \Ngas\ and $|\theta_B -\theta_{\text{fil}}|$ values at the nearest pixel for the known Class~I and~II YSOs, respectively. } 
\label{fig:columndiff} 
\end{figure*}

\begin{figure*}[ht!]
\begin{center}
\includegraphics[width=2\columnwidth]{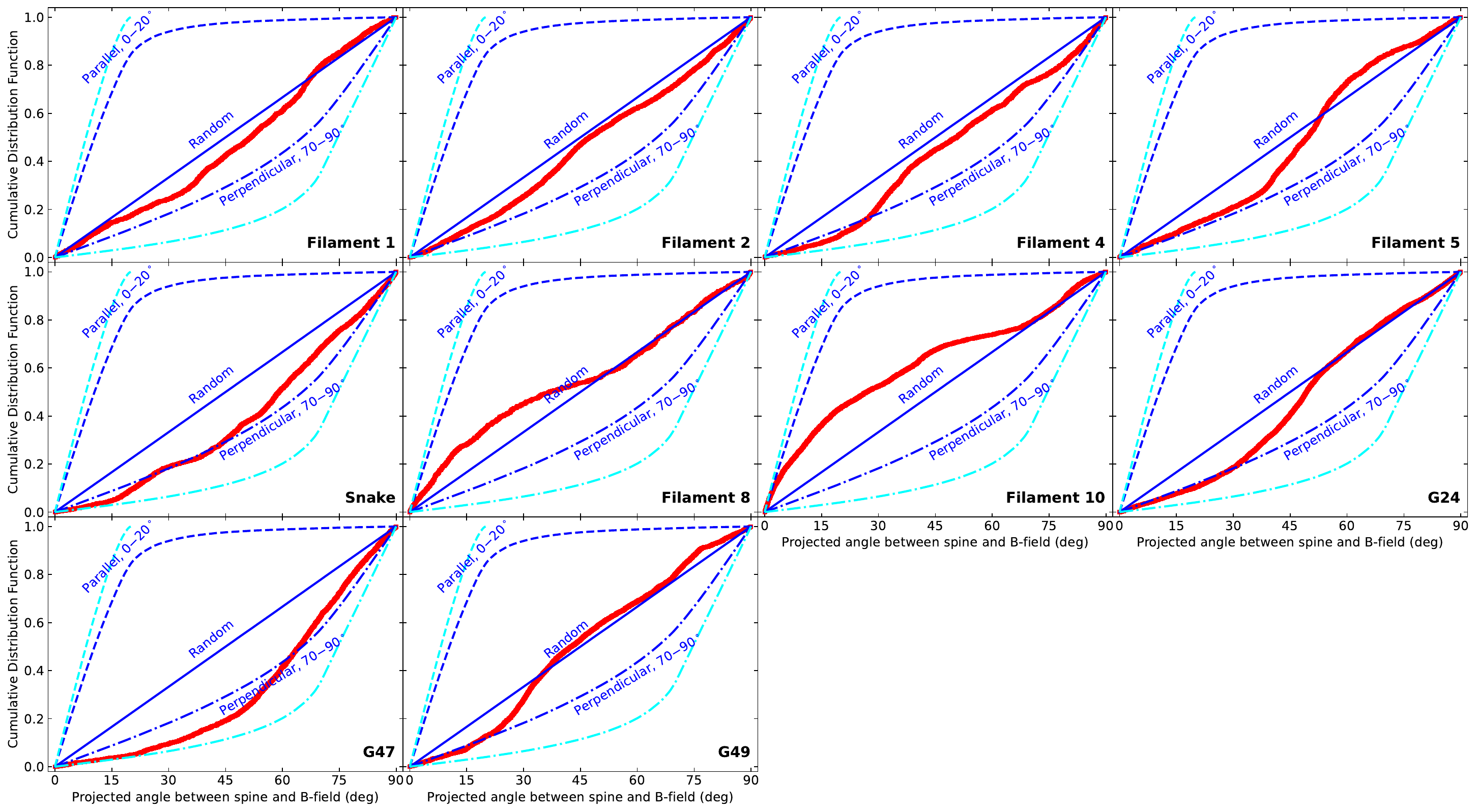}
\end{center}
\caption{Cumulative distribution function of the angle difference between each bone's spine and B-field direction, $|\theta_B -\theta_{\text{fil}}|$. Only pixels within 23$\arcsec$ of the spine are considered. The red line in each panel shows the results for that particular bone. The blue lines in each panel show a Monte Carlo simulation of the expected distribution of projected angles if the spine and B-field directions in 3D are mostly parallel (within $0^\circ$ to $20^\circ$; dashed blue line), mostly perpendicular (within $70^\circ$ to $90^\circ$; dashed-dotted blue line), or random (solid blue line). The cyan lines show the expected projected angles if the bone is fixed in the plane-of-the-sky for mostly parallel (dashed cyan line) and mostly perpendicular (dashed-dotted cyan line).  } 
\label{fig:cdfs} 
\end{figure*}

\section{Field Morphology Along the Bone}\label{sec:orientationbones}
\subsection{Comparing Magnetic Field Orientation, Bone's Morphology, and Density}\label{sec:orientations}
\citetalias{Coude2025} analyzed whether or not inferred B-fields are aligned with the bones, based on both $Planck$ and FIELDMAPS HAWC+ observations. The position angle of the bones was fit via a simple linear fit and then compared to the B-field directions. The $Planck$ B-field maps are sensitive to much of the diffuse emission outside of the bones given its large beam (5$\arcmin$), while the HAWC+ maps are more sensitive to the local magnetic field. For $Planck$, B-fields are parallel to the elongation of the bones, consistent with B-fields being aligned with the spiral arms. However, for HAWC+, B-fields show a slight preference for being perpendicular to the elongation of the bones, although with a significant variance in orientations locally. Here, we analyze the B-field alignment with each bone more extensively, focusing on how it changes locally along the spine of each bone.

As mentioned before, Figure~\ref{fig:fil1vectors} and those in Appendix~\ref{appendix:vectormaps} show the inferred B-field direction for each bone from our HAWC+ polarization data. For each bone, there is no single preferred direction of B-fields relative to the bone's morphology. However, at the same time, fields are usually well-ordered locally and can have small angle dispersions on scales of several pc.

To analyze the general alignment of the B-field orientation with the local morphological direction of each bone, we follow the method described below, which was also used for G47 in \citetalias{Stephens2022}. In this analysis, we first match the pixel-by-pixel inferred B-field directions ($\theta_B$), as determined from SOFIA/HAWC+ observations, to the nearest spine pixel (see Section~\ref{sec:spines}) of the bone. We then take the bone's orientation, $\theta_{\text{fil}}$, at this spine pixel and compute the magnitude of the difference, $|\theta_B - \theta_{\text{fil}}|$. Figure~\ref{fig:diffs1} and~\ref{fig:diffs2} show the resulting maps of $|\theta_B - \theta_{\text{fil}}|$ for all 10 bones. The color-bar is in 6 discrete bins of 15$^{\circ}$ each, where bluer colors indicate locations where the B-fields are more parallel to the bone's spine, while redder colors are more perpendicular. In Figure~\ref{fig:diffs1} and~\ref{fig:diffs2}, the B-fields are observed to be a mix of parallel, perpendicular, and everything in between when following the polynomial spine fit for each bone.

While Figure~\ref{fig:diffs1} and~\ref{fig:diffs2} provide a visual representation of the angle between the spine and the B-field, they do not directly provide a statistical inference of the alignment nor does it account for the projection of these angles in the plane of the sky. In particular, we are interested in alignment with the filament's spine since the spine tends to be higher column density where stars are more likely to form. We first extract a subset of the data to analyze only the B-field vectors close to the spine of the bone. We restrict the data to HAWC+ pixels within 23$\arcsec$ (the size of two 11$\farcs$5 spine pixels) of the bone's spine, as this distance effectively resolves both the spine and the B-field map. Since bones are at different distances, the analysis traces different physical scales, but this method gives consistency in statistics given the resolution of the observations. For the analysis of the $|\theta_B - \theta_{\text{fil}}|$ distributions below, we consider all HAWC+ pixels in this subset with $P_{\text{frac}}$/$\sigma_{P_{\text{frac}}} > 2$.

In \citetalias{Stephens2022}, visual inspection of G47 indicated that B-fields tend to be more perpendicular at higher densities, with these regions also showing a higher concentration of Class~I YSOs. Based on the vector (Figure~\ref{fig:fil1vectors} and Appendix~\ref{appendix:vectormaps}) and difference maps (Figure~\ref{fig:diffs1} and~\ref{fig:diffs2}), this trend is not readily apparent in the other bones. We will first analyze how the alignment $|\theta_B - \theta_{\text{fil}}|$ varies with column density, and then consider the locations of YSOs. The results are plotted in Figure~\ref{fig:columndiff}. We show $|\theta_B - \theta_{\text{fil}}|$ against column density for every other HAWC+ pixel (approximately Nyquist sampled) to capture all the details of the morphology along the spine without excessive oversampling. To help show the general trend of alignment as a function of column density, we divide each plot into 10 equal-sized ranges based on column density\footnote{Points at the minimum and maximum column densities are excluded from the bins due to rounding.}, and a box plot is drawn for all data within each range. No box plot is shown if there are fewer than 4 data points. The box plots show the first and third quartiles in black and the median in orange. For the lowest column density bin where the median of $|\theta_B - \theta_{\text{fil}}|$ is above 60$^\circ$ (a preference for perpendicular alignment), we draw a vertical red line. This line helps guide the eye where perpendicular alignment might transition due to the change in column density. 

Based on Figure~\ref{fig:columndiff}, most bones show a high interquartile range at lower column densities, suggesting that alignment with the bone is random at these column densities. At higher column densities, some bones exhibit a preference for perpendicular alignment between the spine and B-fields. Filament~5 and G47 strongly show this preference, while Filament~1, Filament~2, Filament~4, and the Snake also suggest this preference. In contrast, Filament~8, Filament~10, G24, and G49 do not show a preference for perpendicular alignment at high column densities, with the former two (Filament~8 and Filament~10) even showing preference for parallel alignment. Since all bones span similar ranges of column densities, these varying preferences indicate that there is no definitive column density at which B-field lines become perpendicular to the bones. This fact is supported by the red vertical line in Figure~\ref{fig:columndiff}, which occurs at a variety of column densities. \citet{Planck35} investigated a similar relationship in 10 nearby star-forming clouds but had a lower spatial resolution ($>$0.4\,pc) and therefore could not resolve filaments ($\sim$0.1\,pc in size). Instead, their comparison was with dense elongations. They found that most regions preferred B-fields to be parallel to dense elongations at low column densities (log($N_H [\text{cm}^2]) \sim 21 - 21.5$), with either no alignment or a perpendicular preference at higher column densities. The transitory regime from parallel to non-parallel\footnote{\citet{Planck35} did not find that all regions showed perpendicular alignment, so we use ``not parallel" to describe this range.} varied among these clouds, with an interquartile range for log($N_H [\text{cm}^2])$ of 21.765 to 22.15. Our observations probe the transitory and higher-density regimes rather than the parallel regime noted by \citet{Planck35}. The mix of no alignment and perpendicular alignment at higher densities in nearby star-forming clouds is consistent with our findings for the alignment with the elongation of the bones. However, Filament~8 and Filament~10 preference for some parallel alignment at high column densities was not directly found in \citet{Planck35}, possibly because the resolution of $Planck$ ($>$0.4\,pc) is much larger than the widths of the filaments in those clouds ($\sim$0.1\,pc).

The above analysis does not take into account projection effects. Following previous work (e.g., \citealt{Hull2013,Hull2014,Lee2015}; see \citealt{Stephens2017b} for the methodology we follow), we create a Monte Carlo simulation of a million pairs of random angles on the surface of a unit sphere. From this distribution, we extract angles that are mostly parallel (angle differences $|\theta_B - \theta_{\text{fil}}|$ between 0$^\circ$ and 20$^\circ$) and mostly perpendicular ($|\theta_B - \theta_{\text{fil}}|$ between 70$^\circ$ and 90$^\circ$). We then project all angles onto the plane of the sky to determine the observed angle distributions for mostly parallel, random, and mostly perpendicular cases. In Figure~\ref{fig:cdfs}, we plot the resulting cumulative distribution functions (CDFs) for $|\theta_B - \theta_{\text{fil}}|$ for each bone. The observed distributions are shown in red, and the results of the Monte Carlo simulations are shown in blue. In cyan, we repeat the Monte Carlo simulations, but now we fix the bone in the plane of the sky and randomize the B-field vectors only. We show this because the bones are typically more aligned in the plane of the sky than along the line of sight. Total parallel and perpendicular alignment for the entire bones are likely between the blue and cyan curves. 

Since large-scale Galactic B-fields typically trace spiral arms, a bias may exist for B-fields probed by SOFIA to also lie in the plane of the sky. In this case, projection effects would be minimal, producing CDF curves even farther from the blue and cyan curves for parallel and perpendicular. However, the CDFs and data do not show this preference, and B-field directions in star-forming clouds are usually independent of Galactic location \citep{Stephens2011}. Moreover, \citetalias{Coude2025} showed that the large-scale parallel B-field traced by $Planck$ is very different from the B-fields found with FIELDMAPS. Together, these results imply that the most appropriate assumption in the Monte Carlo simulation is that the B-fields are random on the unit sphere.

Based on Figure~\ref{fig:cdfs}, the B-field vectors do not fit exactly one of the Monte Carlo simulations for parallel, random, or perpendicular alignments. However, the location of the observed CDFs indicates whether B-fields tend to be more parallel or perpendicular compared to random. The alignment of B-fields with the spine is generally more perpendicular than random, especially for G47, whose CDF resembles what would be expected for perpendicular alignment. The two exceptions are Filament~8 and Filament~10, which show a slight tendency to be more parallel. Visual inspection of these two bones in polarimetric maps (Figures~\ref{fig:fil8vectors} and~\ref{fig:fil10vectors}) and their difference maps (Figure~\ref{fig:diffs1} and~\ref{fig:diffs2}) reveal that large areas of these bones have projected B-fields that are quite parallel to the bone's spine. 

Several bones show features of either being somewhat parallel or perpendicular at the beginning of the CDF, and later converges to be along the random line. Filament~5, Filament~8, Filament~10, G24, and G49 especially show these behaviors. In part, this may be due to the fact that we are looking at a range of column densities. In Appendix~\ref{appendix:cdfs}, we show the same CDFs, but now separated in four quartiles based on the column densities at the location of each angle comparison. The results are largely consistent with what is presented in Figure~\ref{fig:columndiff}, as discussed above.

Furthermore, there may be some bias on the results based on Galactic locations. Based on \citet{Reid2019}, if bones follow spiral arms, the majority of the bones would be mostly in the plane-of-the-sky. The exceptions are G47 and G49, which lie in the Sagittarius--Carina Arm roughly at the tangent point of our line of sight. Given the fact that G47 and G49 have large spatial extents, they may be spurs off the main spiral arms. Moreover, \citet{Zucker2015, Zucker2018b} identified Filament~4 and G24 to likely be interarm bones. Overall, our analysis does not find anything substantially different with these spur-like or interarm bones, albeit our sample size in these categories are too small to draw any firm conclusions. Finally, the two bones that have a substantial amount of B-fields that are parallel to the spine, Filament~8 and Filament~10, happen to be the only bones that we observed in Quadrant 4 of the Galaxy. Under the assumption that exactly two of the ten bones have parallel B-fields assigned at random, the probability that both Quadrant 4 bones (Filament~8 and~10) are parallel while none of the eight Quadrant 1 bones are is 1/45 ($\sim$2\%). However, given that only two Quadrant bones were observed in FIELDMAPS and that Filament~8 was not fully mapped, we cannot make a firm conclusion about a Galactic bias here.

\begin{deluxetable}{lccccc}
\tablecolumns{6}
\tabletypesize{\footnotesize}
\tablewidth{0pt}
\tablecaption{Median Column Densities for Bones and their YSOs \label{tab:YSOcolumn}}
\tablehead{\colhead{Bone} & All & \multicolumn{2}{c}{\underline{Class I YSOs}} & \multicolumn{2}{c}{\underline{Class II YSOs}} \\ 
\colhead{Name} & \Ngas& \colhead{Total} & \colhead{\Ngas} & \colhead{Total} & \colhead{\Ngas} 
}
\startdata
Filament 1 & 0.72 & 5 & 0.83 & 0 & -- \\
Filament 2 & 1.1 & 10 & 1.4 & 18 & 1.1 \\
Filament 4 & 1.2 & 6 & 1.4 & 4 & 1.4 \\
Filament 5 & 1.5 & 7 & 3.1 & 2 & 2.6 \\
Snake & 2.0 & 8 & 4.0 & 14 & 2.7 \\
Filament 8 & 1.8 & 5 & 3.1 & 2 & 1.7 \\
Filament 10 & 1.2 & 24 & 2.1 & 10 & 1.6 \\
G24 & 1.5 & 15 & 2.1 & 13 & 1.7 \\
G47 & 0.87 & 13 & 1.5 & 8 & 1.2 \\
G49 & 1.9 & 9 & 2.9 & 7 & 2.0 \\
\enddata
\tablecomments{All values of \Ngas\ are medians and are in units of $\times$10$^{22}$\,cm$^{-2}$. }
\end{deluxetable}

\subsection{Bone-Field Alignment at Locations of YSOs}
We also investigate where the YSOs are forming and whether their locations depend on $|\theta_B - \theta_{\text{fil}}|$. First, we examine whether the locations of YSOs correlate with \Ngas. To do this, we compare the typical \Ngas\ across each bone to the \Ngas\ column density at the locations of Class~I and Class~II YSOs. We mask out all \Ngas\ pixels and YSOs more than 23$\arcsec$ from the spine to focus on regions closest to the bone. For each bone, we calculate the median \Ngas\ of the bone and compare it to the median \Ngas\ at the exact locations of Class~I and Class~II YSOs. The results are summarized in Table~\ref{tab:YSOcolumn}. In all bones, Class~I YSOs are more likely to be found at higher \Ngas\ values than the median \Ngas\ for their bone. Class~II YSOs also tend to be found at higher \Ngas\ values, though not as high as Class~I YSOs. The median \Ngas\ for Class~I YSOs is typically 1.6 times higher than the median \Ngas\ for the entire bone, while Class~II YSOs are 1.2 times higher. Class~I YSOs are generally located in the upper quartile of the column density distribution, whereas Class~II YSOs are not. These results suggest that as YSOs evolve, they may migrate away from the bone and/or the bone itself may disperse.

\begin{deluxetable}{lccc}
\tablecolumns{4}
\tabletypesize{\footnotesize}
\tablewidth{0pt}
\tablecaption{Field and Spine Alignment at YSO locations \label{tab:YSOalignment}}
\tablehead{ & \colhead{Total} & \multicolumn{2}{c}{$|\theta_B -\theta_{\text{fil}}|$}  \vspace{-8pt} \\
\colhead{Group of Bones} & \colhead{YSOs} & \colhead{$>$45$^\circ$} & \colhead{$>$60$^\circ$} 
}
\startdata
All bones, Class I & 102 & 44\%\,$\pm$\,5\% & 29\%\,$\pm$\,5\%\\
All bones, Class II & 78 & 56\%\,$\pm$\,6\% & 40\%\,$\pm$\,6\%\\ 
All Pixels, all bones & & 59.8\%\,$\pm$\,0.6\% & 40.3\%$\pm$\,0.6\% \\
All perp. bones, Class I  & 49 & 61\%\,$\pm$\,7\% & 43\%\,$\pm$\,7\%\\ 
All perp. bones, Class II  &  46 & 63\%\,$\pm$\,7\% & 43\%\,$\pm$\,7\%\\ 
All perp. pixels & & 63.1\%\,$\pm$\,0.8\% & 43.0\%$\pm$\,0.8\% \\
Very perp. bones, Class I  & 20 & 85\%\,$\pm$\,8\% & 60\%\,$\pm$\,11\%\\ 
Very perp. bones, Class II  & 10 & 60\%\,$\pm$\,15\% & 40\%\,$\pm$\,15\%\\ 
All very perp. pixels & & 72.2\%\,$\pm$\,1.3\% & 46.8\%$\pm$\,1.4\% 
\enddata
\tablecomments{``Total YSOs" indicates the number of YSOs within 23$\arcsec$ of the bone's spine for each particular group. ``All perp. bones" includes Filament~1, Filament~2, Filament~4, Filament~5, the Snake, and G47, while ``Very perp. bones" include only Filament~5 and G47. For all bones, all perp. bones, and all very perp. bones, we also provide the group's B-field-spine alignment statistics for all pixels within 23$\arcsec$ of its spine.}
\end{deluxetable}

To assess whether or not star formation depends on $|\theta_B - \theta_{\text{fil}}|$, in Figure~\ref{fig:columndiff} we plot Class~I and Class~II YSOs within 23$\arcsec$ of the spine, using their corresponding column densities and $|\theta_B - \theta_{\text{fil}}|$ differences at these locations. No obvious correlation is apparent across all bones. We quantify the frequency of B-field-spine perpendicularity at the locations of Class~I and Class~II YSOs in Table~\ref{tab:YSOalignment} for three categories: (1) all bones, (2) bones with mostly perpendicular B-fields at high column densities (Filament~1, Filament~2, Filament~4, Filament~5, the Snake, and G47; ``All perp. fields"), and (3) bones with very perpendicular B-fields at high column densities (Filament~5 and G47; ``Very perp. fields"). We evaluate both ``any” perpendicularity (angles $>$45$^\circ$) and ``moderate” perpendicularity (angles $>$60$^\circ$). Table~\ref{tab:YSOalignment} also provides the percentages of ``any” and ``moderate” perpendicularity for all pixels within 23$\arcsec$ of the spine in each of the three groups of bones. These percentages represent the expected values if the YSOs were distributed randomly within each group. The errors for each percentage $p$ are calculated via $\sqrt{p(1-p)/s_n}$, where $s_n$ is the number of YSOs for each subsample.

If YSOs tend to form at locations of perpendicular B-fields, we would expect to see a higher degree of perpendicularity for Class~I YSOs compared to Class~II YSOs, as younger YSOs should be closer to their birthplace. However, for the entire sample of bones, there is no significant preference for YSOs to form at locations of B-field perpendicularity. In the “All perp. fields” category, there is a slight tendency towards perpendicularity, and this preference is somewhat stronger in the “Very perp. fields” category. Despite this, the difference in perpendicularity between Class~I and Class~II YSOs across all three groups is not statistically significant, showing no more than $\sim$1$\sigma$ deviation from random. 

These results suggest that YSOs do not show a strong preference for the alignment of B-fields and filaments. YSOs form at high column densities, and if B-fields indicate the direction of gas flow, the location of YSOs are independent of whether the flows are perpendicular or parallel to bones.

\subsection{Percentage of Light Polarized along Spines}\label{sec:Pfrac}
\begin{deluxetable}{lcccc}
\tablecolumns{6}
\tabletypesize{\footnotesize}
\tablewidth{0pt}
\tablecaption{Median Percentage of Polarized Light for Column Density Cutoffs \label{tab:Pfrac}}
\tablehead{\colhead{Bone} & No Cutoff & $>$1\,$\times$\,10$^{22}$  & $>$1.5\,$\times$\,10$^{22}$  & $>$2\,$\times$\,10$^{22}$   \vspace{-8pt} 
}
\startdata
Filament 1 & 3.0 & 2.0 & 2.0 & 1.9 \\
Filament 2 & 3.5 & 3.1 & 2.1 & 1.1 \\
Filament 4 & 3.7 & 3.4 & 3.0 & 3.4 \\
Filament 5 & 4.6 & 4.4 & 2.9 & 2.0 \\
Snake\tablenotemark{a} & 4.2 & 4.1 & 4.0 & 3.9 \\
G24 & 2.8 & 2.8 & 2.5 & 2.1 \\
G47 & 3.8 & 3.3 & 2.4 & 2.1 \\
G49 & 3.4 & 3.4 & 2.8 & 2.5 \\
Filament 8 & 4.0 & 3.2 & 3.6 & 3.5 \\
Filament 10 & 3.3 & 3.1 & 2.9 & 2.7 
\enddata
\tablecomments{All values are median P$_\%$, the percent of light that is polarized across the bone. Only pixels within 23$\arcsec$ of the bone's spine are considered and every other pixel is skipped. Different cutoffs are based on \Ngas, with units of cm$^{-2}$. }
\tablenotetext{a}{The Snake uses chop-nod, while the rest of the bones on-the-fly mapping.} 
\end{deluxetable}

One potential indicator of the inclination of the magnetic field in the plane of the sky is the polarization fraction \pfrac. If the magnetic field is in the plane of the sky, dust grains will be maximally polarized, as the long axis of the grain is also in the plane of the sky. On the other hand, if the magnetic field is along the line of sight, the long-axis of a spinning dust grain will draw circles in the sky, causing minimal polarization. \citetalias{Coude2025} presents \pfrac\ across the entire images for each bone and discusses the results in the context of grain alignment. Here, we focus specifically on the polarization fraction near the bones using the same pixels presented in Figure~\ref{fig:columndiff} (i.e., every other pixel and only pixels within 23$\arcsec$ of the bone's spine). Table~\ref{tab:Pfrac} gives the median polarization percentage, $P_\%$\,=\,\pfrac \,$\times$\,100\%, for pixels near the bone spine, detected at a 2$\sigma$ level (i.e., the level used for the sliding box analysis). The statistics of  $P_\%$ values at lower column densities can suffer from lack of completeness due to limits on the signal to noise, and they can be artificially inflated due to spatial filtering (see Section~\ref{sec:sofia_obs}). As such, Table~\ref{tab:Pfrac} also provides the median $P_\%$ values for different column density cutoffs at each pixel.

Median polarization percentages vary from 2\% to 4.4\% for \Ngas\,$>$1\,$\times$\,10$^{22}$\,cm$^{-2}$ and 1.1\% to 3.9\% for \Ngas\,$>$2\,$\times$\,10$^{22}$\,cm$^{-2}$. The difference in median $P_\%$ from bone to bone indicates there could indeed be projection effects of fields in the plane of the sky. However, the median $P_\%$ can vary significantly depending on the \Ngas\ cutoff, which is especially evident for Filament~2 and Filament~5.  We also caution interpreting the larger values for the Snake in the context of the other bones, as the mapping strategy (chop-nod) was different for this bone.

Besides magnetic field projection effects, there are numerous factors that can also affect the observed polarization fractions calculated for each bone. Among these include grain alignment efficiency, spatial filtering for the telescope, beam-averaging, optical depth effects, confusion along the line of sight, and stellar feedback (e.g., outflows). Beam-averaging is especially complex, as the bones are at varying distances, and fields are likely more complex at smaller scales, especially for the highest column densities. Moreover, there is likely some low-level $P_\%$ that are below our 2$\sigma$ cutoff for this analysis, which may reduce the median $P_\%$ for some of the bones.  Due to these confounding variables, a general conclusion about the inclination of the magnetic field is an exceedingly difficult task, beyond the scope of this paper. However, what can be said is that $P_\%$ does not vary greatly from bone to bone, meaning it is unlikely that severe projection effects across the entirety of bones are causing a significant effect on the overall conclusions of this paper.

\section{Magnetic Field and Virial Estimates via Sliding Box Analysis}\label{sec:sliding}
In this section, we analyze the relative importance of magnetic fields and gravity in the bones using a sliding box analysis similar to \citetalias{Stephens2022}. In short, we slide a box along the spine of each bone and estimate the average B-field strength within this box and evaluate its importance compared to gravity via the critical ratio $\lambda$. Additionally, we compute the virial parameter, $\alpha_{\text{vir}}$, and combine it with $\lambda$ to evaluate the potential for collapse within the bone.


\subsection{Sliding Box Method}
We calculate the B-field of each bone following the sliding box procedure outlined for G47 in \citetalias{Stephens2022}, with some modifications. We slide a rectangular box along the spine of each bone and compute the magnetic B-field for the box using the DCF technique. This box has a length and width of about 4 and 3 HAWC+ beams (72$\farcs$8\,$\times$\,54$\farcs$6, or 16\,$\times$\,12 HAWC+ pixels), respectively, which is similar to but slightly different from the dimensions used in \citetalias{Stephens2022} due to differences in pixel size (see Section~\ref{sec:sofia_obs}). On a spatial scale, the box size corresponds to approximately 0.5\,pc\,$\times$\,0.35\,pc for the closest bone (Filament~8) and 1.8\,pc\,$\times$\,1.4\,pc for the farthest bone (G49).  This box size was chosen so that there are enough beams to provide a sufficient determination of the angle dispersion.  Specifically, the rectangular area of the box divided by the circular area of the HAWC+ beam (assuming the radius is half the FWHM) is 14.5. Given that our underlying assumption when using the DCF technique (discussed below) is that the unperturbed B-field is uniform, we do not use a larger box as the uniform B-field assumption becomes less accurate. The rectangular box is aligned along the spine, and we slide it by one spine pixel (11$\farcs$5; see Section~\ref{sec:spines}) at a time. The angle of the box along the spine is determined from the instantaneous slope of the polynomial fit of the spine of each bone, $\theta_{\text{fil}}$ (Section~\ref{sec:spines}). Details on which pixels fall within a tilted box are provided in the appendix of \citetalias{Stephens2022}.

Within each rectangular box, we calculate the plane-of-sky B-field strength using the modified DCF technique from \citet{Ostriker2001}:
\begin{equation}
B_{\text{pos}} =  Q \sqrt{4\pi \bar{\rho}}\, \frac{\delta v_{\text{los}}}{\delta \theta} ,
\end{equation}
where $Q$ is a projection factor (taken to be 0.5; \citealt{Ostriker2001}), $\bar{\rho}$ is the mean density, $\delta v_{\text{los}}$ is the line-of-sight velocity dispersion, and $\delta \theta$ is the angle dispersion within the box. Typical median $\delta v_{\text{los}}$ for each sliding box measured for \nht\ vary slightly from bone to bone, from $\sim$0.4\,\kms\ to 1\,\kms. Although $\delta v_{\text{los}}$ technically refers to the non-thermal component alone, we do not subtract the thermal component in this analysis as it is negligible; the temperatures of these bones are usually less than 20\,K (\citetalias{Coude2025}), implying thermal linewidths of $<$0.1\,\kms. \citetalias{Stephens2022} used the \citet{Skalidis2021} version of the DCF technique, but we opt to use the \citet{Ostriker2001} version because it is used more frequently and can be more directly compared to other studies. Additionally, the statistical estimates from \citet{Ostriker2001} exhibit smaller deviations from corresponding simulation estimates compared to the \citet{Skalidis2021} study \citep{Myers2024}. The B-field strength estimate from \citet{Skalidis2021} differs from \citet{Ostriker2001} by a factor of $\sqrt{2\delta \theta}$, where $\delta \theta$ is in radians. Consequently, our B-field estimates in regions with lower angular dispersion may be up to a factor of $\sim$2 larger than the \citet{Skalidis2021} version of the DCF technique.


The value for $\bar{\rho}$ in each box is calculated from the \Ngas\ maps following the appendix in \citetalias{Stephens2022}, which assumes bones are cylindrical with radii provided in Table~\ref{tab:fits}. For calculating $\bar{\rho}$, we specifically use the median value of \Ngas\ in the box as it removes potential outliers without considerably changing it from the mean value \citepalias{Stephens2022}. To determine $\delta v_{\text{los}}$, we take the velocity dispersion maps discussed in Section~\ref{sec:molecularlines}, which may be (and often is) a mix of low and high-density tracers. If 39 (which corresponds to $\sim$20\%) of the 16\,$\times$\,12 sliding box pixels have a valid fit with a high-density tracer (equivalent to the area of a circle with a RAMPS beam's FWHM as the diameter), we take the median velocity dispersion of the high-density tracer only. Otherwise, we take the median of the mix of low and high-density tracers. We find that if we do not use the 39~pixel criterion, there are several instances where the low density tracer provides excessively high velocity dispersions compared to the low velocity dispersions, causing large $B_{\text{pos}}$ that we believe are likely not as accurate.


The intrinsic standard deviation, $\delta \theta$, is affected by observational errors. Many studies correct for the intrinsic dispersion via $\delta \theta = \sqrt{\delta \theta_{\text{obs}}^2-\sigma_{\theta}^2}$ \citep[e.g.,][]{Girart2006,Rao2009,Stephens2013}, where, $\delta \theta_{\text{obs}}$ is the standard deviation of angles in the box, and $\sigma_{\theta}$ is the observational error. This methodology was used for G47 in \citetalias{Stephens2022}. However, this formula assumes that the telescope beam does not significantly smooth out the intrinsic angle dispersion within the box. Whether the HAWC+ significantly does this or not is unclear, but we suspect that since we are estimating the dispersion of a larger-scale field within the box, this affect is not large. The correction also assumes all errors are Gaussian. For this study we assume that $\delta \theta$ = $\delta \theta_{\text{obs}}$ rather than applying the correction. We choose not to correct for two reasons: (1) we want to avoid potentially overestimating the B-field, and (2) there are many locations where the field is quite structured, and $\sigma_{\theta}$ happens to be larger than $\delta \theta_{\text{obs}}$; such a result implies these are statistical anomalies or the observational errors of HAWC+ are overestimated. In general, we find that applying the correction has a minimal effect on the median B-field strength across the bone for the sliding box analysis, typically increasing the strength by less than 10\%.

To calculate the importance of B-fields compared to gravity, we determine the critical ratio, $\lambda$, which compares the observed mass-to-flux ratio with the mass-to-flux ratio required for gravitational collapse, as described by \citep{Crutcher2004}:
\begin{equation}\label{eq:lambda}
\lambda = \frac{(M/\Phi)_{\text{observed}}}{(M/\Phi)_{\text{crit}}},
\end{equation}
where $(M/\Phi)_{\text{crit}}$ is assumed to be $1/(2\pi\sqrt{G})$ \citep{McKeeOstriker2007}. In the context of the sliding box analysis, as outlined in the appendix of \citetalias{Stephens2022}, we consider the 3D geometry of the sliding box, which is modeled as a cutout of the cylindrical bone. The direction of the B-field is taken as the median angle within the sliding box, and the B-field is assumed to be inclined in the sky by the statistical average of 38\fdg2 \citep{Crutcher2004}.

We also calculate the virial parameter within each sliding box using the cylindrical formulation of the virial parameter \citep{LiPS2022} which is
 \begin{equation}
 \alpha_{\text{vir}} = \frac{2 \delta v_{\text{los}}^2}{G M_l} ,
 \end{equation}
where $\delta v_{\text{los}}$ is again the line of sight velocity dispersion (determined the same way it was for $B_{\text{pos}}$ above), and $M_l$ is the mass per unit length of the bone within the box. The mass within the box is computed from the box size and \Ngas\ maps, assuming a mean molecular weight of $\mu_p = 2.37$ \citep{Kauffmann2008}. $M_l$ is simply this mass divided by the length of the box. An ideal, long cylindrical filament can be supported against collapse if $\alpha_{\text{vir}} > 1$.

We then calculate what we call the ``equilibrium index," $\epsilon$, which accounts for the fact that in order for a cloud to collapse, it must not only have gravity dominating over the B-fields but also sufficient gravity to overcome other forms of support. The equilibrium index is defined as
\begin{equation}
\epsilon = \sqrt{\lambda^{-2} +  \alpha_{\text{vir}}^2}.
\end{equation} 
An ideal, long cylindrical filament is supported against collapse by a B-field and thermal/turbulent motions if $\epsilon$\,$>$\,1 \citep{LiPS2022}.

\begin{figure*}
      \begin{minipage}[b]{1\linewidth}
         \includegraphics[width=\textwidth]{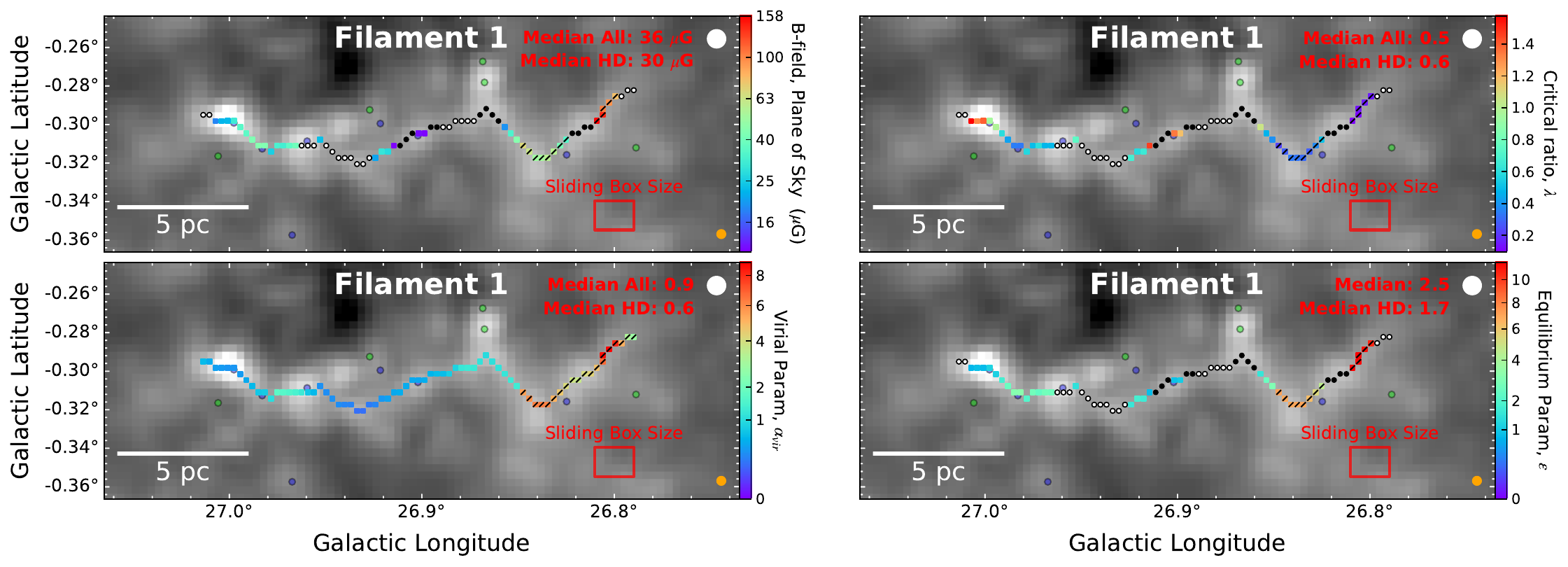}
         \includegraphics[width=\textwidth]{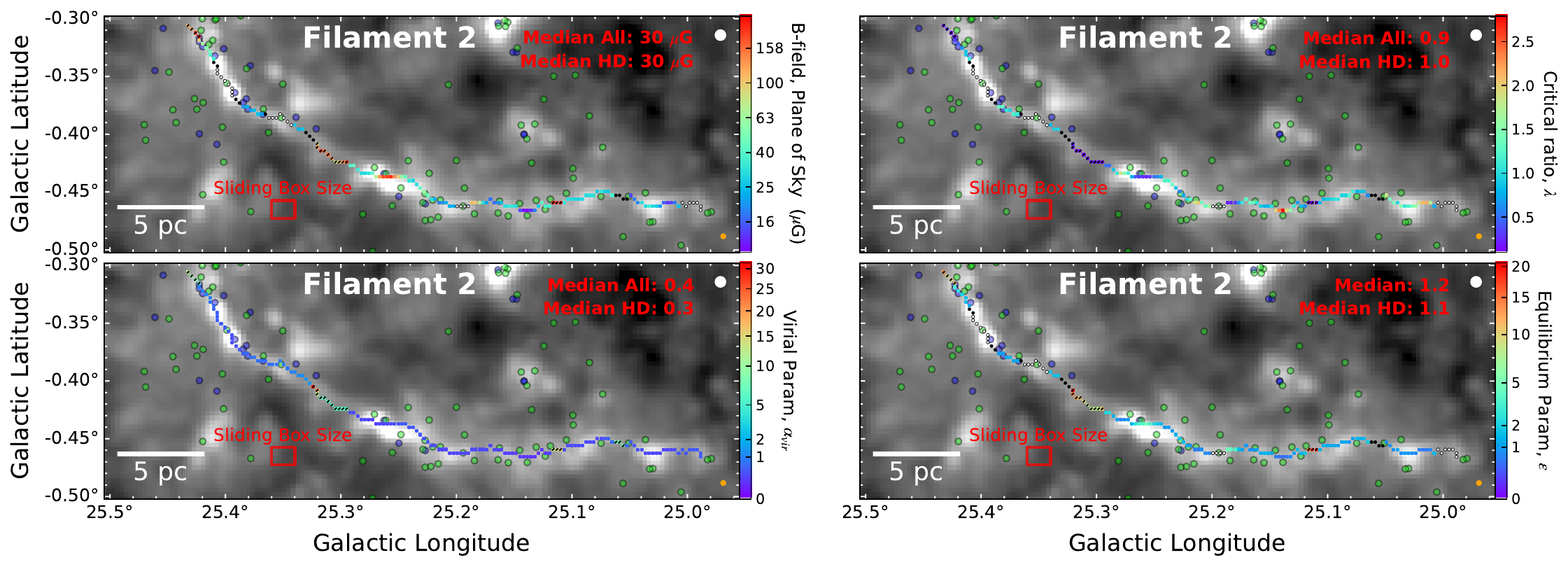}
         \includegraphics[width=\textwidth]{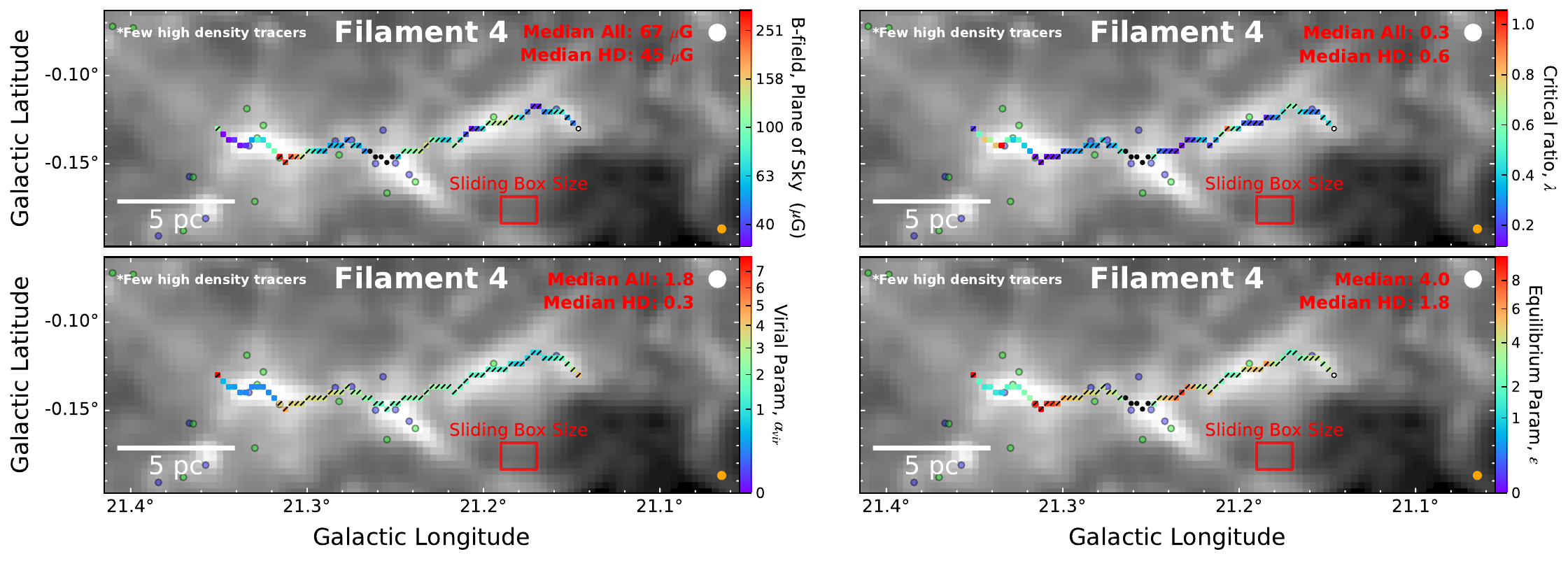}

            \end{minipage}
\end{figure*}

\begin{figure*}
      \begin{minipage}[b]{1\linewidth}
         \includegraphics[width=\textwidth]{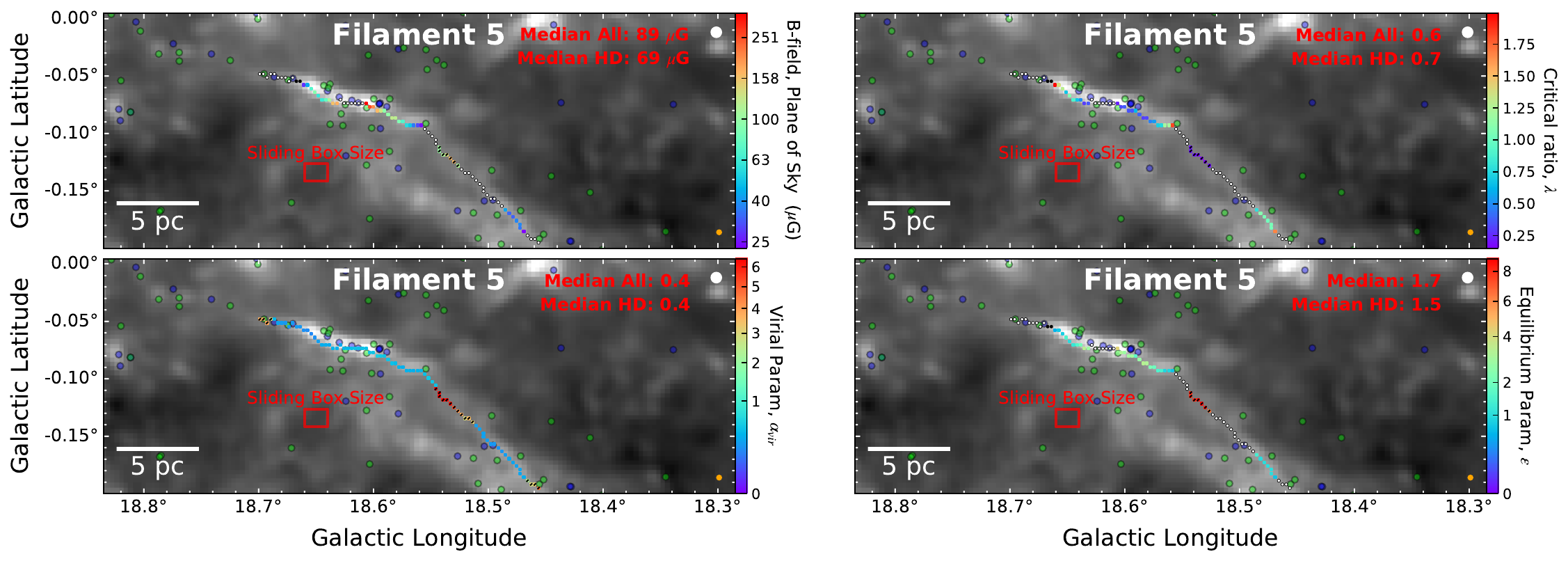}
         \includegraphics[width=\textwidth]{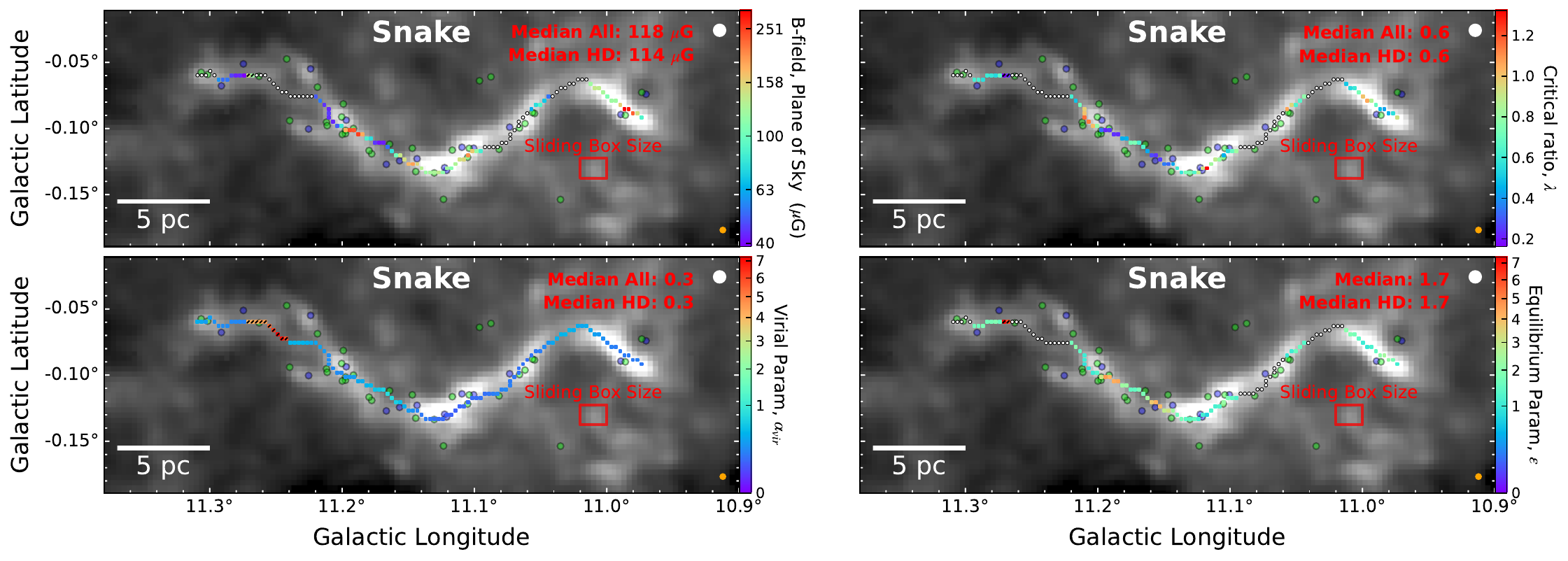}
	 \includegraphics[width=\textwidth]{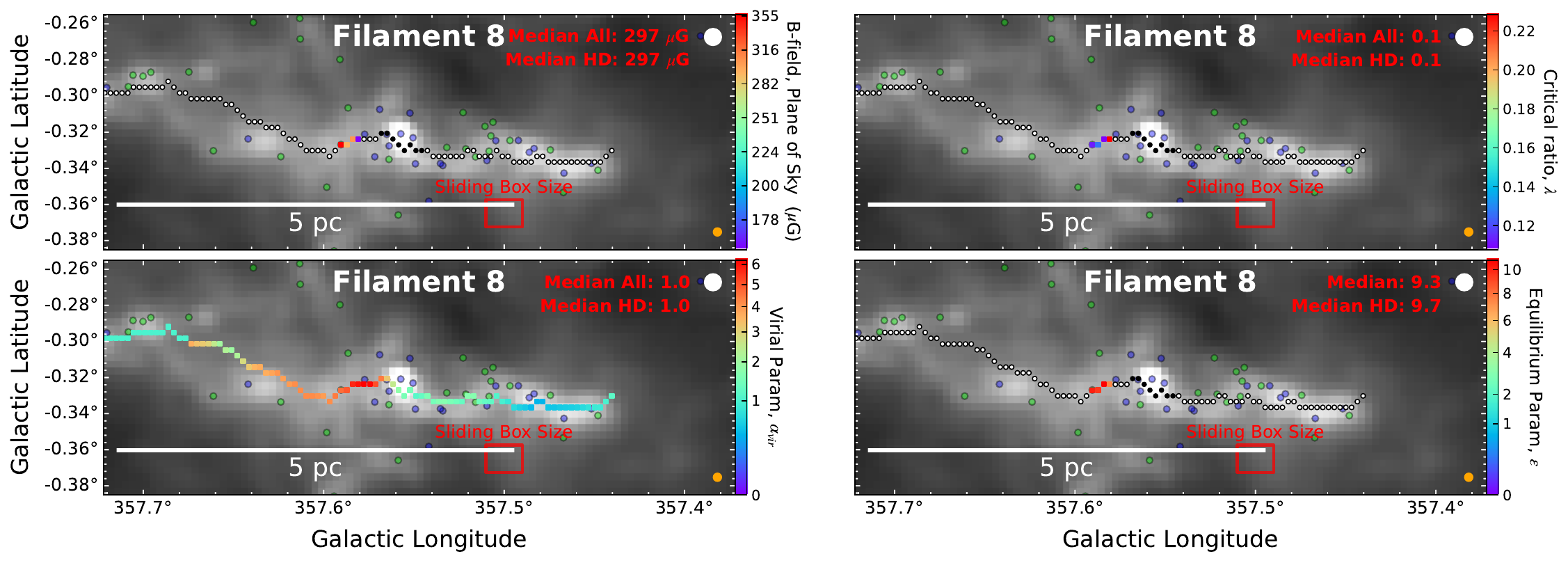}
            \end{minipage}
\end{figure*}

\begin{figure*}
      \begin{minipage}[b]{1\linewidth}
          \includegraphics[width=\textwidth]{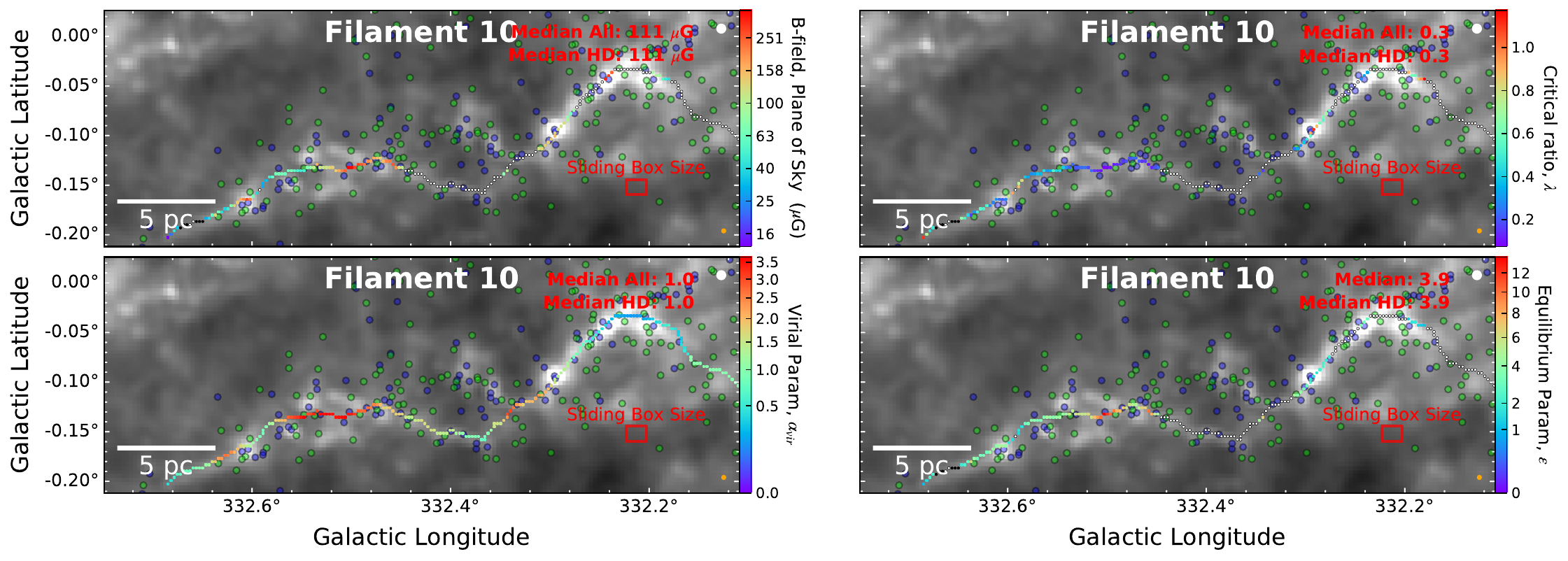}
         \includegraphics[width=\textwidth]{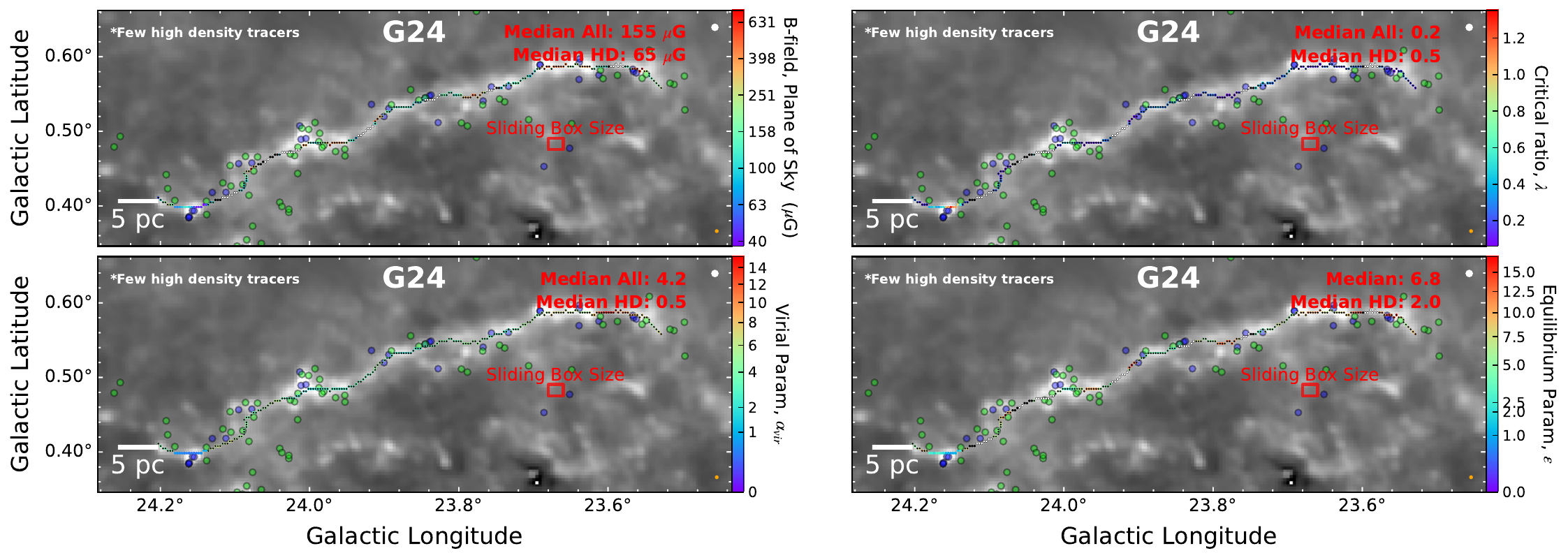}
        \includegraphics[width=\textwidth]{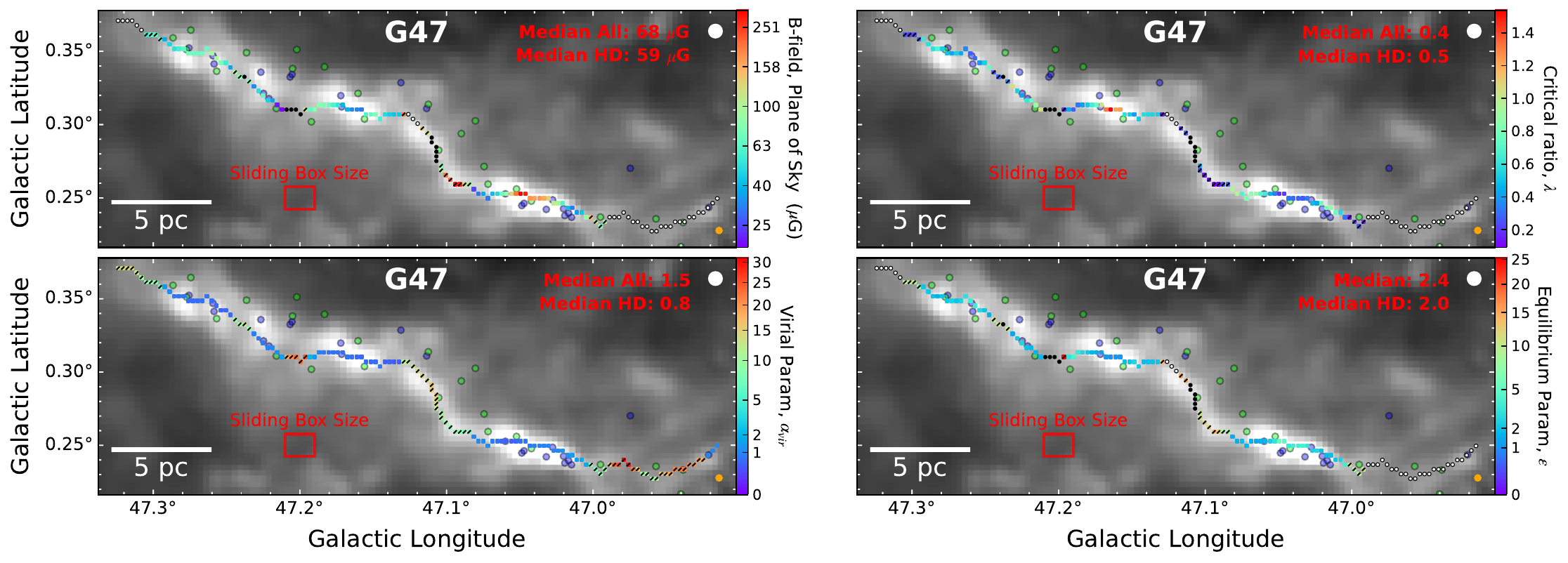}
            \end{minipage}
\end{figure*}

\begin{figure*}
      \begin{minipage}[b]{1\linewidth}
        \includegraphics[width=\textwidth]{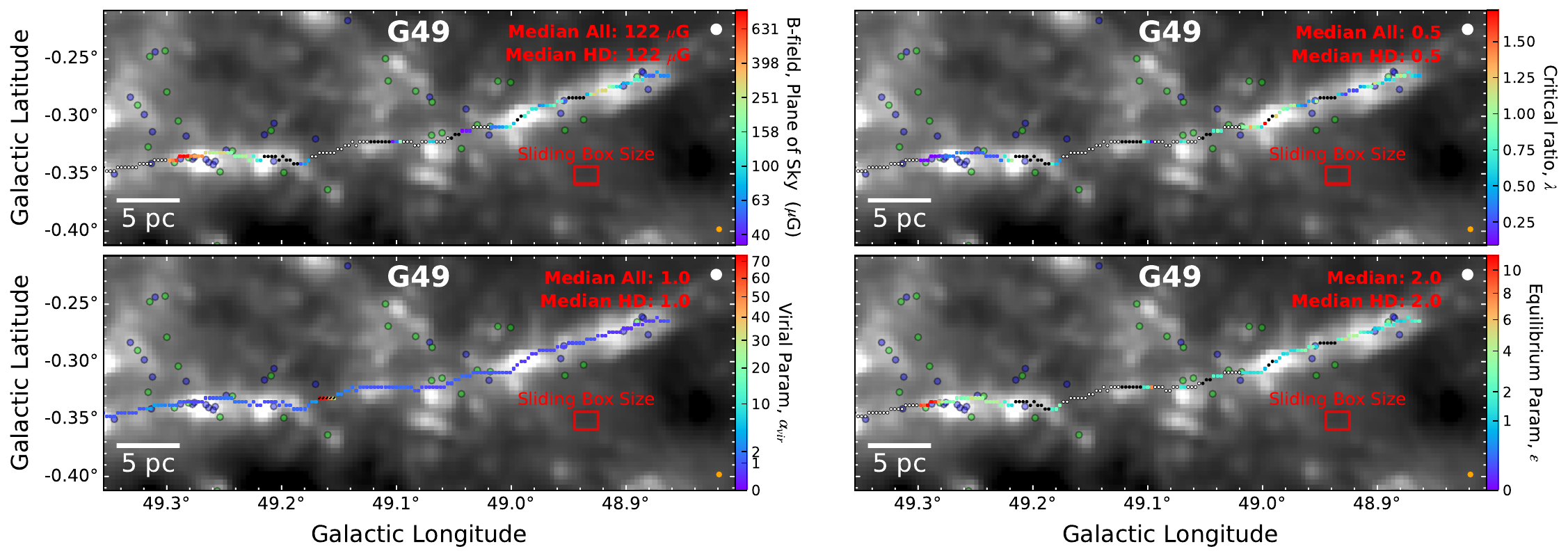}
            \end{minipage}
    \caption{Sliding box results for each bone, where $B_{\text{pos}}$, $\lambda$, $\alpha_{\text{vir}}$, and $\epsilon$ for the sliding box along the spine are shown in color-scale, and the gray-scale shows the gas column density, \Ngas. Color-scales for $\lambda$ is a linear scale, while the others are on a square root scale. A black forward slash `/' through a point indicates that the velocity dispersions is likely overestimated, as they were measured with a low-density tracer (see Section~\ref{sec:molecularlines}). Black and white points are locations where we were unable to make estimates (see text). The box size is shown in each panel, which was allowed to rotate as it slid across the spine. The top right of each panel shows the median for all points and for those using only high-density (HD) tracers. Figure~\ref{fig:fil1vectors} gives the beam sizes, and Figure~\ref{fig:fil1vectors} and the figures in Appendix~\ref{appendix:vectormaps} quantify the gray-scale for each bone.  } \label{fig:4panel}
\end{figure*}

The results of the sliding box analysis for each bone are shown in Figure~\ref{fig:4panel}. For each bone, there are 4 panels which show  $B_{\text{pos}}$, $\lambda$, $\alpha_{\text{vir}}$, and $\epsilon$ along the bone's spine, with the color-scale showing the values. If a point has a black forward slash, the parameter was estimated using low-density tracers.

In the panels for $B_{\text{pos}}$, $\lambda$, and $\epsilon$, points that are black in their entirety indicate where the angle dispersion $\delta \theta > 25^\circ$, as the DCF technique no longer applies in these cases \citep{Ostriker2001}. At these locations, (1) the turbulence may be super-Alfv\'enic, (2) the box is at a location where there is a transition of two distinct B-field morphologies, and/or (3) the assumption of a uniform field is poor. White points mark locations with a limited number of significant B-field vectors in the box. Specifically, these are areas where less than 70\% of the box has vectors detected at a 2$\sigma$ level, or less than 30\% of the box has vectors at a 3$\sigma$ level. Notably, for some extended regions of the bones, such as Filament~8 and Filament~10, large areas were not mapped, resulting in a considerable number of white points. The panel showing $\alpha_{\text{vir}}$ does not have white or black points since it is independent of the B-field morphology. Lastly, some areas along the spine are blank due to difficulties in accurately fitting a well-constrained velocity dispersion to any pixels within the box. These blank regions are infrequent but are most apparent on the western (right) side of Filament~1.

Table~\ref{tab:IQRstats} further quantifies the range of each sliding box parameter, reporting the median and interquartile range (i.e., 25 and 75 percentile) of the data used in the sliding box analysis. For all parameters, we only report values where we have a valid B-field strength that was estimated using the velocity dispersion of a high-density tracer. Note that some of the median $\alpha_{\text{vir}}$ values reported here differs from those reported in Figure~\ref{fig:4panel} because the medians in the figure includes $\alpha_{\text{vir}}$ values measured at locations with no B-field measurements. As seen, B-field measurements tend to be about 30--150\,$\mu$G, $\lambda$ and $\alpha_{\text{vir}}$ are $\sim$0.5, and $\epsilon$ is $\sim$2. Filament~8 and~10 have higher $\alpha_{\text{vir}}$, which is likely due to using \ceoto\ as the high-density tracer, which probably overestimates the velocity dispersion.

\begin{figure*}[ht!]
\begin{center}
\includegraphics[width=2\columnwidth]{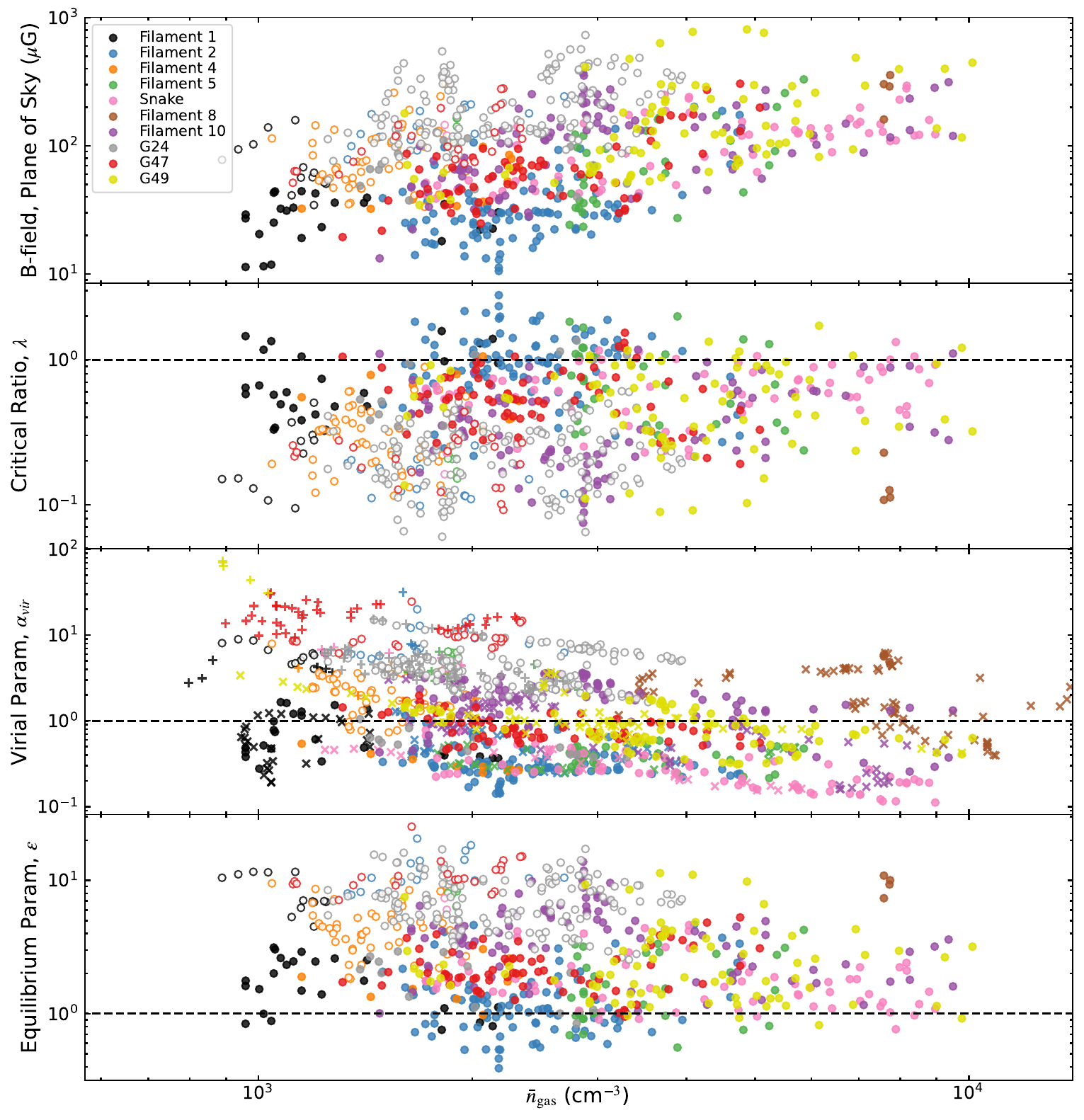}
\end{center}
\caption{Values of $B_{\text{pos}}$, $\lambda$, $\alpha_{\text{vir}}$, and $\epsilon$ for the sliding box analysis plotted against their respective gas density, \smallngas, color-coded for each bone. Solid and empty circles show where the velocity dispersion was estimated by high- and low-density tracers, respectively. $\alpha_{\text{vir}}$ is shown as circles where there is also a $B_{\text{pos}}$ estimate, and as $\times$ or $+$ symbols where there is no $B_{\text{pos}}$ estimate, yet some SOFIA data still exists in the sliding box. The $\times$ and  $+$ symbols are for $\alpha_{\text{vir}}$ parameters measured with high- and low-density tracers, respectively. Dashed lines in the bottom three panels are at $y=1$. Bones are typically magnetically subcritical (fields are dominant), but have virial parameters that are above 1, implying gravity along much of the bones may be insufficient for collapse.  
 } 
\label{fig:fourpanels_n} 
\end{figure*}

\begin{deluxetable*}{lccccccccccccccc}
\tablecolumns{16}
\tabletypesize{\footnotesize}
\tablewidth{2\textwidth}
\tablecaption{Sliding Box Parameters with High-Density Tracer and B-field Measurements \label{tab:IQRstats}}
\tablehead{
\colhead{Bone} & \colhead{Total} & \colhead{$n/1000$} & \colhead{$\delta v_{\text{los}}$} & \colhead{$B_{\text{pos}}$} & \colhead{$\lambda$} & \colhead{$\alpha_{\text{vir}}$} & \colhead{$\epsilon$} & \\
\colhead{Name} & \colhead{count} & \colhead{[cm$^{-3}$]} & \colhead{(\kms)} &\colhead{($\mu$G)} & \colhead{} & \colhead{} & \colhead{} &
}
\startdata
Filament~1 & 24 & 1.1$^{(1.4)}_{(1.0)}$ & 0.49$^{(0.63)}_{(0.39)}$ & 30$^{(36)}_{(22)}$ & 0.6$^{(1.1)}_{(0.5)}$ & 0.5$^{(1.2)}_{(0.4)}$ & 1.7$^{(2.5)}_{(1.1)}$ \\
Filament~2 & 94 & 2.3$^{(2.7)}_{(2.0)}$ & 0.39$^{(0.48)}_{(0.34)}$ & 30$^{(38)}_{(22)}$ & 1.0$^{(1.3)}_{(0.8)}$ & 0.3$^{(0.4)}_{(0.3)}$ & 1.1$^{(1.5)}_{(0.8)}$ \\
Filament~4 & 10 & 2.1$^{(2.3)}_{(1.7)}$ & 0.47$^{(0.47)}_{(0.46)}$ & 45$^{(76)}_{(35)}$ & 0.6$^{(0.8)}_{(0.4)}$ & 0.3$^{(0.3)}_{(0.3)}$ & 1.8$^{(2.4)}_{(1.4)}$ \\
Filament~5 & 33 & 3.0$^{(3.9)}_{(2.9)}$ & 0.54$^{(0.62)}_{(0.52)}$ & 69$^{(119)}_{(37)}$ & 0.7$^{(1.1)}_{(0.4)}$ & 0.4$^{(0.5)}_{(0.3)}$ & 1.5$^{(2.8)}_{(1.0)}$ \\
Snake & 67 & 4.0$^{(6.4)}_{(2.7)}$ & 0.53$^{(0.59)}_{(0.48)}$ & 114$^{(147)}_{(53)}$ & 0.6$^{(0.8)}_{(0.5)}$ & 0.2$^{(0.3)}_{(0.2)}$ & 1.7$^{(2.2)}_{(1.2)}$ \\
Filament~8 & 4 & 7.7 & 1.25 & 297 & 0.1 & 5.2 & 9.7 \\
Filament~10 & 92 & 2.9$^{(4.2)}_{(2.2)}$ & 1.01$^{(1.15)}_{(0.75)}$ & 111$^{(166)}_{(61)}$ & 0.3$^{(0.5)}_{(0.2)}$ & 1.9$^{(2.8)}_{(1.2)}$ & 3.9$^{(5.6)}_{(2.3)}$ \\
G24 & 14 & 1.9$^{(2.7)}_{(1.5)}$ & 0.74$^{(0.78)}_{(0.67)}$ & 65$^{(83)}_{(47)}$ & 0.5$^{(1.1)}_{(0.4)}$ & 0.5$^{(0.6)}_{(0.4)}$ & 2.0$^{(2.4)}_{(1.1)}$ \\
G47 & 67 & 2.3$^{(3.2)}_{(2.0)}$ & 0.65$^{(0.76)}_{(0.61)}$ & 59$^{(79)}_{(40)}$ & 0.5$^{(0.8)}_{(0.4)}$ & 0.8$^{(1.0)}_{(0.6)}$ & 2.0$^{(2.6)}_{(1.7)}$ \\
G49 & 79 & 3.7$^{(5.1)}_{(3.2)}$ & 1.01$^{(1.29)}_{(0.84)}$ & 122$^{(230)}_{(70)}$ & 0.5$^{(0.9)}_{(0.3)}$ & 0.8$^{(1.2)}_{(0.6)}$ & 2.0$^{(3.4)}_{(1.4)}$ \\
\enddata
\tablecomments{Statistics for each parameter are reported only at locations with valid B-field measurements where a high-density tracer was used to calculate the velocity dispersion, with the total sample count listed in Column (2). The table also includes the volume density $n$ and $\delta v_{\text{los}}$ used for each measurement in the sliding box analysis. Columns (3)--(8) give the median values of each parameter, with superscript and subscript values indicating the upper and lower quartiles (25th and 75th percentiles), respectively. Quartiles are not reported for Filament~8 due to its small sample size.
}
\end{deluxetable*}

\subsection{Analyzing the Sliding Box Results}
Here we analyze the sliding box results of Figure~\ref{fig:4panel} more thoroughly. To assist with the interpretation, we first review key aspects and limitations of the sliding box approach. Firstly, each value along the spine represents an average within the rectangular sliding box, which is rotated to align with the spine.  The size of the box is 72$\farcs$8\,$\times$\,54$\farcs$6 (16\,$\times$\,12 HAWC+ pixels), providing a consistent angular scale, but the spatial scales vary due to the differing distances of the bones. Therefore, comparisons between bones should consider the different spatial extents probed for each bone. In many regions, the B-field strengths could not be estimated, mainly due to excessive angular dispersion ($\delta \theta > 25^\circ$) or insufficient B-field data. These regions could be potential locations where B-field strengths are low and thus more easily tangled, but they can also be places where there are multiple B-field components within each box along the line of sight and/or the plane of the sky. These regions could also be locations where most of the B-field is along the line of sight. Moreover, we primarily focus on measurements that are derived from velocity dispersions of the high-density tracers, as the ones from low-density tracers are likely overestimated (see Appendix~\ref{appendix:linewidths}). G24 and Filament~4 have very few velocity dispersions measured with high-density tracers.

Additionally, there are significant uncertainties arising from assumptions such as bone geometry, diffuse emission, the choice of the DCF method, and the assumption of a uniform B-field. These factors cause considerable difficulty in determining accurate B-field uncertainties at each location, and thus we do not explicitly estimate them in this paper. The B-field strength and $\lambda$ estimates could be off by a factor of 2 or even 3. Some of the factors may cause an overestimate of the B-field strength (e.g., beam smearing of polarization and kinematic data), while other assumptions may cause an underestimate of the B-field strength (e.g., assumed uniform structure in each box and not correcting for observational errors). Because of these confounding factors, it is difficult to ascertain whether B-fields are generally overestimated or underestimated. However, the errors due to the assumptions are at least somewhat correlated, which allows us to make valid comparisons of the relative field strengths between bones, and along the spines of each bones themselves.

To gather a general sense of the values across all figures, in Figure~\ref{fig:fourpanels_n} we plot all $B_{\text{pos}}$, $\lambda$, and $\alpha_{\text{vir}}$ as a function of number density, \smallngas, which complements the information presented in Figure~\ref{fig:4panel}. For Figure~\ref{fig:fourpanels_n}, we will focus the discussion primarily on the solid circles, as these have more accurate velocity dispersions measured by high density tracers. Across the bones, most B-field strengths along the spine are a few 10s of $\mu$G to a few hundred $\mu$G. The largest estimated B-fields are for G49, which reach almost 1 mG. A weak but positive correlation is found between number density and volume density. This topic will be further explored in Section~\ref{sec:discussion}. 

The critical ratio, $\lambda$, is typically found to be less than 1, indicating that B-fields are generally strong enough to counteract gravitational collapse in the bones. In most cases, $\lambda$ is 0.5 or less, and values of $\sim$0.2 is not atypical. Even if $\lambda$ is underestimated (e.g., by overestimating $B_{\text{pos}}$), the B-fields are still likely playing a significant role in supporting the bones against gravitational collapse. Critical ratios appears to be independent of volume density.

\begin{deluxetable*}{lccccc|ccccc|ccccc|}
\tablecolumns{16}
\tabletypesize{\footnotesize}
\tablewidth{2\textwidth}
\tablecaption{YSO Nearby Median Sliding Box Parameters \label{tab:YSOstats}}
\tablehead{
\colhead{} &
\multicolumn{5}{c}{Class I+II} &
\multicolumn{5}{c}{Class I} &
\multicolumn{5}{c}{Class II} \\
\colhead{Bone} &
\colhead{YSO} & \colhead{$B_{\text{pos}}$} & \colhead{$\lambda$} & \colhead{$\alpha_{\text{vir}}$} & \colhead{$\epsilon$} &
\colhead{YSO} & \colhead{$B_{\text{pos}}$} & \colhead{$\lambda$} & \colhead{$\alpha_{\text{vir}}$} & \colhead{$\epsilon$} &
\colhead{YSO} & \colhead{$B_{\text{pos}}$} & \colhead{$\lambda$} & \colhead{$\alpha_{\text{vir}}$} & \colhead{$\epsilon$}  \\ 
\colhead{Name} &
\colhead{count} & \colhead{($\mu$G)} & \colhead{} & \colhead{} & \colhead{} &
\colhead{count} & \colhead{($\mu$G)} & \colhead{} & \colhead{} & \colhead{} &
\colhead{count} & \colhead{($\mu$G)} & \colhead{} & \colhead{} & \colhead{}
}
\startdata
Filament 1  & 3  & 30  & 0.95 & 0.47 & 1.1 & 3  & 30  & 0.95 & 0.47 & 1.1 & 0  & --  & --   & --   & --   \\
Filament 2  & 22 & 31  & 0.99 & 0.34 & 1.1 & 7  & 35  & 1.1  & 0.65 & 1.1 & 15 & 30  & 0.99 & 0.27 & 1.1 \\
Filament 4  & 2  & 58  & 0.72 & 0.26 & 1.8 & 1  & 37  & 1.1  & 0.25 & 0.98& 1  & 79  & 0.39 & 0.27 & 2.6 \\
Filament 5  & 5  & 201 & 0.37 & 0.50 & 2.8 & 3  & 201 & 0.37 & 0.50 & 2.8 & 2  & 141 & 1.1  & 0.46 & 1.9 \\
Snake & 20 & 152 & 0.55 & 0.19 & 1.8 & 7 & 150 & 0.55 & 0.19 & 1.8 & 13 & 163 & 0.55 & 0.19 & 1.8 \\
Filament 8  & 1  & 356 & 0.11 & 4.6  & 10  & 0  & --  & --   & --   & --  & 1  & 356 & 0.11 & 4.6  & 10  \\
Filament 10 & 28 & 137 & 0.22 & 2.0  & 4.9 & 20 & 142 & 0.25 & 1.9  & 4.5 & 8  & 132 & 0.22 & 2.3  & 5.1 \\
G24         & 0  & --  & --   & --   & --  & 0  & --  & --   & --   & --  & 0  & --  & --   & --   & --  \\
G47         & 19 & 64  & 0.58 & 0.76 & 2.0 & 13 & 64  & 0.62 & 0.83 & 2.0 & 6  & 61  & 0.55 & 0.73 & 2.0 \\
G49 & 14 & 152 & 0.44 & 0.63 & 2.4 & 8 & 235 & 0.63 & 0.61 & 2.1 & 6 & 152 & 0.43 & 0.88 & 2.7 \\
\enddata
\tablecomments{The median values in this table are not at the YSO itself, but rather the nearest spine pixel to each YSO. The table only considers YSOs within 23$\arcsec$ of the spine, and the sliding box analysis uses a high density tracer and has an estimated B-field strength.}
\end{deluxetable*}


As measured by the high-density tracers, virial parameters are almost always less than 2, indicating they are at least self-gravitating, but they are usually even less than 1, which indicates collapse in the absence of B-fields. The only exceptions are Filament~8 and Filament~10, but these virial parameters are likely overestimated due to using \ceoto\ as the high-density tracer rather than \nht\ (see Appendix~\ref{appendix:linewidths}). Since Filament~8 and Filament~10 likely have overestimated velocity dispersions, they likely have overestimated B-field strengths as well. Finally, the virial parameters as measured by low-density tracers do have virial parameters often much more than 2. While their velocity dispersions are likely overestimated, some calculations exceed 10, indicating there may be portions of the bones that are unbound.




Despite the low virial parameters, the strong B-fields caused the bones to have $\epsilon$ values above~1, indicating that B-fields are helping to support them against collapse, even when the virial parameter is small. Only Filament~2 has a substantial portion of its bone with $\epsilon$ below 1. Filament~2, which is already has a large population of YSOs, is likely to continue to collapse and form additional stars. Given the uncertainties, its possible that other bones have parts likely to collapse, but nevertheless, B-fields likely play a significant role. Moreover, high $\alpha_{\text{vir}}$ values are often found in the presence of strong B-fields ($\lambda < 1$), which may suggest that B-fields may be important for the dissipation of bones, which will discussed more in Section~\ref{sec:discussion}. 

Given that YSOs have been identified toward these bones, despite these high B-field strengths, there has been a history of star formation, and a reasonable assumption would be that star formation would continue in the future. To determine the approximate sliding box parameters ($B_{\text{pos}}$, $\lambda$, $\alpha_{\text{vir}}$, and $\epsilon$) at the location of each YSO, we first match each YSO to the nearest sliding box pixel. We only consider YSOs within 23$\arcsec$ of the bone's spine and that have a valid B-field strength as measured by a high-density tracer. For each bone, we then take the median value of all the YSO-matched sliding box results, which are presented in Table~\ref{tab:YSOstats}. The results are separated in all Class~I and~II YSOs, Class~I by themselves, and Class~II by themselves. There is no obvious difference in the sliding box results for Class~I versus Class~II results, which is partially due to low number statistics. Moreover, when comparing the YSO results to the sliding box results across the entire bones (Figure~\ref{fig:4panel}), the medians in general are similar. These results thus suggest no correlation between YSO location and the sliding box results.

\section{Discussion}\label{sec:discussion}
In Section~\ref{sec:sliding}, we find evidence that magnetic fields are important for collapse and perhaps dissipation of the bones. The critical ratio $\lambda$ typically much less than 1, and even given the large uncertainties in B-field measurements, the bones appear mostly subcritical. Moreover, low $\alpha_{\text{vir}}$ values suggest that much of the bones would collapse in the absence of B-fields. However, there are some low-density locations of the bones that may be unbound.



The substantially large virial parameters traced by the low-density tracers may suggest that parts of the bones are not gravitationally bound. Large virial parameters have also been observed in the majority of giant molecular clouds, both in the Milky Way and in other galaxies \citep{Dobbs2011,Evans2021}. \citet{Contreras2017} calculated the virial parameter for intermediate and high-mass clumps (pc scale) probed by MALT90 \citep{Jackson2013} in the Galaxy, and found that approximately 30\% of them are unbound (virial parameters of 2 or more). \citet{Myers2025} also analyzed the high virial parameters toward star-forming clouds, and suggested that clouds do not globally collapse, but rather locally collapse. In the simulation by \citet{DuarteCabral2017}, they analyzed the time evolution of giant filaments, and found they are not globally gravitationally bound, yet they may be confined by external pressure. \citet{Zucker2019b} found that bones can form in both the arm and interarm regions through galactic dynamics alone (i.e., ignoring feedback, local self-gravity, and B-fields). Those in spiral arms are potentially formed due to mass compressing as it enters the spiral potential, and those in the interarms are potentially formed due to differential rotation. Because of the Galactic dynamics, bones can be constantly forming and re-forming. In locations where the virial parameters are high, the structure of the bones will likely persist for at least a crossing time. The velocity dispersions $\delta v_{\text{los}}$ in the more diffuse areas vary but are mostly in the range of 1 to 4\,\kms. If we approximate the bone's minimum lifetime as the crossing time, $R/\delta v_{\text{los}}$, with $R \sim 1$\,pc, then even diffuse parts of the bone could last for at least 0.25 to 1\,Myr. \citet{Zucker2018b} suggest that bones typically exist for a few million years before they are destroyed by internal feedback. B-fields, along with external pressure from, for example, the weight of the Galaxy (gas + stars), may slow this dissipation and help guide the flow of gas. The models from \citet{Fiege2000} showed that for helical B-fields, the toroidal component can help confine the gas by effectively squeezing the cloud. Given the complexity of B-fields in each bone, we certainly cannot rule out a toroidal component in some regions. Moreover, simulations by \citet{Hix2023} show that strong B-fields in quasi-cylindrical clouds can help confine gas in a direction perpendicular to the magnetic field, ultimately allowing gas to be retained for longer periods even with feedback. Additionally, bones are for the most part near the local minima in the Galaxy's spiral potential, which could cause mass to accumulate in these areas. It is possible that these bones will even accrete much more mass in the future. For additional discussion on filament accretion and fragmentation, see \citet{Hacar2023}.



Despite the large virial parameters in diffuse areas, most of the bones have small virial parameters indicative of collapse in the absence of magnetic fields. To further assess the importance of B-fields in the bones, we calculate the energy density of B-fields and non-thermal kinetic energy \citep[e.g.,][]{Pattle2017}. The magnetic energy density of a cloud is $U_B = B^2/(8\pi)$, and the total energy is $E_B = U_B V$, where $V$ is the volume of the bone within the sliding box. $V$ is simply the area of the box times the average path length along the line of sight \citepalias[see Appendix of][]{Stephens2022}. The non-thermal kinetic energy within the box is $E_K = 1.5M (\delta v_{\text{los}})^2$ (the thermal component is negligible; $\lesssim$0.1\,\kms). Repeating the sliding box analysis to estimate these parameters, and only considering locations where we have a high-density tracer, we find that $E_B$ and $E_K$ typically have the same order of magnitude. The median value of $E_B$/$E_K$ for each bone varies from $\sim$1--5, with 2.5 being the most typical. If we take into account locations traced by the low-density tracer, the results are similar, with 2.3 being the most typical. Given the large uncertainties, especially since $E_B$ depends on $B^2$, and the overall simplicity of this energy analysis, we do not interpret the results in more detail. Nevertheless, it shows that B-fields need to be considered in both the collapse and dissipation of these bones.


Despite the strong B-fields, there are young YSOs identified in the bones, which indicates that B-fields certainly have not stopped collapse altogether. In fact, we find no correlation between YSO location and collapse likelihood of the bone. Our estimates reflect what will happen in the future and are not indicative of the past. Moreover a region that is magnetically subcritical can still accrete along field lines to become supercritical, allowing local collapse in the future \citet[e.g.,][]{McKeeOstriker2007}. Furthermore, it is possible that within our boxes, there exists high-density pockets that will collapse (or have collapsed already) to form stars. The sliding boxes used are $\sim$1\,pc in size, which is much larger than the YSO's core size of $\sim$0.1\,pc, and thus criticality for collapse may happen at a smaller scale. Moreover, these high-density pockets may have formed before or even simultaneously with the accumulation of mass in the bone, and the bone may serve as a reservoir for additional mass accretion. 



In some bones such as Filament~2, the cylindrical virial parameter at some locations is significantly less than unity.  In simple models this property implies a filament in global radial collapse.  However, such collapse is inconsistent with observed line profiles, with the clumpy structure of their column density, and with their relatively low star formation efficiency, typically a few percent \citep{ZhangMM2019}. Instead, these properties resemble those in turbulent MHD simulations of molecular clouds, which evolve from an initial virial parameter of $\sim$2 as their turbulence is allowed to dissipate on a cloud-crossing time scale. Simulations by \citet{Grudic2019} along with STARFORGE simulations by \citet{Grudic2021,Grudic2022} suggest a hierarchical disordered collapse, and clouds have relatively little global collapse motion. Instead their local motions form filaments, cores, and protostars for several Myr, with virial parameter values close to 1, until feedback from massive stars disperses most of the cloud gas. The strong B-fields of bones may further slow collapse and dispersal. We suggest that some bones, or subregions of bones with low virial/equilibrium parameter values, may undergo the kind of regional local collapse seen in these simulations.

We can also compare the B-field strengths of bones with other studies probing B-fields at similar densities \citep{Crutcher2010,Pattle2023}. We find that the B-field strengths vary from tens of $\mu$G to a few hundred $\mu$G, though many have higher values, especially at higher densities. The medians shown in Figure~\ref{fig:4panel} and Table~\ref{tab:IQRstats} indicate that the B-fields are most commonly in the 30-150\,$\mu$G range. The median strengths are likely closer to the lower ranger of these values since B-fields are probably weaker in regions where we were unable to make estimates with the sliding box. The bones, at the scales probed in the sliding box analysis, have typical densities of 10$^3$\,cm$^{-3}$ to 10$^4$\,cm$^{-3}$. As a comparison, \ion{H}{1} and OH Zeeman observations \citep[][and references therein]{Crutcher2010}, which measure the line-of-sight B-field, $B_{\text{los}}$, have shown that at these densities, B-fields are typically only $\sim$10\,$\mu$G, with maximum values around 100\,$\mu$G. However, \citet{Pattle2023} collected DCF estimates from the literature and found B-field strengths quite consistent with what we observe in the bones, with some values reaching over 1\,mG at $\sim$10$^4$\,cm$^{-3}$. \citet{Pattle2023} and references therein discuss the discrepancy between Zeeman $B_{\text{los}}$ measurements and the DCF $B_{\text{pos}}$ estimates, but said there is no clear resolution to this discrepancy. It is worth noting that the polarization morphology toward the bones (Figure~\ref{fig:fil1vectors} and Appendix~\ref{appendix:vectormaps}) often changes rapidly in the plane of the sky, so it would be reasonable to believe that fields also change directions frequently within the large beam (usually $>$1$\arcmin$) used in the Zeeman observations, lowering the Zeeman signals compared to the uni-direction B-field case.

\citet{Crutcher2010} and \citet{Pattle2023} show a relation between $B$ and \smallngas, and the relation is also apparent in Figure~\ref{fig:fourpanels_n}, albeit, with significant scatter. However, the relations observed by \citet{Crutcher2010} and \citet{Pattle2023} span many orders of magnitude, whereas we are analyzing the relationship over just a single order of magnitude. Within this narrower range, the dispersion of B-field strengths with density appears to be similar to the above studies.


\citet{Zhao2024} produced galactic MHD zoom-in simulations with supernova feedback, finding that fields can be made more orderly and strengthened due to expanding shells. They found that fields are both parallel and perpendicular to filaments. In our study, four bones exhibited little perpendicular alignment at high column densities: G24, G49, Filament~8, and Filament~10 (Section~\ref{sec:orientations}). Filament~8 and Filament~10 even showed a preference for parallel alignment. Additionally, three bones had virial parameters greater than 1 throughout their entire length, indicating that these bones are not globally collapsing at pc scales: G24, Filament~8, and Filament~10 (Figure~\ref{fig:fourpanels_n}). This suggests that bones without perpendicularly aligned B-fields generally coincide with those having high virial parameters. G49 is an exception, as it contains many regions where the virial parameter is below 1. Therefore, while there is some evidence that perpendicular B-fields may indicate bones more likely to collapse, the current sample size is not large enough to provide strong statistical support for this conclusion. Whether known Galactic shells impact such alignment, as seen in the \citet{Zhao2024} simulations, is left for future work. 

The lack of global alignment, either parallel and perpendicular, throughout the entirety of most bones may indicate that B-fields were not the dominant force in their formation. For example, galactic forces such as galactic potential of spiral arms and/or differential rotation might be their origins \citep{Zucker2019b}. Nevertheless, the present-day B-fields are strong and frequently very structured on a local clump (few pc) scales. If B-fields were also strong during the formation of the bones, the present-day disordered B-fields may indicate that bones originally formed as a strand of clumps within a gravitational potential well. If that is the case, cylindrical geometry for the entire bone filaments may be a poor assumption if the clumps do not coalesce. On the contrary, it is also possible that a variety of feedback mechanisms, both within and outside of the bones, have changed fields from originally aligned to less aligned. 

Finally, the analysis in this paper has a selection bias of long structures that are parallel to the Galactic plane. Whether similar analyses of other Galactic structures would yield the same results remains unclear.

\section{Summary}\label{sec:summary}
In this paper, we analyze the inferred magnetic field morphology for all 10 bones in the FIELDMAPS survey, using a method similar to that applied to G47 in \citetalias{Stephens2022}. The images reveal that the B-fields are structured, but morphologically diverse across the bones. We first examine the alignment of magnetic fields with the spines of the bones, as well as the locations of high column density and young stellar objects (YSOs). We find the following:

\begin{enumerate}
\item Along each bone, there is no dominant preferred orientation (e.g., parallel or perpendicular) between the spine and the B-fields. However, all bones except Filament~8 and Filament~10 show a slight preference for perpendicular alignment between the spine and the B-fields. Filament~8 and Filament~10, on the other hand, display a slight preference for parallel alignment.
\item At lower column densities (\Ngas\,$\sim$\,0.5\,--\,1\,$\times$\,10$^{22}$\,cm$^{-2}$), B-field alignment with the bone is random. This is consistent with \citet{Planck35}, as these column densities represent where they saw the alignment of B-fields transition from mostly parallel with the elongations to not parallel.
For bones showing a preference for perpendicular alignment, this tendency generally becomes more pronounced at higher column densities. In contrast, the only two bones observed in Quadrant~4 of the Galaxy, Filament~8 and Filament~10, exhibit a greater degree of parallel alignment at higher column densities. 
\item Class~I and Class~II YSOs are usually found at locations of higher column densities. Class~I YSOs are more likely at locations of higher column density than Class~II YSOs, indicating YSOs either migrate or the bone evolves significantly during their evolution. However, they are not significantly more likely to form at places with perpendicular or parallel alignment. Assuming the YSO migration rate and dissipation rate is small, the lack of alignment suggests that YSOs form at high column densities, but are ignorant of the direction of the local magnetic fields direction or their potential guiding of gas flow on the $\sim$0.1 -- 1 pc scales probed by our observations.
\end{enumerate}

Next, we slide a box of approximately 1\,pc in size down the spine of each bone, allowing to rotate along the direction of the spine. Within the box, we estimate the average B-field and investigate its importance relative to gravity. We find the following:
\begin{enumerate}
\item The median B-field strengths differ by up to a factor of a few between the bones, yet B-fields can vary significantly across bones. Nevertheless, the pc-scale B-field strengths for the bones are typically 30--150\,$\mu$G. There exists a slightly positive correlation between density and B-field strengths.
\item We compare the calculated mass-to-flux ratio with the mass-to-flux ratio needed for gravitational collapse using the critical ratio, $\lambda$. We find that $\lambda$ values are usually much less than 1. The low values of $\lambda$ indicate that even if we are overestimating the B-field strengths, B-fields likely provide at least some support against gravitational collapse in the bones.
\item The cylindrical virial parameter $\alpha_{\text{vir}}$ is frequently found to be less than 1, suggesting that the bones would likely collapse in the absence of magnetic fields. The low-density tracers indicate high virial parameters at some locations of the bones, suggesting gravitationally bound bones may be embedded in unbound gas. Given that $\lambda$ is generally low everywhere and the magnetic energy density is high, B-fields may be important for both the collapse and dissipation of mass from the bones.
\item Few locations across any of the bones show low $\alpha_{\text{vir}}$ and high $\lambda$, which we quantify with the equilibrium parameter, $\epsilon$. Given the uncertainties in the estimates, some bones may still have locations likely to collapse. Filament~2 appears to be the most likely bone to form additional stars now or in the near future.
\end{enumerate}s

We suggest that bones may form via mass collecting in the spiral potential. Magnetic fields are significant, and may help aid the lifetime of the bones by slowing collapse and dissipation. Given that there are many known YSOs across the bones, yet much of the bones appear they are not critical to collapse, parts of the bones may have been denser in the past, or dense pockets at smaller scales (e.g., cores) can collapse to form more YSOs. Moreover, YSOs may have resulted from high-density cores that formed either concurrently with or even before the bones.


\bigskip
We thank an anonymous referee for critical comments that greatly improved the paper. This paper is based on observations made with the NASA/DLR Stratospheric Observatory for Infrared Astronomy (SOFIA). SOFIA was jointly operated by the Universities Space Research Association, Inc. (USRA), under NASA contract NNA17BF53C, and the Deutsches SOFIA Institut (DSI) under DLR contract 50 OK 0901 to the University of Stuttgart. Financial support for this work was provided by NASA through awards  \#05\_0109 and \#08\_0186 issued by USRA.
The Green Bank Observatory is a facility of the National Science Foundation operated under cooperative agreement by Associated Universities, Inc.
This publication is based on data acquired with the Atacama Pathfinder Experiment (APEX) under programmes 092.F-9315 and 193.C-0584. APEX is a collaboration among the Max-Planck-Institut fur Radioastronomie, the European Southern Observatory, and the Onsala Space Observatory. The processed data products are available from the SEDIGISM survey database located at \url{https://sedigism.mpifr-bonn.mpg.de/index.html}, which was constructed by James Urquhart and hosted by the Max Planck Institute for Radio Astronomy." We thank James Urquhart for early help with the data products.
L.W.L. acknowledges support from NSF AST-1910364 and NSF AST-2307844.
T.G.S.P. gratefully acknowledges support by the National Science Foundation under grant No. AST-2009842 and AST-2108989 and by NASA award \#09-0215 issued by USRA.
We thank Michael Gordon for his effort in setting up the on-the-fly maps for the FIELDMAPS project and Sachin Shenoy for his for helping us with the SOFIA pipeline for data reduction.
We thank Leah Zuckerman for early efforts on analyzing FIELDMAPS data.


\facilities{SOFIA (HAWC+), Herschel, APEX, GBT, FCRAO}
\software{APLpy \citep{Robitaille2012}, 
Astropy \citep{Astropy2013, Astropy2018},
MAGNETAR \citep{Soler2013},
Reproject \citep{Robitaille2020}
}

\appendix

\section{Images of the magnetic field morphology for all bones.}\label{appendix:vectormaps}
\begin{figure*}
\begin{center}
 \includegraphics[width=0.9\textwidth]{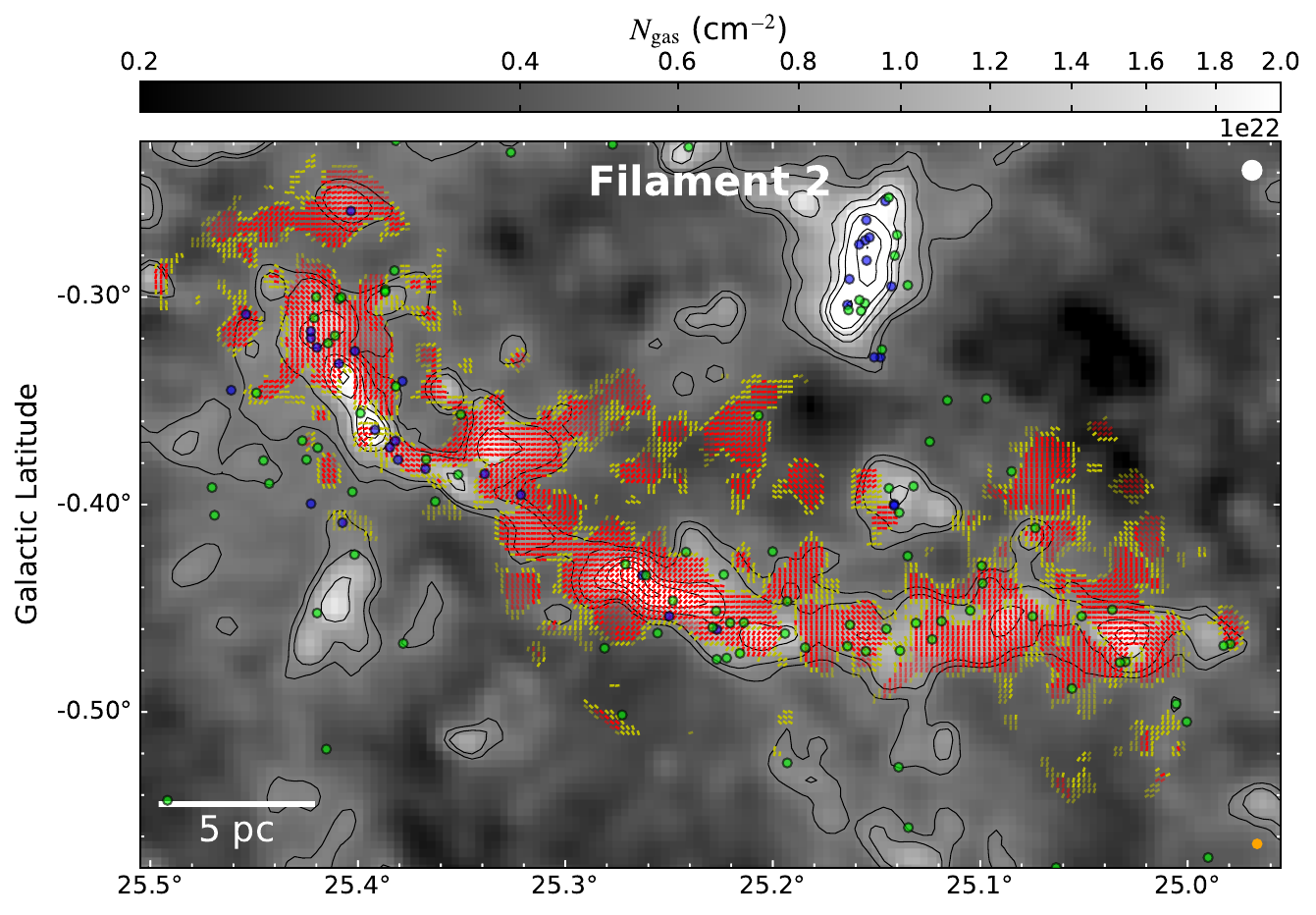}
 \includegraphics[width=0.9\textwidth]{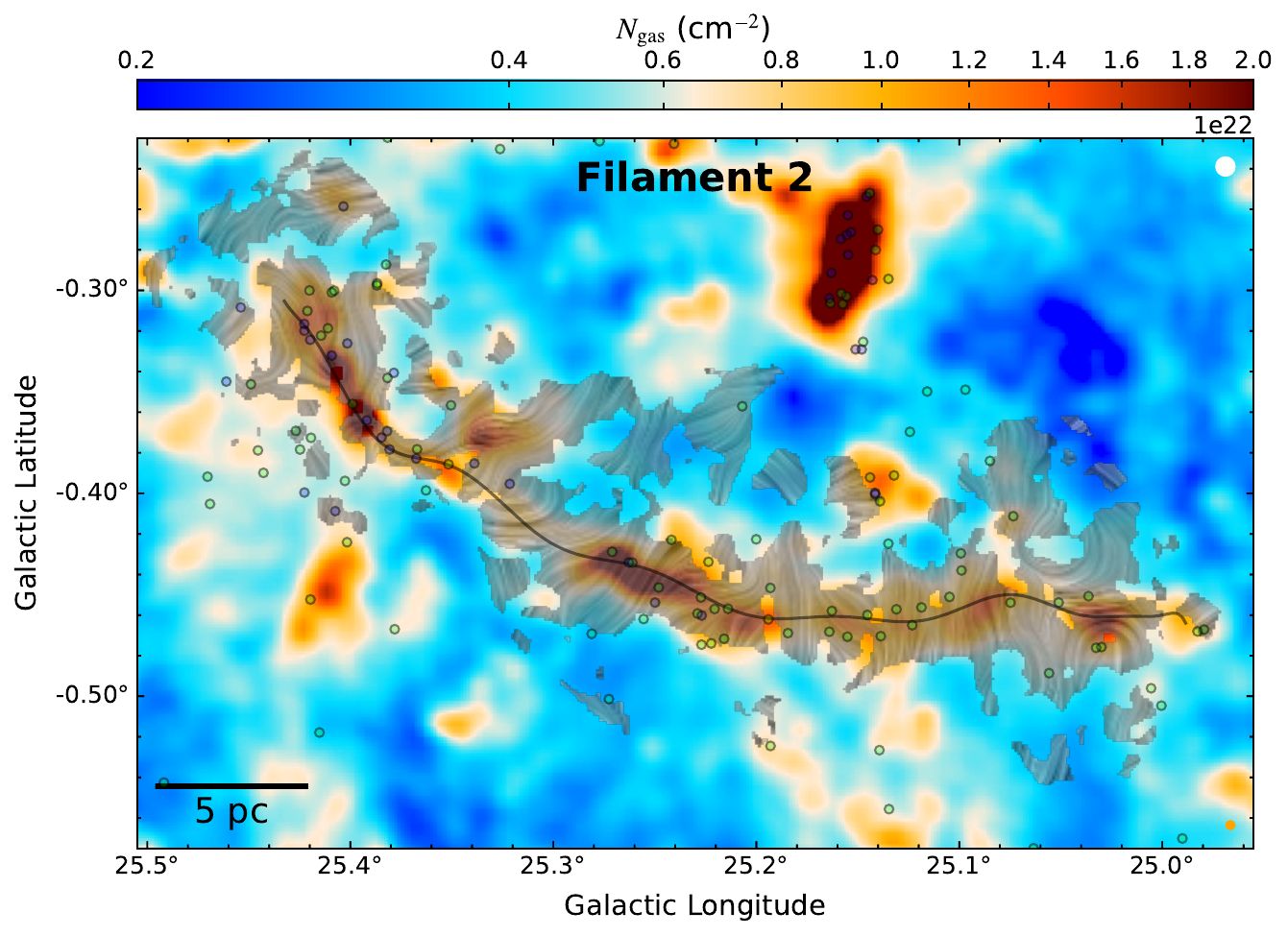}
\end{center} \vspace{-16pt}
 \caption{Figure Caption is the same as \ref{fig:fil1vectors} except now for Filament 2.  \label{fig:fil2vectors}}
 \end{figure*}

\begin{figure*}
\begin{center}
 \includegraphics[width=0.85\textwidth]{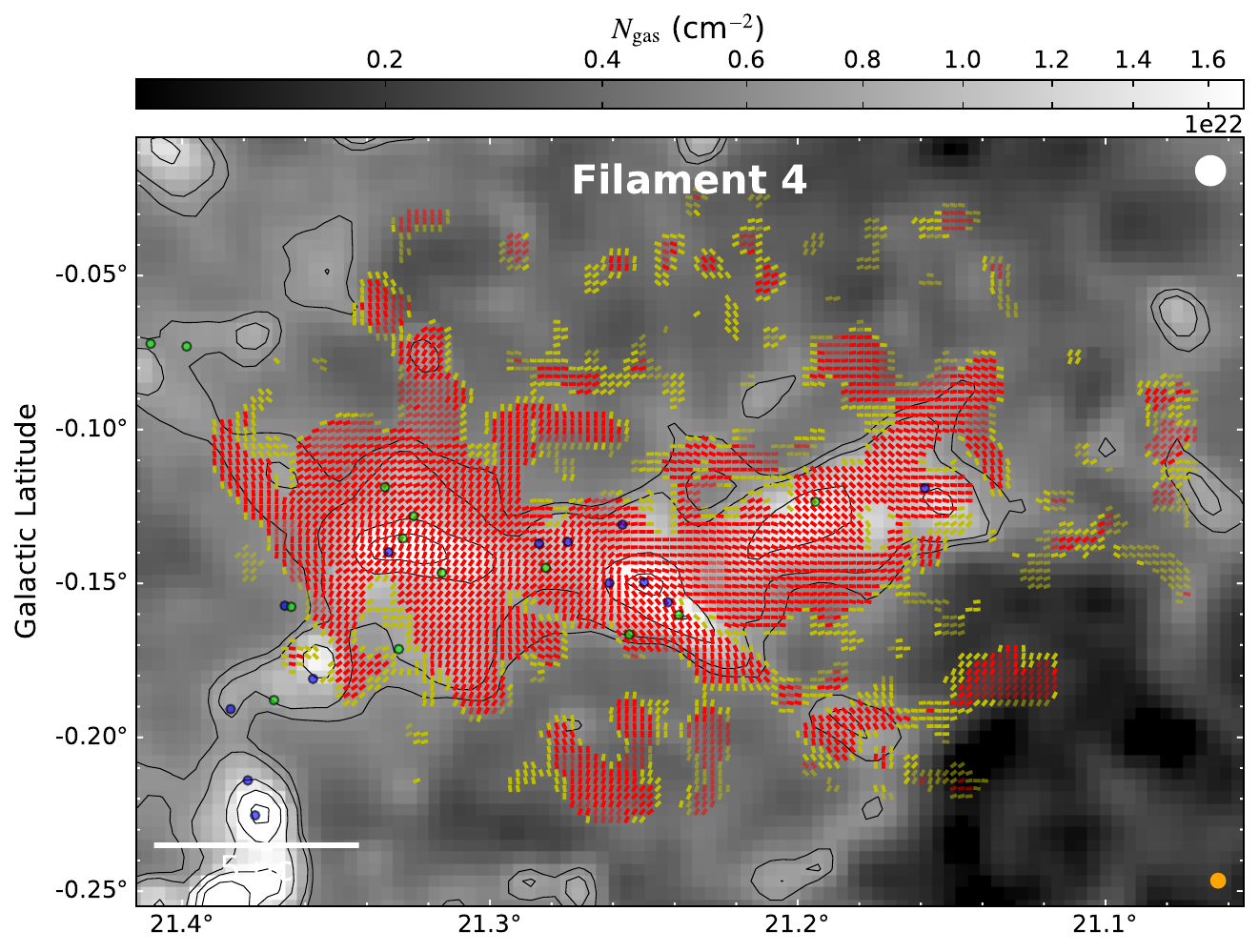}
 \includegraphics[width=0.85\textwidth]{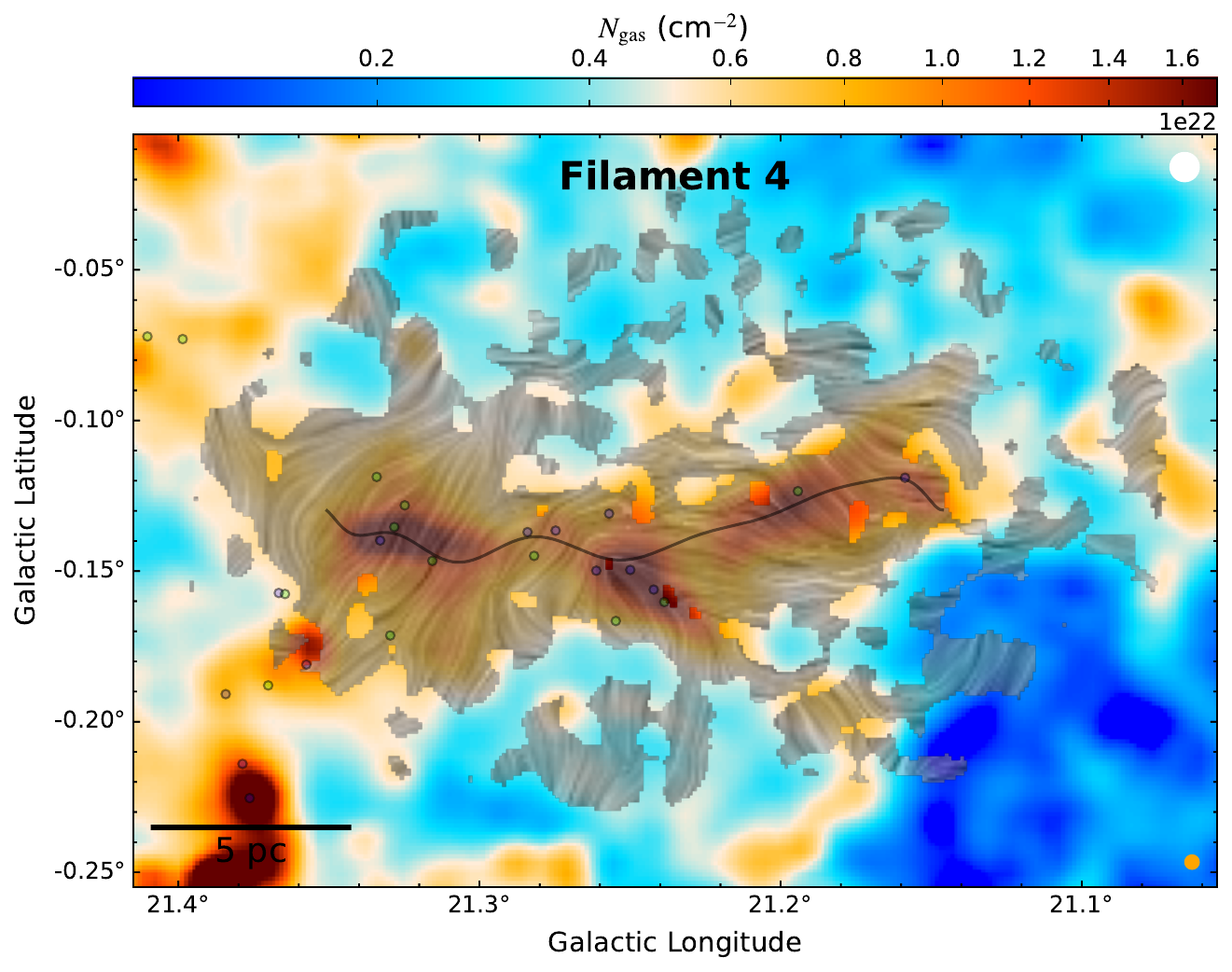}
 \end{center} \vspace{-16pt}
 \caption{Figure Caption is the same as \ref{fig:fil1vectors} except now for Filament 4.  \label{fig:fil4vectors}}
 \end{figure*}

\begin{figure*}
 \begin{center}
 \includegraphics[width=0.9\textwidth]{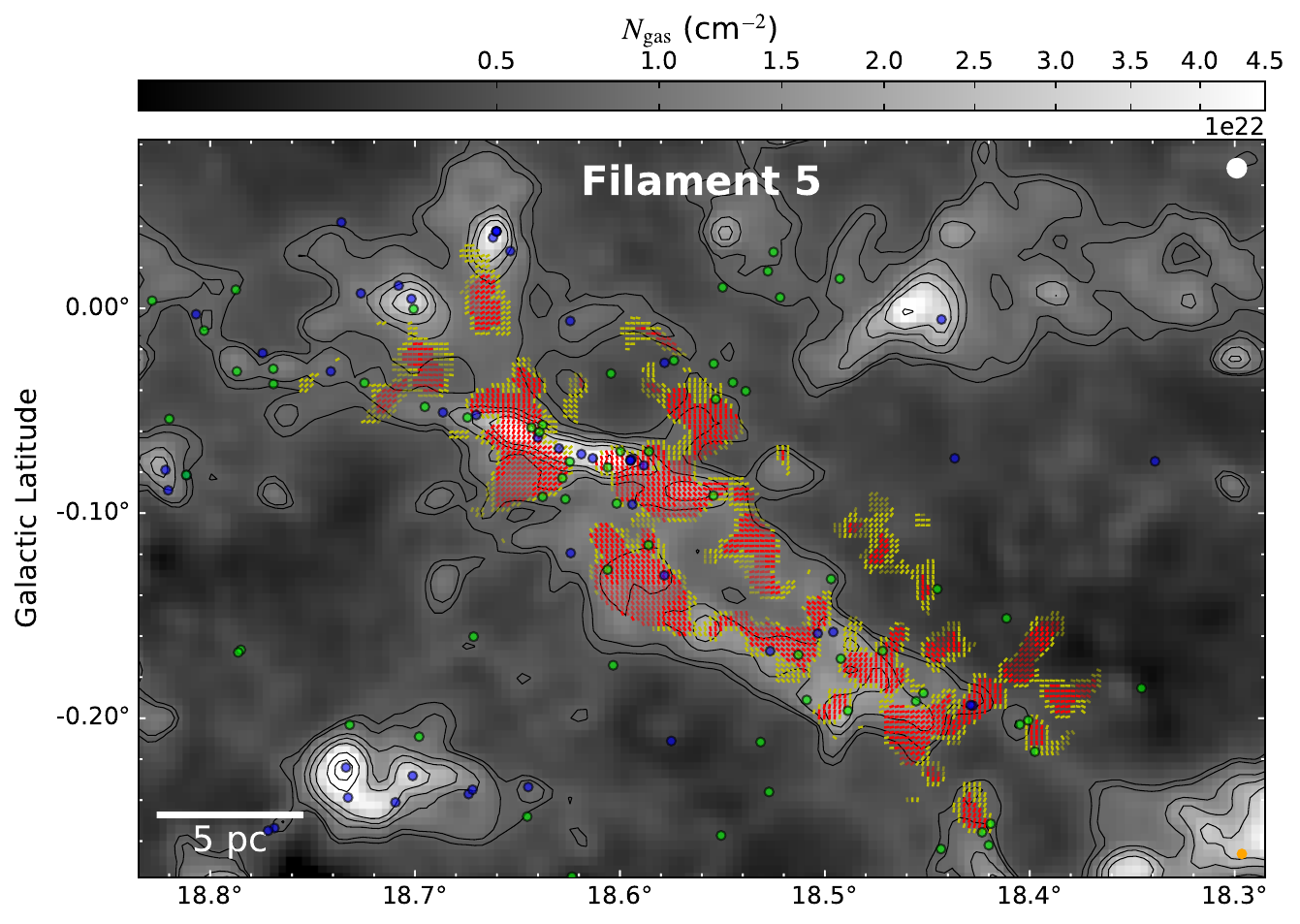}
 \includegraphics[width=0.9\textwidth]{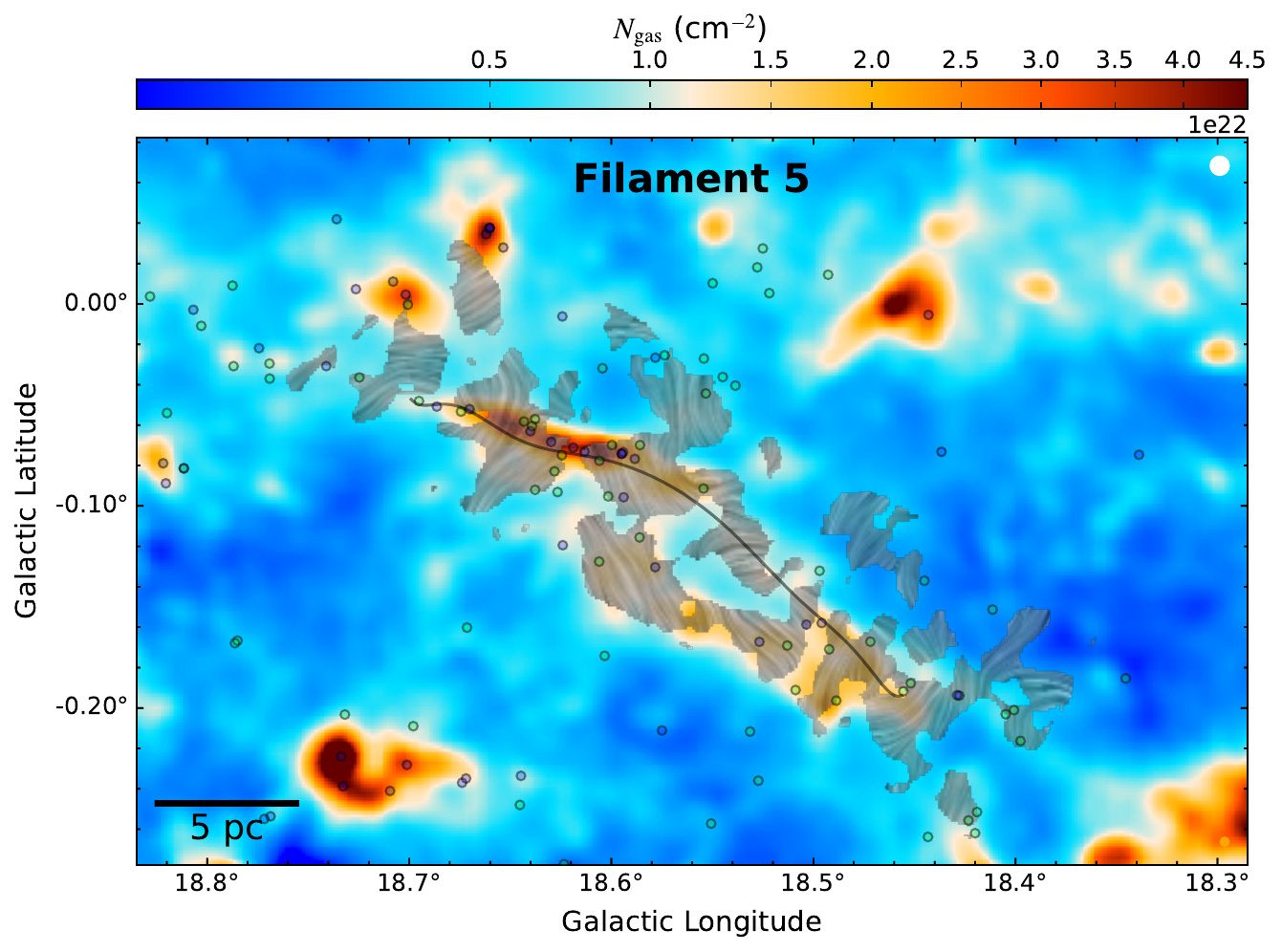}
 \end{center} \vspace{-16pt}
 \caption{Figure Caption is the same as \ref{fig:fil1vectors} except now for Filament 5.  \label{fig:fil5vectors}}
 \end{figure*}

\begin{figure*}
 \begin{center}
 \includegraphics[width=\textwidth]{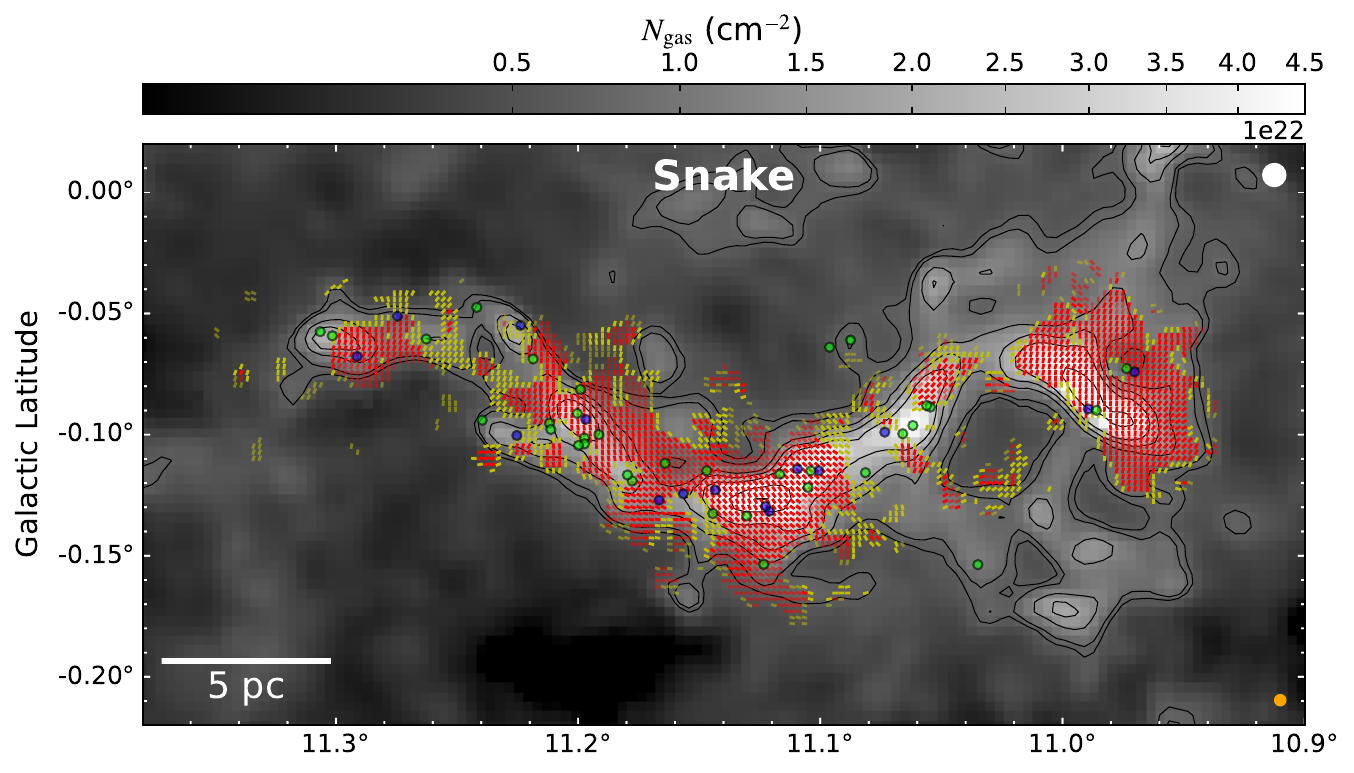}
 \includegraphics[width=\textwidth]{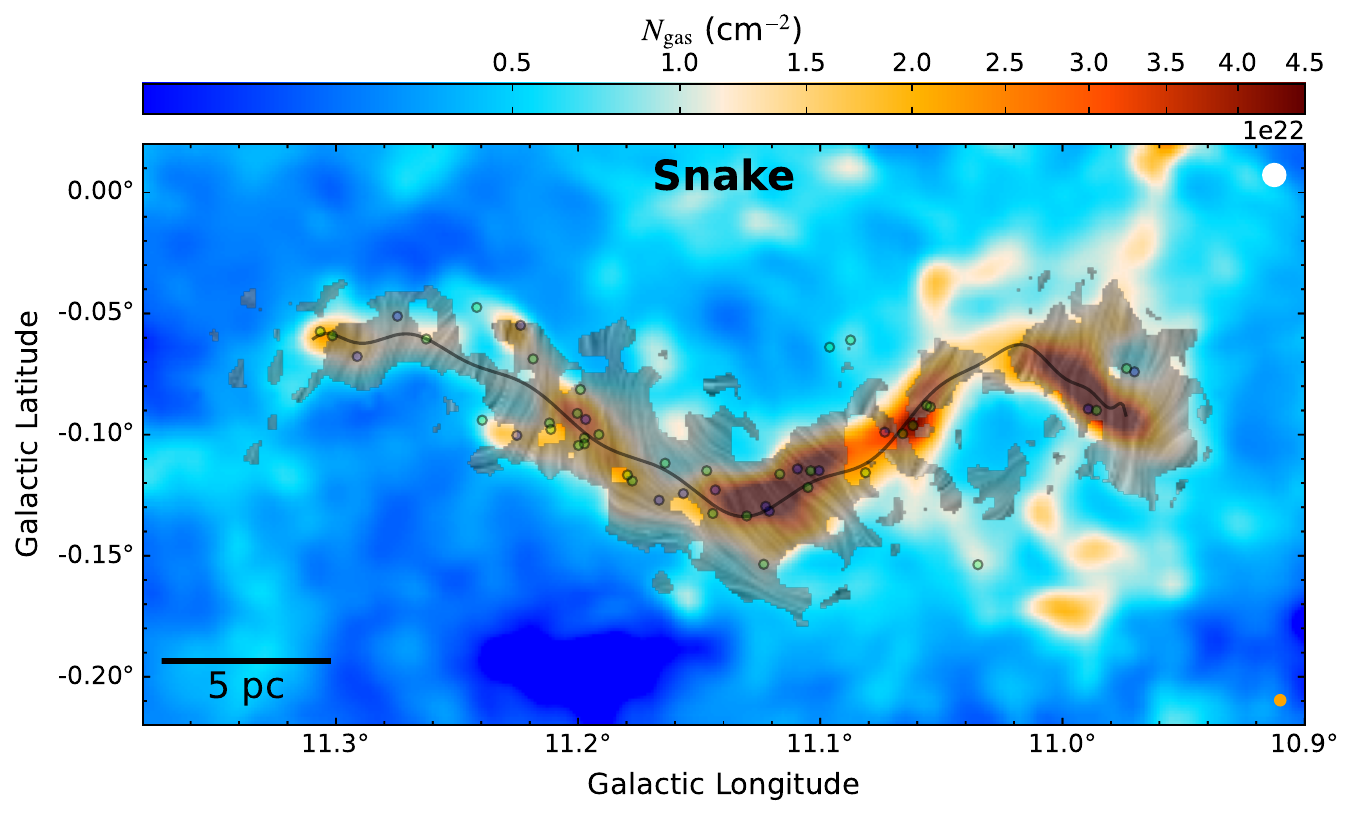}
 \end{center} \vspace{-16pt}
 \caption{Figure Caption is the same as \ref{fig:fil1vectors} except now for the Snake.  \label{fig:snakevectors}}
 \end{figure*}

\begin{figure*}
 \begin{center}
 \includegraphics[width=\textwidth]{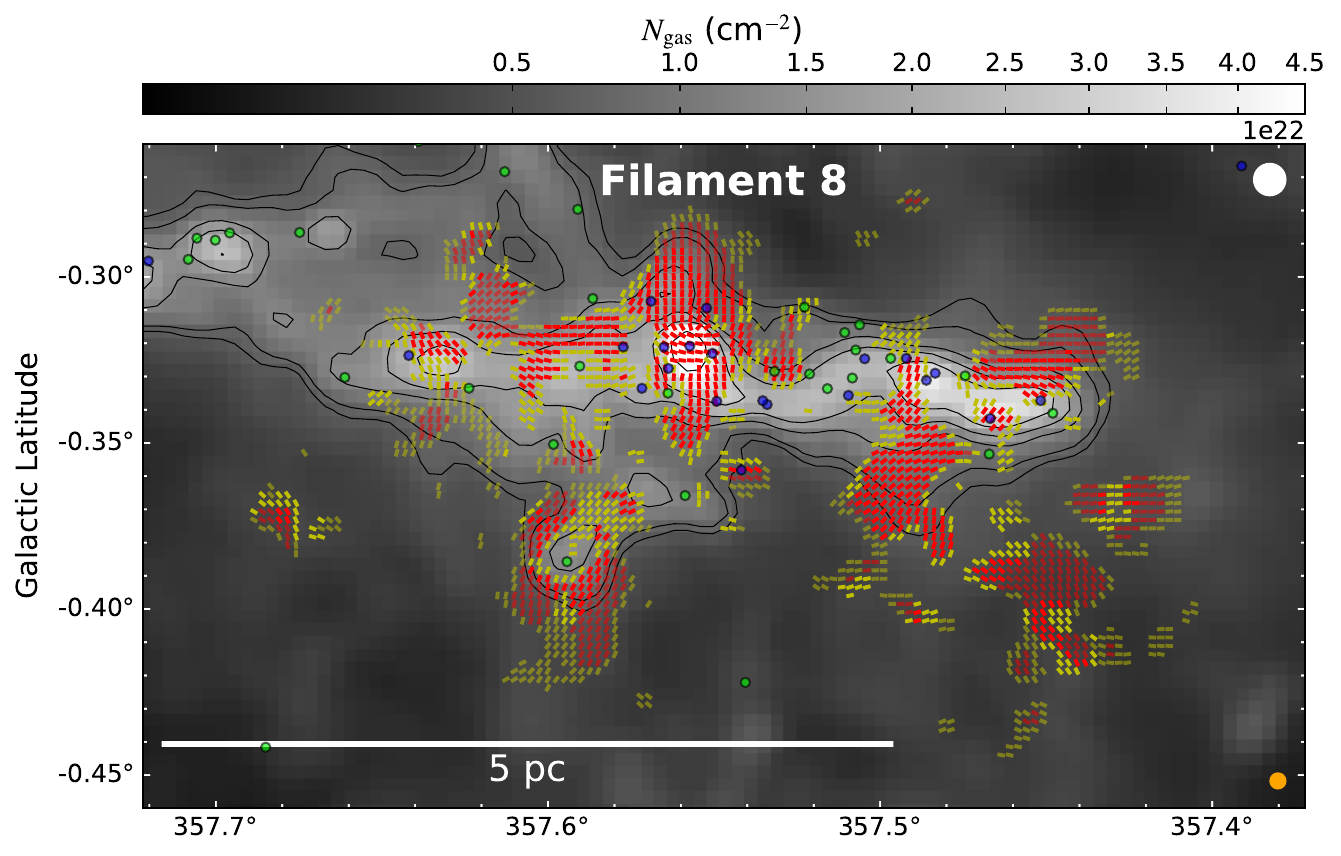}
 \includegraphics[width=\textwidth]{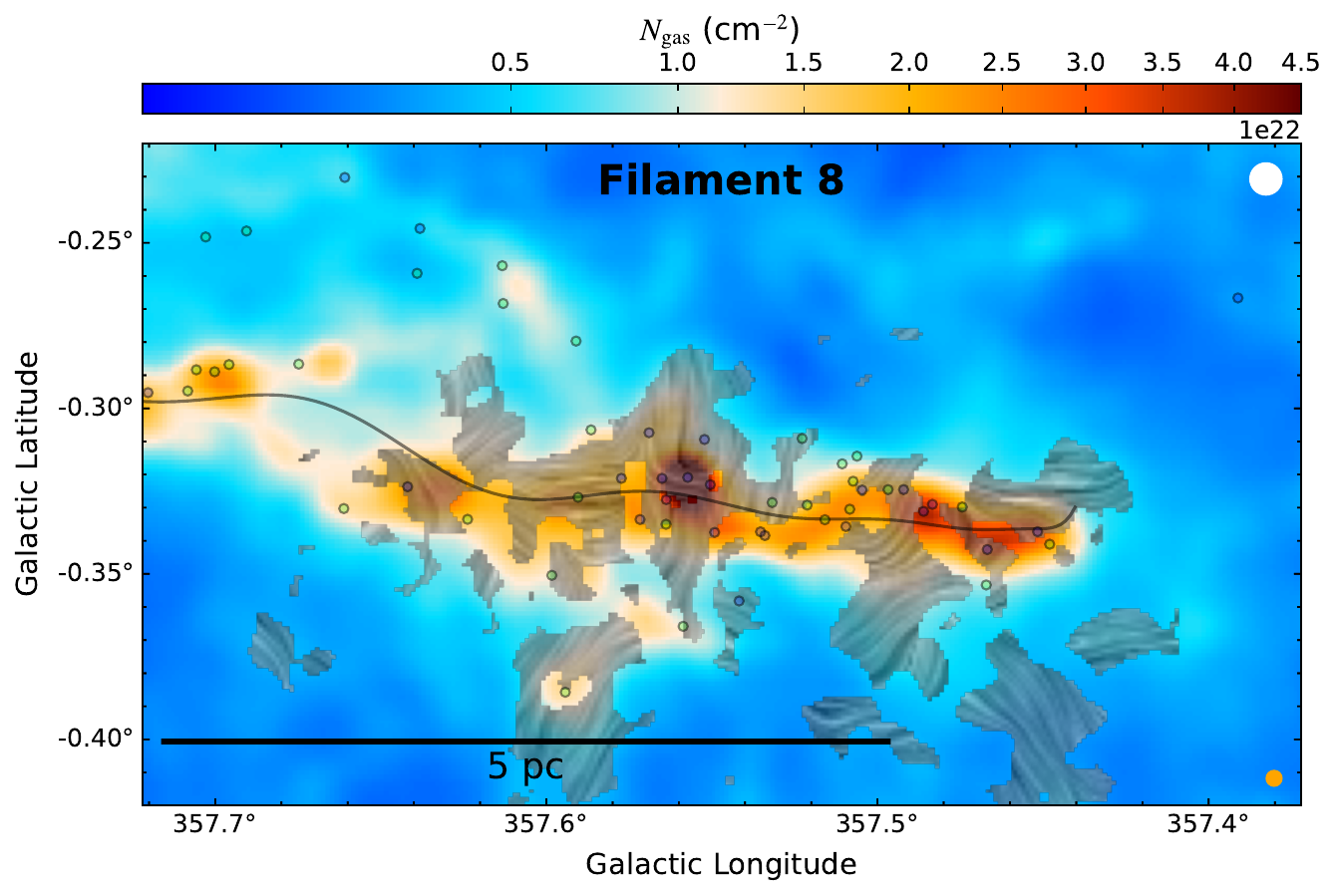}
 \end{center} \vspace{-16pt}
 \caption{Figure Caption is the same as \ref{fig:fil1vectors} except now for Filament 8.  \label{fig:fil8vectors}}
 \end{figure*}

\begin{figure*}
 \begin{center}
 \includegraphics[width=\textwidth,angle=90]{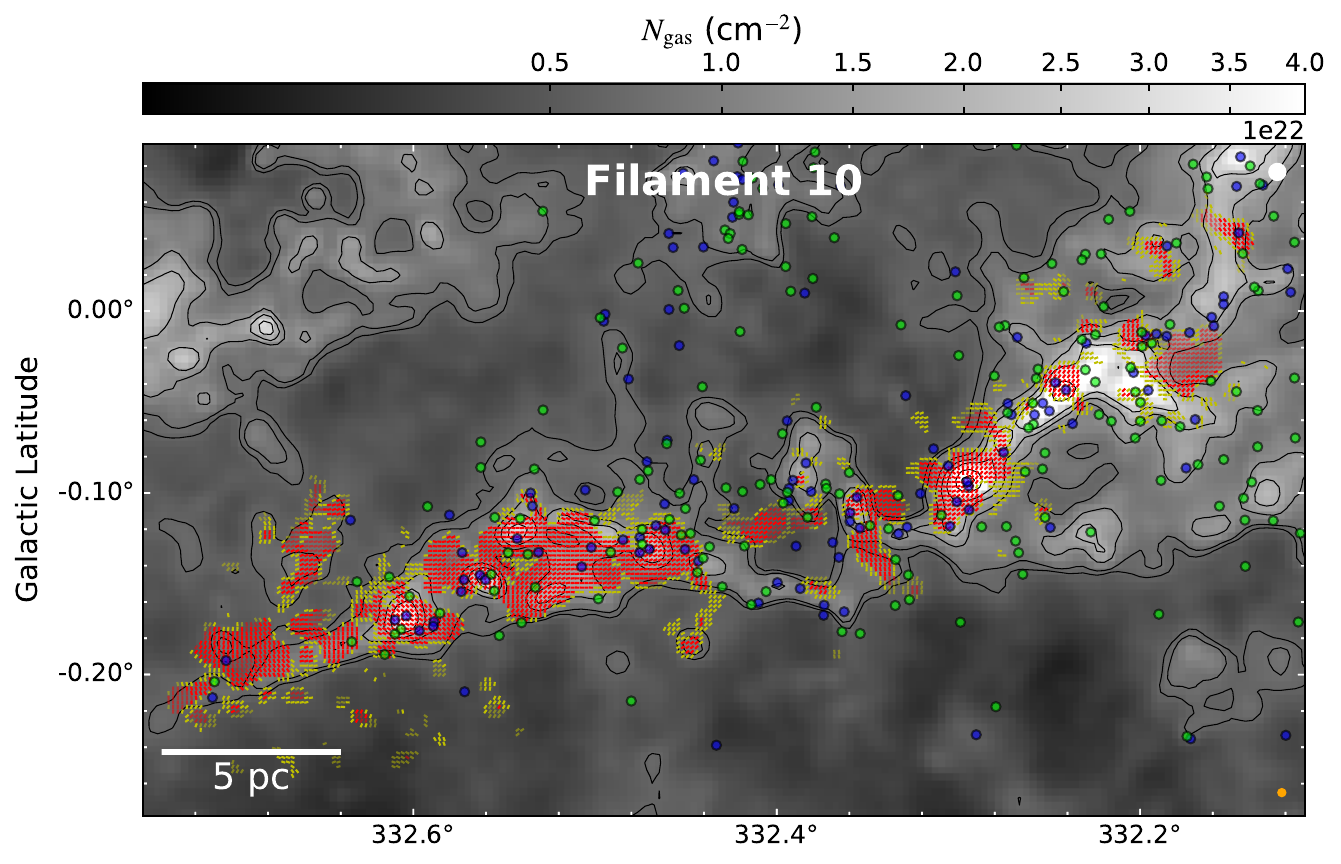}
 \end{center} \vspace{-16pt}
 \caption{Figure Caption is the same as the Top Panel of \ref{fig:fil1vectors} except now for Filament 10.  \label{fig:fil10vectors}}
 \end{figure*}

\begin{figure*}
 \begin{center}
 \includegraphics[width=\textwidth,angle=90]{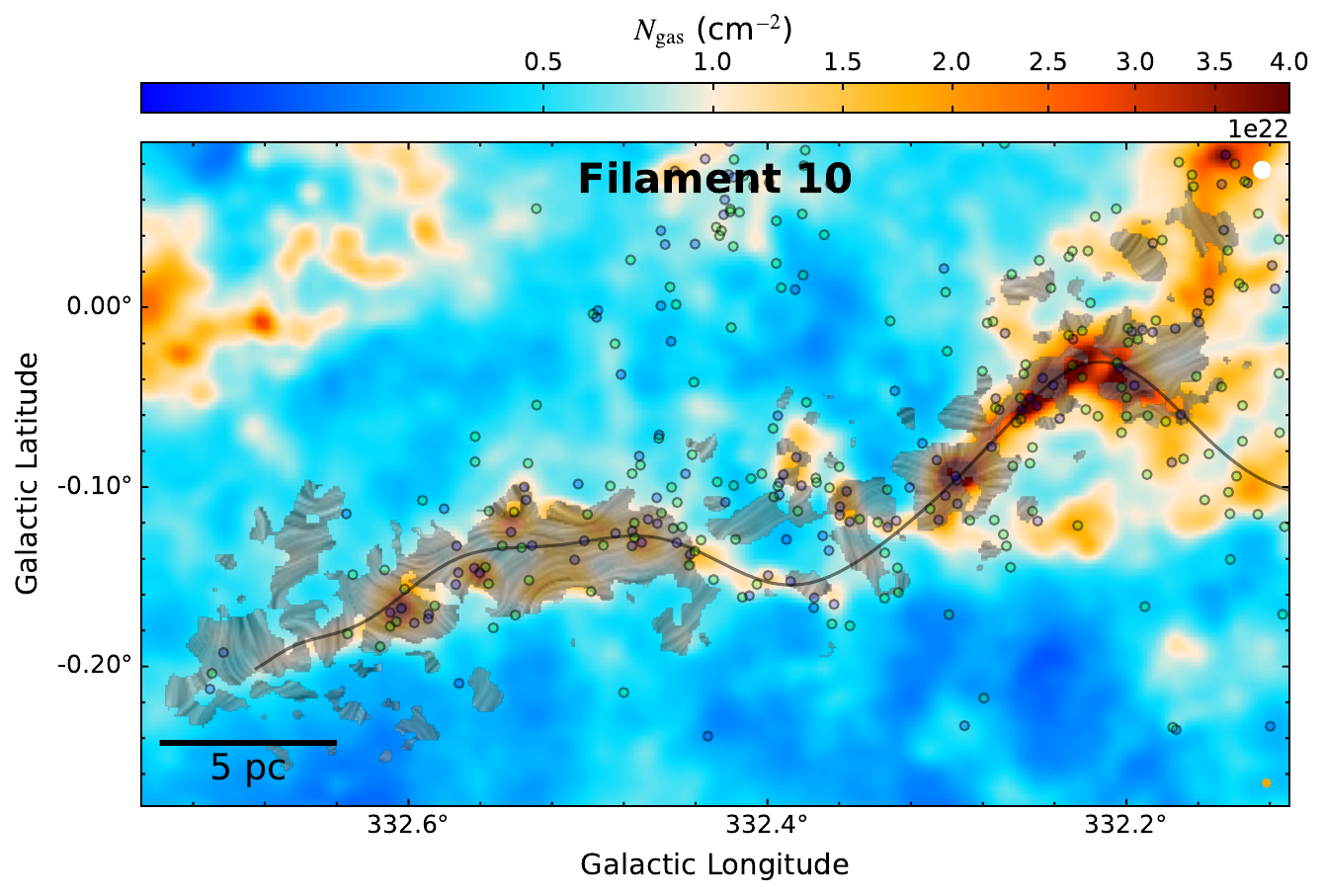}
 \end{center} \vspace{-16pt}
 \caption{Figure Caption is the same as the Bottom Panel of \ref{fig:fil1vectors} except now for Filament 10.  \label{fig:fil10vectors_b}}
 \end{figure*}

\begin{figure*}
\begin{center}
\includegraphics[width=\textwidth,angle=90]{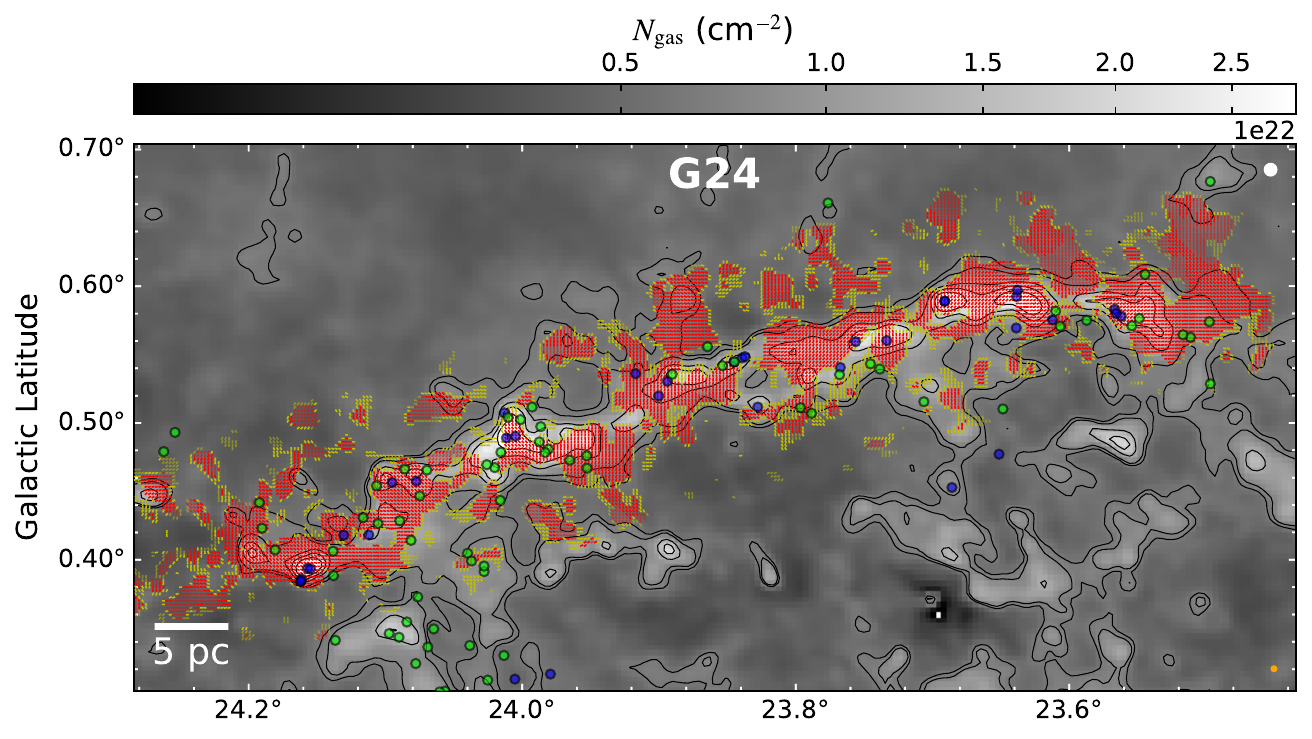}
\end{center} \vspace{-16pt}
\caption{Figure Caption is the same as the Top Panel of \ref{fig:fil1vectors} except now for G24.  \label{fig:g24vectors}}
\end{figure*}

\begin{figure*}
\begin{center}
\includegraphics[width=\textwidth,angle=90]{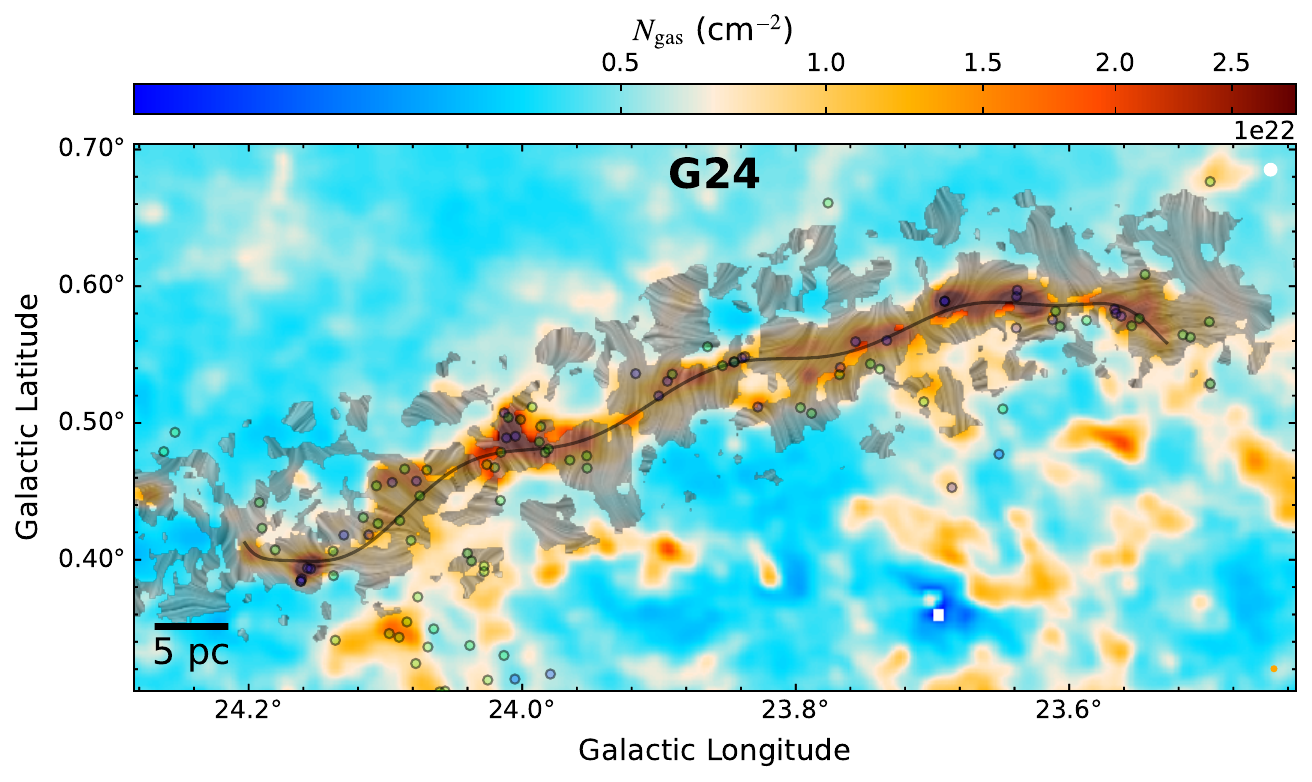}
\end{center} \vspace{-16pt}
\caption{Figure Caption is the same as the Bottom Panel of \ref{fig:fil1vectors} except now for G24.  \label{fig:g24vectors_b}}
\end{figure*}

\begin{figure*}
 \begin{center}
 \includegraphics[width=0.9\textwidth]{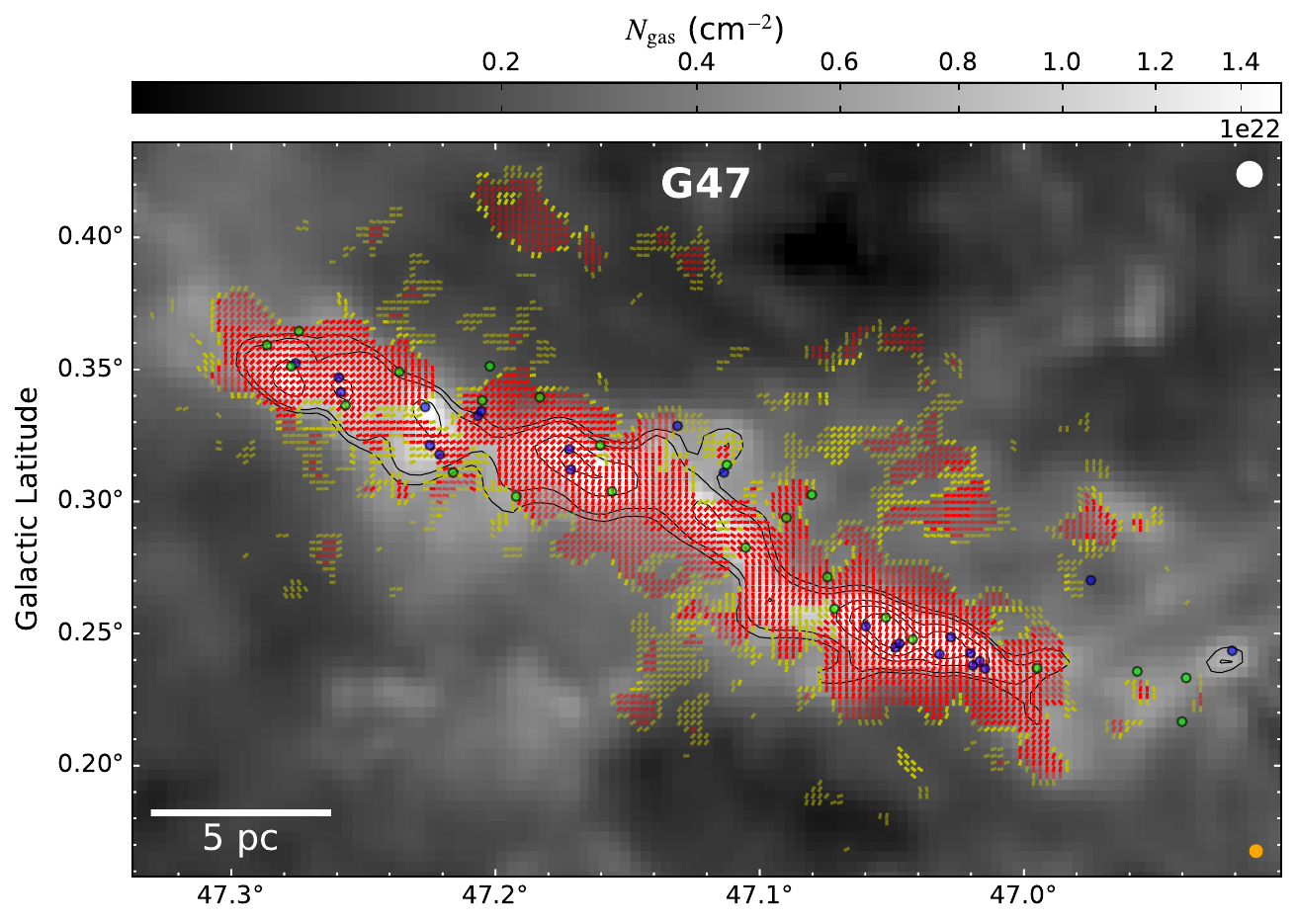}
 \includegraphics[width=0.9\textwidth]{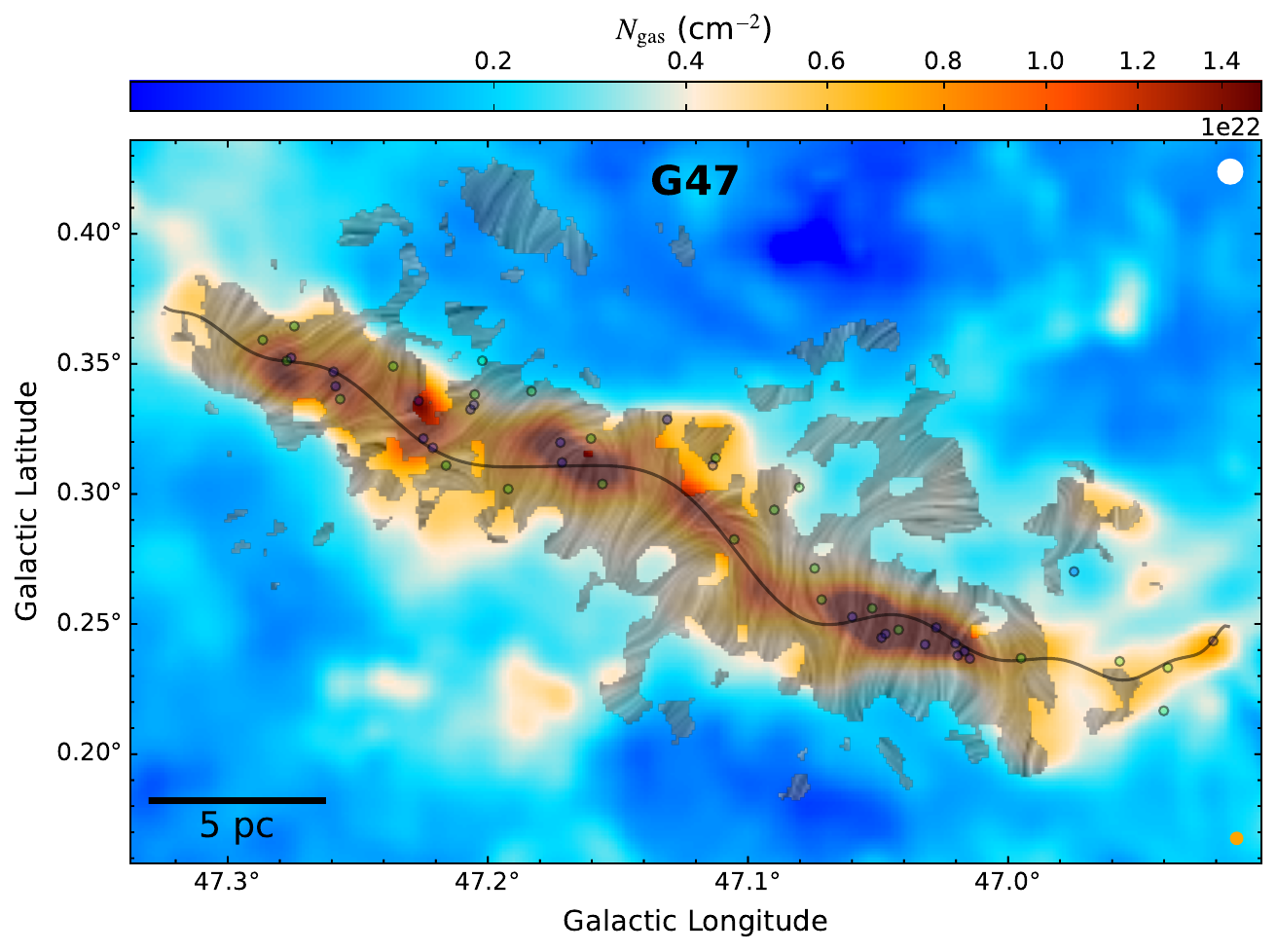}
 \end{center} \vspace{-16pt}
 \caption{Figure Caption is the same as \ref{fig:fil1vectors} except now for G47.  \label{fig:g47vectors}}
 \end{figure*}

\begin{figure*}
  \begin{center}
 \includegraphics[width=\textwidth]{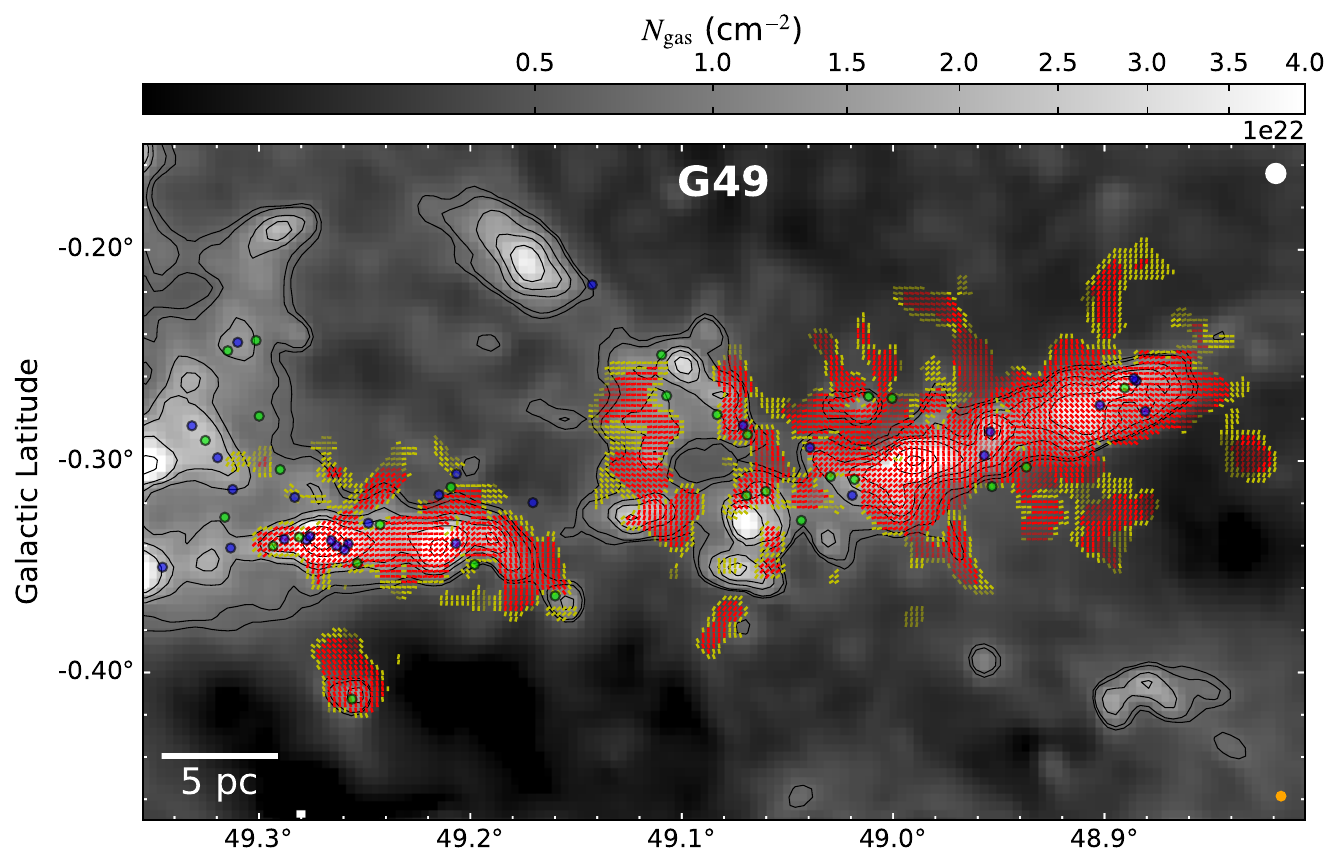}
 \includegraphics[width=\textwidth]{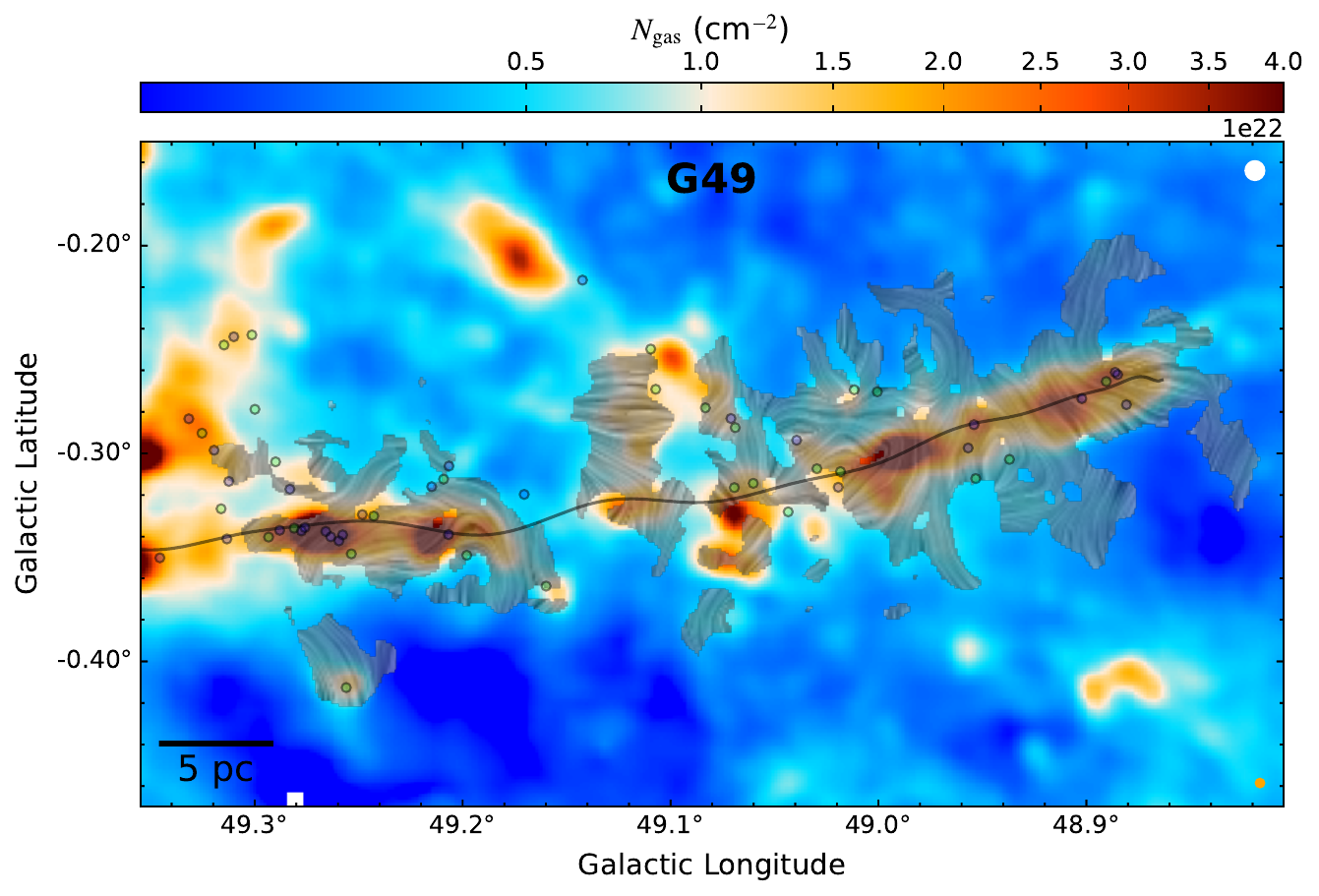}
 \end{center} \vspace{-16pt}
 \caption{Figure Caption is the same as \ref{fig:fil1vectors} except now for G49.  \label{fig:g49vectors}}
 \end{figure*}

Figures~\ref{fig:fil2vectors} -- \ref{fig:g49vectors} present the vector and LIC maps for all bones, as only Filament~1 is shown in the main text (Figure~\ref{fig:fil1vectors}). Spacing for the top panels are shown for every two pixels, which is approximately Nyquist sampling. We choose to display vectors this way, as it shows the most robust detail possible of the magnetic field morphology. Some readers may prefer more spacing between vectors for some bones to make the morphology more clearer. Those sort of maps are shown in \citetalias{Coude2025}.

%

\section{Velocity Dispersion Comparisons}\label{appendix:linewidths}
\begin{figure*}
\begin{center}
 \includegraphics[width=0.32\textwidth]{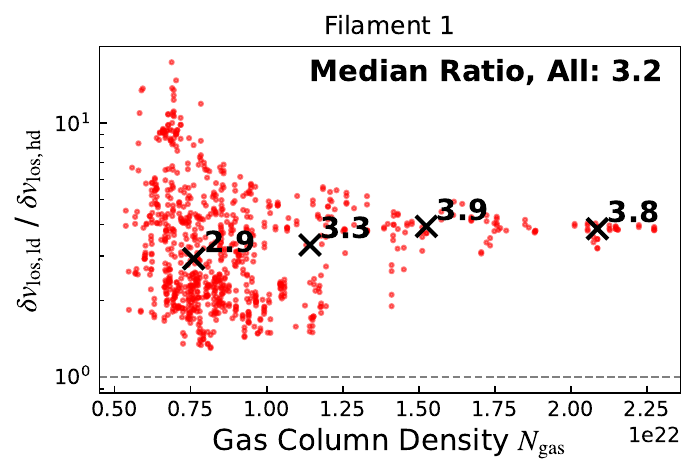}
 \includegraphics[width=0.32\textwidth]{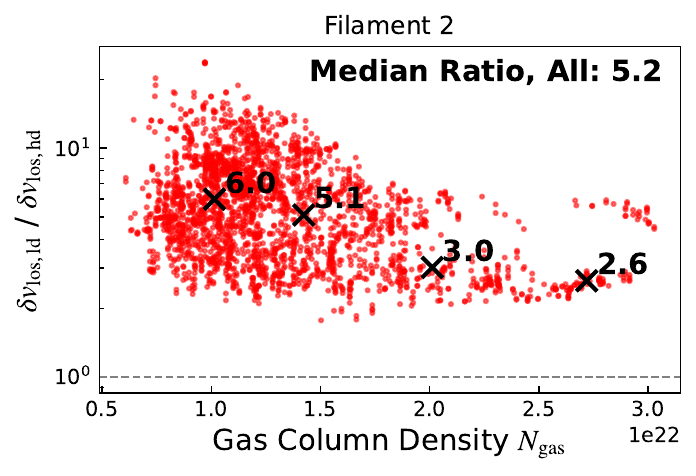}
 \includegraphics[width=0.32\textwidth]{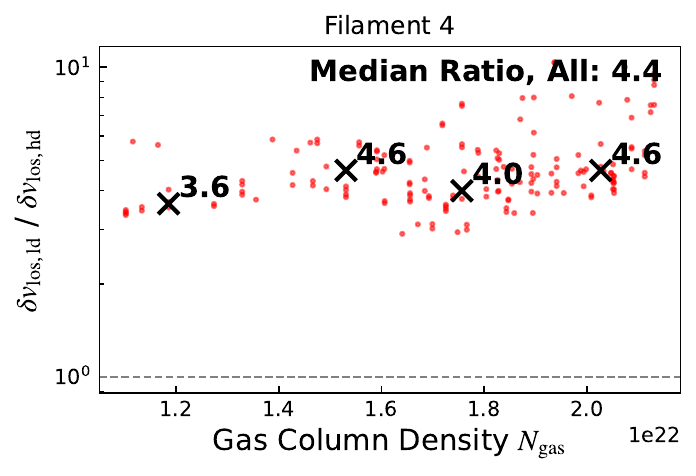}
 \includegraphics[width=0.32\textwidth]{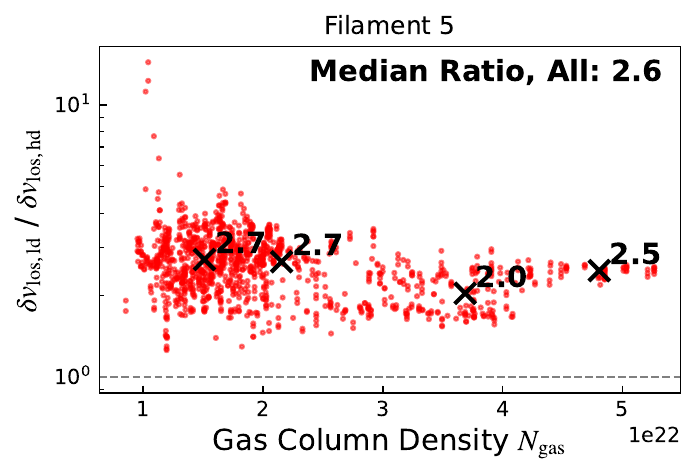}
 \includegraphics[width=0.32\textwidth]{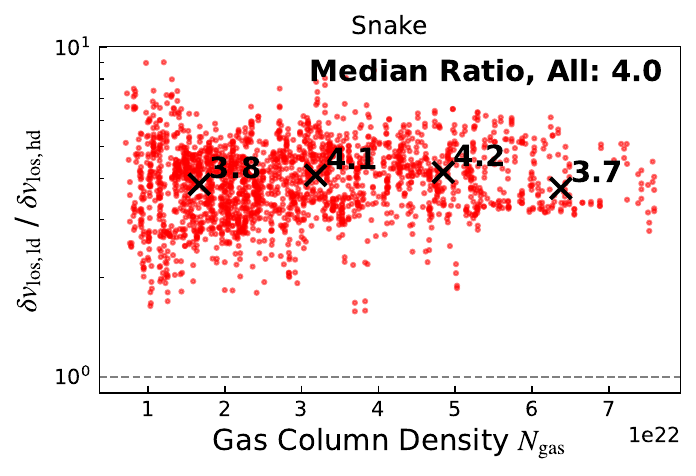}
 \includegraphics[width=0.32\textwidth]{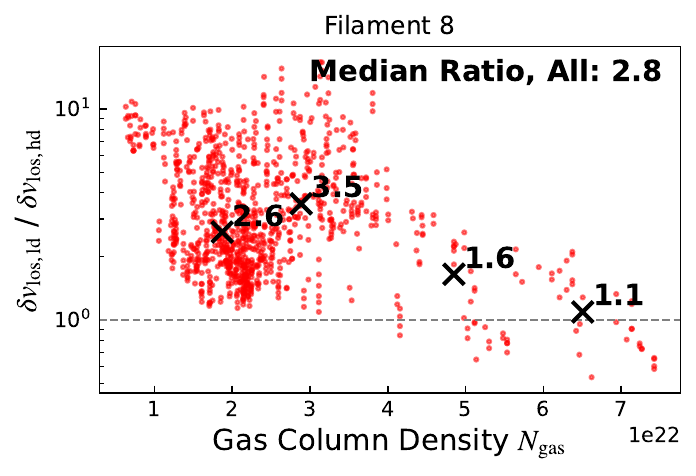}
 \includegraphics[width=0.32\textwidth]{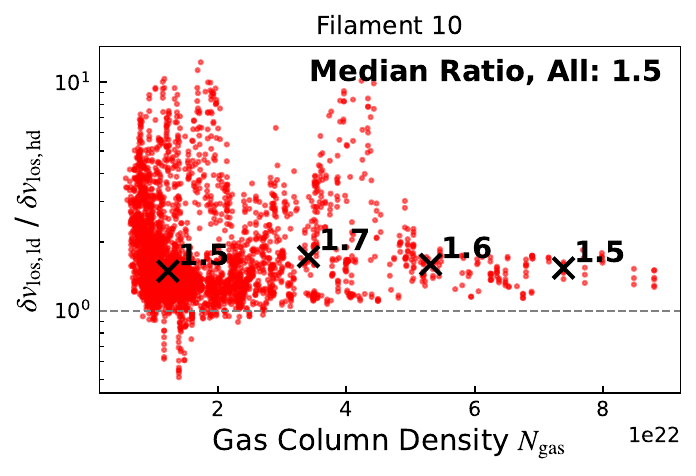}
 \includegraphics[width=0.32\textwidth]{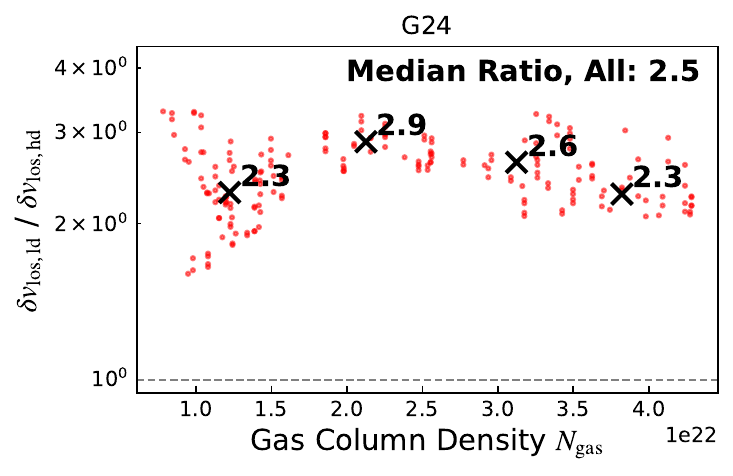}
 \includegraphics[width=0.32\textwidth]{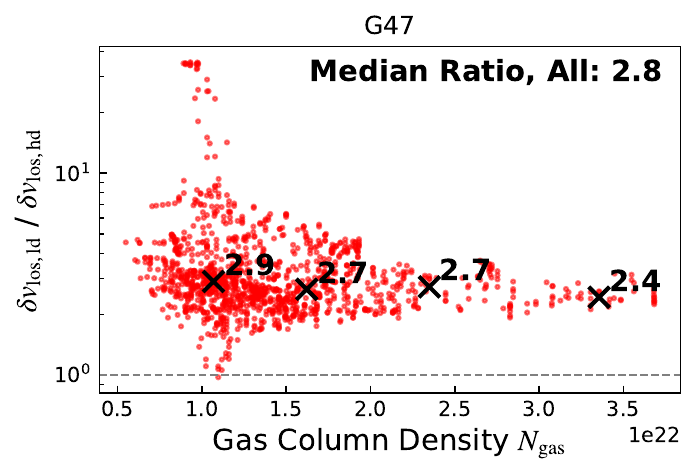}
 \includegraphics[width=0.32\textwidth]{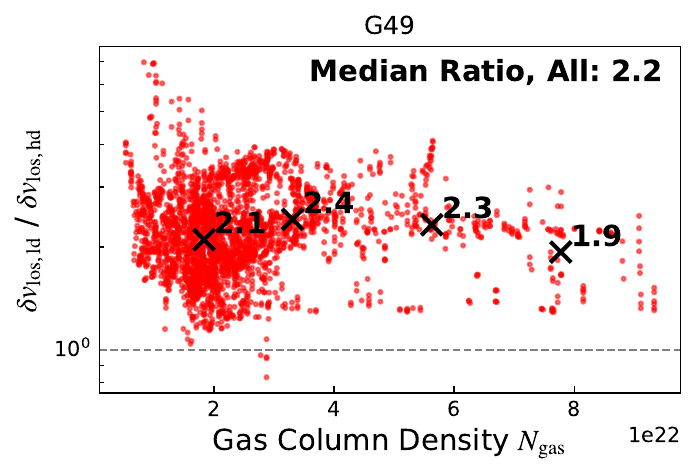}
\end{center}
 \caption{Ratio in measured velocity dispersions of the high-density tracer ($\delta v_{\text{los,ld}}$)  and low-density tracer ($\delta v_{\text{los,hd}}$) as a function of gas column density $N_{\mathrm{gas}}$ for each bone. Data are binned in four equal-width intervals in \Ngas\, with black crosses show the median ratio per bin. The dashed line marks a ratio of unity. The overall median ratio is indicated in the upper right of each panel. Comparisons are only shown where \pfrac/\sigpfrac\,$>$\,2 and pixels are no more than 23$\arcsec$ from the spine. \label{fig:lineratios}.}
 \end{figure*}
 
In Table~\ref{tab:lines}, we report the low and high density line tracers used for the paper. In summary, all bones use \nht\ for their high density tracer, except Filament~8 and Filament~10, which use \ceoto. For low-density tracers, all bones use \ttcooz, except the Snake, Filament~8, and Filament~10, which use \ttcoto. For the sliding box analysis (Section~\ref{sec:sliding}), we are specifically interested in measuring the line-of-sight velocity dispersion $\delta v_{\text{los}}$. Here, we do a point by point comparison of the velocity dispersions measured near the spine of the bone by choosing only those pixels within 23$\arcsec$ of the polynomial fit of the spine of the bone (see Section~\ref{sec:spines}), and where we detect polarization at a level \pfrac/\sigpfrac\,$>$\,2. Figure~\ref{fig:lineratios} shows the ratio in velocity dispersions of the high-density tracer ($\delta v_{\text{los,ld}}$)  and low-density tracer ($\delta v_{\text{los,hd}}$) as a function of column density. Points in the plot are heavily oversampled, as the values use the velocity dispersion maps have been regridded to the SOFIA resolution since the regridded maps are used for the sliding box analysis. Since median values are reported, they are largely unaffected by the oversampling.

$\delta v_{\text{los,ld}}$ is consistently higher than $\delta v_{\text{los,hd}}$, usually by a factor of $\sim$2 or more, and the median values are mostly consistent with column density. The obvious exception is Filament~10, and the higher column density data for Filament~8. These two bones uses different line tracers from the SEDIGISM data, which is different than the rest of the sample. These values may be expected to be more close, as the tracers are taken at the same angular resolution (30$\arcsec$), while the \nht\ and \ttcooz\ are at different resolutions (32$\arcsec$ vs 45$\arcsec$), and the higher $J$ transitions trace higher densities. However, we caution that \ceoto, which is used for Filament~8 and Filament~10, may not be as good as a high-density tracer as \nht; the velocity dispersions ratios for the Snake, which compares \nht\ to \ttcoto, are higher than that of these two bones, indicating that \nht\ spectra likely have smaller velocity dispersions. The fact that \ceoto\ may overestimate the linewidth is reflected in Figure~\ref{fig:fourpanels_n}, as Filament~8 and Filament~10 tend to show higher virial parameters than the other bones.

The velocity dispersion ratios are exceptionally high for Filament~2. Visual inspection of the the spectra across the bone show that \ttcooz\ has a severely flattened peak which is a result of a very optically thick line. In turn, fitting spectra to such optically thick lines result in a large velocity dispersions.

\section{Cumulative Distribution Functions of Alignment}\label{appendix:cdfs}
\begin{figure*}[ht!]
\begin{center}
\includegraphics[width=2\columnwidth]{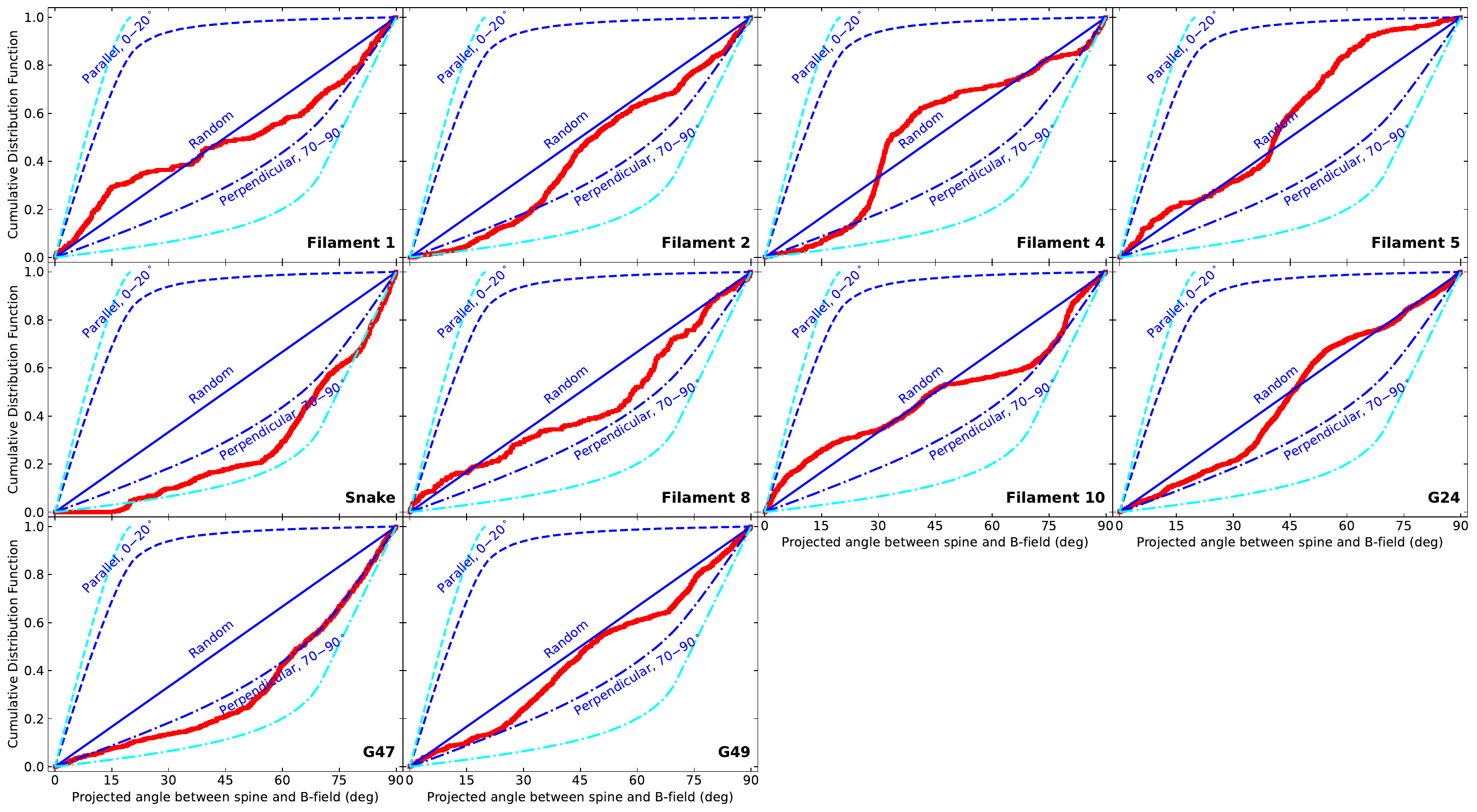}
\end{center}
\caption{Caption is the same as Figure~\ref{fig:cdfs}, but now only showing data in the first quartile (lowest values) for column density.} 
\label{fig:cdfs1} 
\end{figure*}

\begin{figure*}[ht!]
\begin{center}
\includegraphics[width=2\columnwidth]{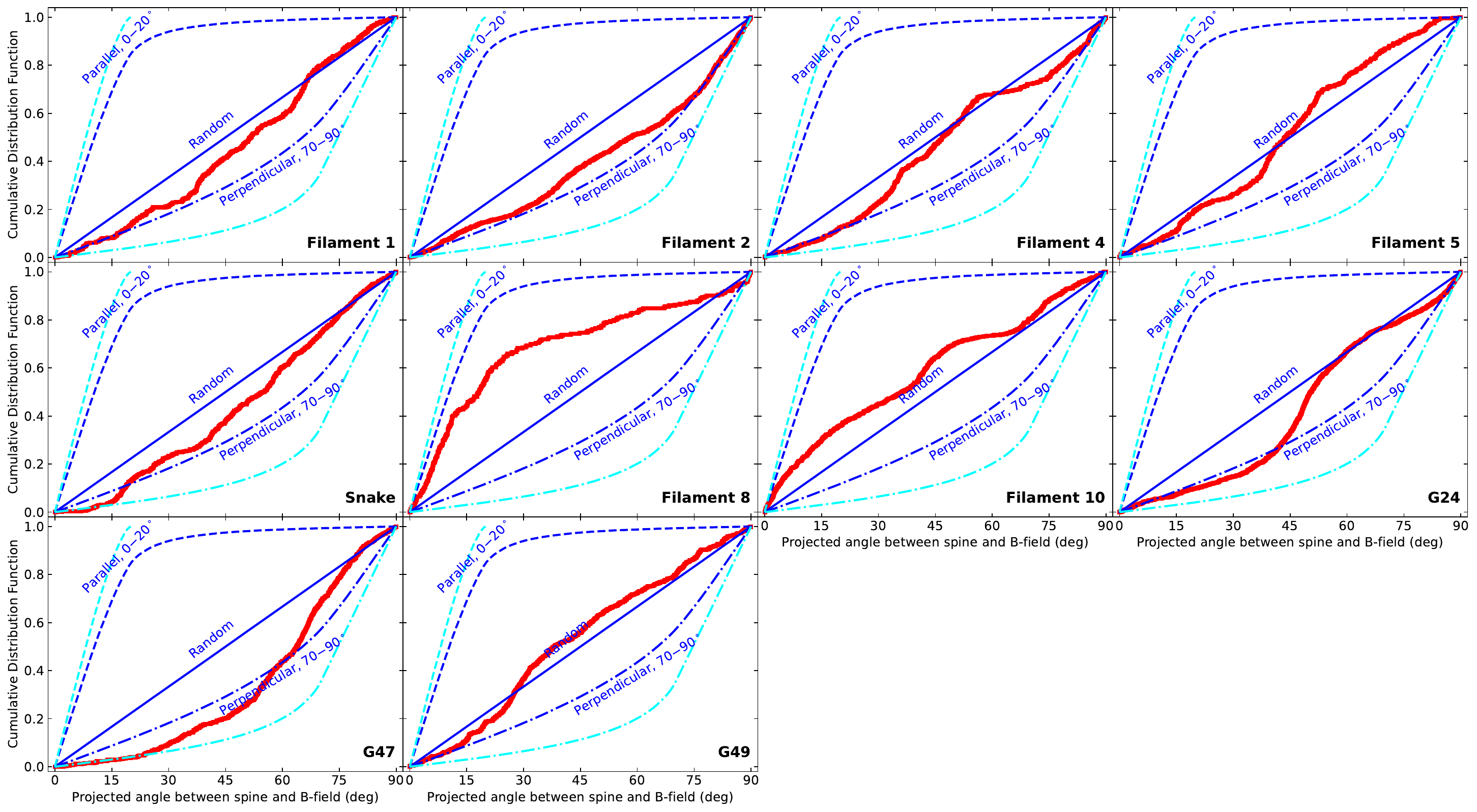}
\end{center}
\caption{Caption is the same as Figure~\ref{fig:cdfs}, but now only showing data in the second quartile for column density.} 
\label{fig:cdfs2} 
\end{figure*}

\begin{figure*}[ht!]
\begin{center}
\includegraphics[width=2\columnwidth]{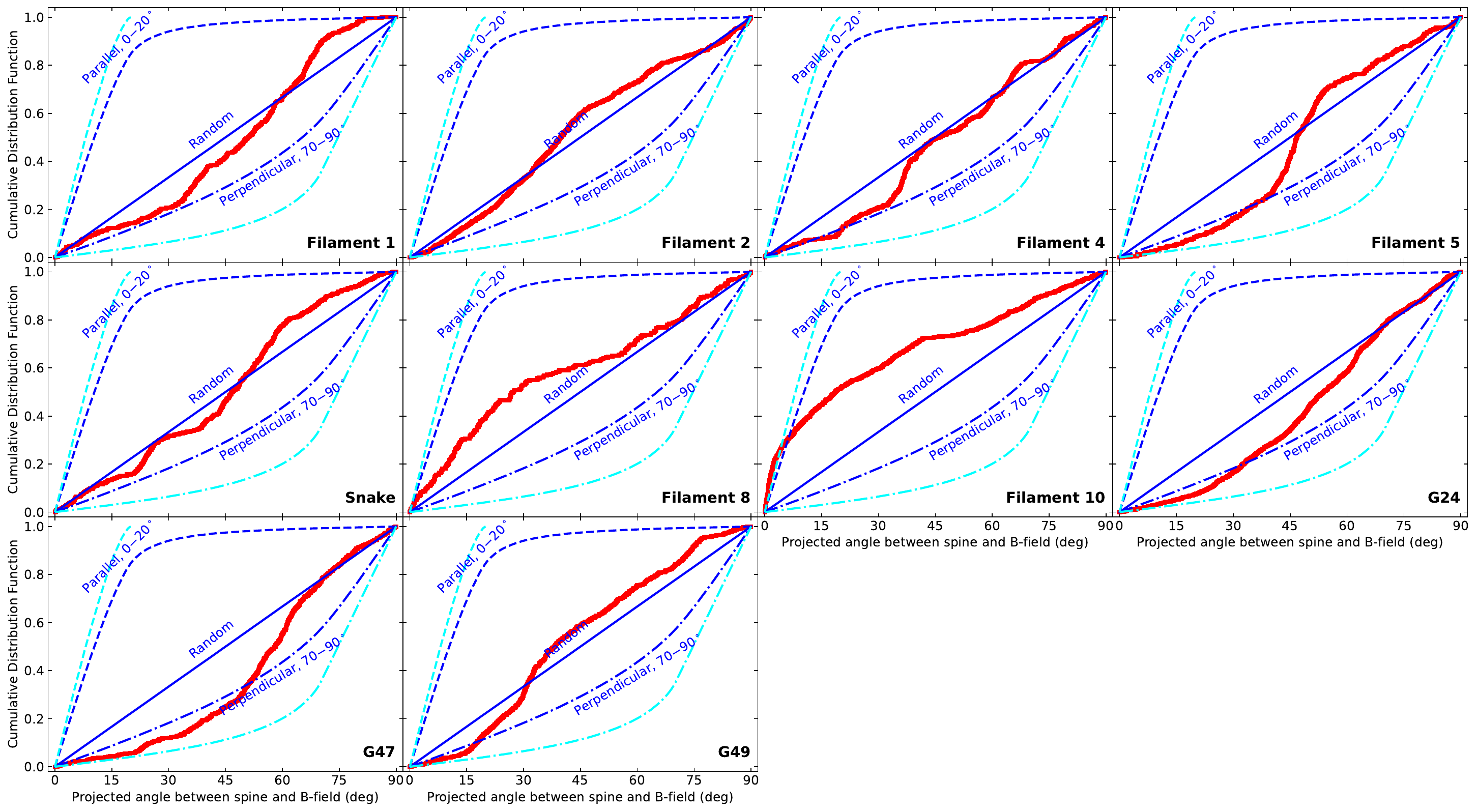}
\end{center}
\caption{Caption is the same as Figure~\ref{fig:cdfs}, but now only showing data in the third quartile for column density.} 
\label{fig:cdfs3} 
\end{figure*}

\begin{figure*}[ht!]
\begin{center}
\includegraphics[width=2\columnwidth]{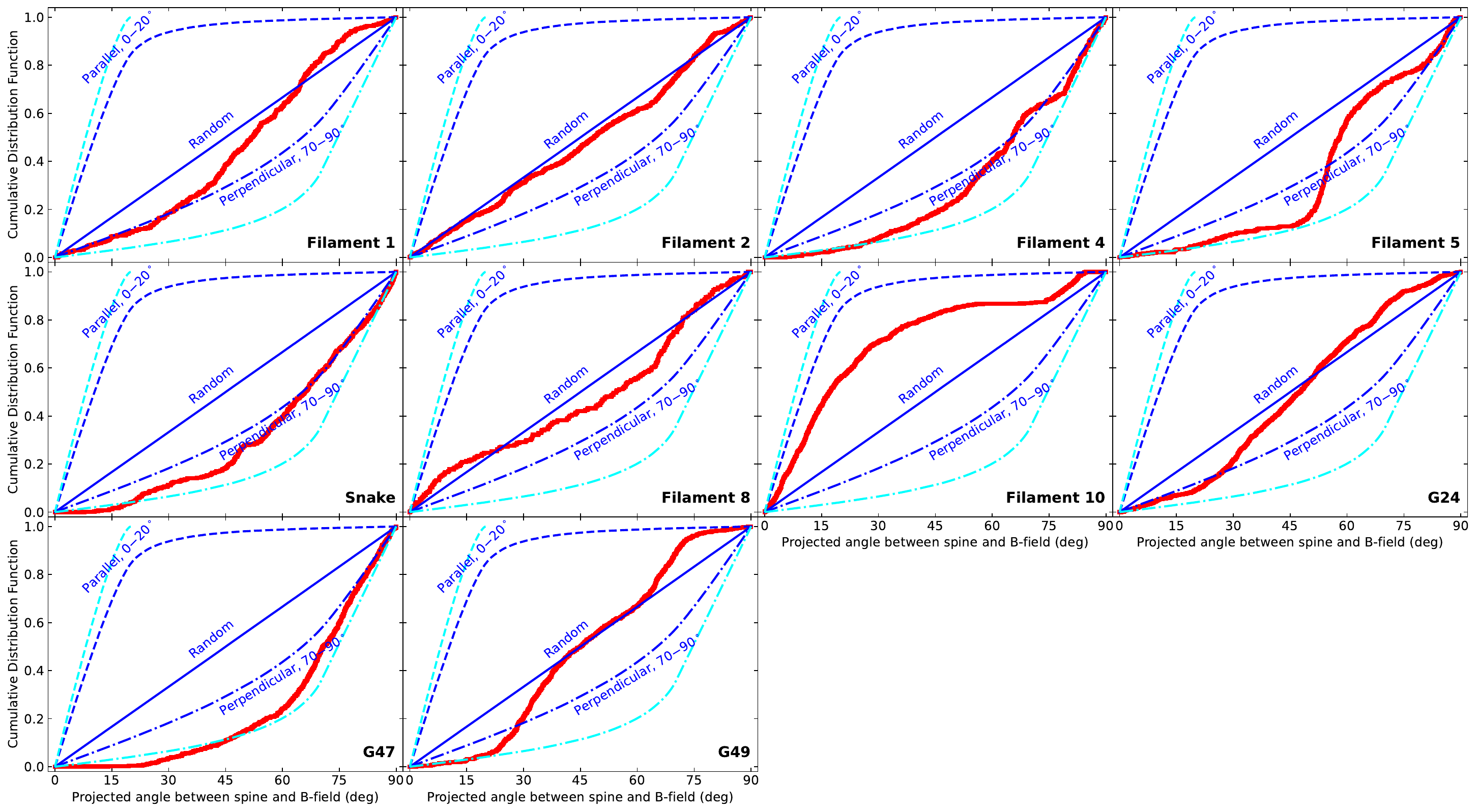}
\end{center}
\caption{Caption is the same as Figure~\ref{fig:cdfs}, but now only showing data in the fourth quartile (highest values) for column density.} 
\label{fig:cdfs4} 
\end{figure*}

This appendix provides additional figures of cumulative distribution functions presented in Section~\ref{sec:orientationbones}, which compares the distribution of the projection angle difference between the bone's spine and the B-field position angle for each pixel. Figures~\ref{fig:cdfs1} through~\ref{fig:cdfs4} separates the data presented in Figure~\ref{fig:cdfs} in four equal-sized quartiles based on the column density of the angle comparison.

There is no general trend that can be seen across all bones for these particular cumulative distribution function, other than that match the general findings seen in Figure~\ref{fig:columndiff}, as discussed in Section~\ref{sec:orientationbones}. Notably, Filament~4, Filament~5, the Snake, and G47 spine B-field angle differences appear even more perpendicular at the highest column densities (Figure~\ref{fig:cdfs4}). G47 in particular appears quite perpendicular at the highest column densities, matching closely with the cyan line, which is the case where G47 is perfectly in the plane of the sky and the B-fields are projected at random. Filament~4 and Filament~5 show more parallel features at the lowest column densities (Figure~\ref{fig:cdfs1}) when comparing to all the data (Figure~\ref{fig:cdfs}), while G47 is similar. The Snake on the other hand is a bit of an oddity, as it appears more perpendicular at both lower and higher column densities, and is slightly less perpendicular in the second quartile, and has roughly random and even slightly parallel in the third quartile. Filament~10 at the highest column densities (Figure~\ref{fig:cdfs4}) spine/B-field alignment appears even more parallel, which is also evident in Figure~\ref{fig:columndiff}.

\clearpage
\bibliography{stephens_bib}

\begin{thebibliography}{}
\expandafter\ifx\csname natexlab\endcsname\relax\def\natexlab#1{#1}\fi

\bibitem[{{Andersson} {et~al.}(2015){Andersson}, {Lazarian}, \&
  {Vaillancourt}}]{Andersson2015}
{Andersson}, B.-G., {Lazarian}, A., \& {Vaillancourt}, J.~E. 2015, \araa, 53,
  501

\bibitem[{{Andr{\'e}} {et~al.}(2010){Andr{\'e}}, {Men'shchikov}, {Bontemps},
  {K{\"o}nyves}, {Motte}, {Schneider}, {Didelon}, {Minier}, {Saraceno},
  {Ward-Thompson}, {di Francesco}, {White}, {Molinari}, {Testi}, {Abergel},
  {Griffin}, {Henning}, {Royer}, {Mer{\'{\i}}n}, {Vavrek}, {Attard},
  {Arzoumanian}, {Wilson}, {Ade}, {Aussel}, {Baluteau}, {Benedettini},
  {Bernard}, {Blommaert}, {Cambr{\'e}sy}, {Cox}, {di Giorgio}, {Hargrave},
  {Hennemann}, {Huang}, {Kirk}, {Krause}, {Launhardt}, {Leeks}, {Le Pennec},
  {Li}, {Martin}, {Maury}, {Olofsson}, {Omont}, {Peretto}, {Pezzuto}, {Prusti},
  {Roussel}, {Russeil}, {Sauvage}, {Sibthorpe}, {Sicilia-Aguilar}, {Spinoglio},
  {Waelkens}, {Woodcraft}, \& {Zavagno}}]{Andre2010}
{Andr{\'e}}, P., {Men'shchikov}, A., {Bontemps}, S., {et~al.} 2010, \aap, 518,
  L102

\bibitem[{{Astropy Collaboration} {et~al.}(2013){Astropy Collaboration},
  {Robitaille}, {Tollerud}, {Greenfield}, {Droettboom}, {Bray}, {Aldcroft},
  {Davis}, {Ginsburg}, {Price-Whelan}, {Kerzendorf}, {Conley}, {Crighton},
  {Barbary}, {Muna}, {Ferguson}, {Grollier}, {Parikh}, {Nair}, {Unther},
  {Deil}, {Woillez}, {Conseil}, {Kramer}, {Turner}, {Singer}, {Fox}, {Weaver},
  {Zabalza}, {Edwards}, {Azalee Bostroem}, {Burke}, {Casey}, {Crawford},
  {Dencheva}, {Ely}, {Jenness}, {Labrie}, {Lim}, {Pierfederici}, {Pontzen},
  {Ptak}, {Refsdal}, {Servillat}, \& {Streicher}}]{Astropy2013}
{Astropy Collaboration}, {Robitaille}, T.~P., {Tollerud}, E.~J., {et~al.} 2013,
  \aap, 558, A33

\bibitem[{{Astropy Collaboration} {et~al.}(2018){Astropy Collaboration},
  {Price-Whelan}, {Sip{\H{o}}cz}, {G{\"u}nther}, {Lim}, {Crawford}, {Conseil},
  {Shupe}, {Craig}, {Dencheva}, {Ginsburg}, {VanderPlas}, {Bradley},
  {P{\'e}rez-Su{\'a}rez}, {de Val-Borro}, {Aldcroft}, {Cruz}, {Robitaille},
  {Tollerud}, {Ardelean}, {Babej}, {Bach}, {Bachetti}, {Bakanov}, {Bamford},
  {Barentsen}, {Barmby}, {Baumbach}, {Berry}, {Biscani}, {Boquien}, {Bostroem},
  {Bouma}, {Brammer}, {Bray}, {Breytenbach}, {Buddelmeijer}, {Burke},
  {Calderone}, {Cano Rodr{\'\i}guez}, {Cara}, {Cardoso}, {Cheedella}, {Copin},
  {Corrales}, {Crichton}, {D'Avella}, {Deil}, {Depagne}, {Dietrich}, {Donath},
  {Droettboom}, {Earl}, {Erben}, {Fabbro}, {Ferreira}, {Finethy}, {Fox},
  {Garrison}, {Gibbons}, {Goldstein}, {Gommers}, {Greco}, {Greenfield},
  {Groener}, {Grollier}, {Hagen}, {Hirst}, {Homeier}, {Horton}, {Hosseinzadeh},
  {Hu}, {Hunkeler}, {Ivezi{\'c}}, {Jain}, {Jenness}, {Kanarek}, {Kendrew},
  {Kern}, {Kerzendorf}, {Khvalko}, {King}, {Kirkby}, {Kulkarni}, {Kumar},
  {Lee}, {Lenz}, {Littlefair}, {Ma}, {Macleod}, {Mastropietro}, {McCully},
  {Montagnac}, {Morris}, {Mueller}, {Mumford}, {Muna}, {Murphy}, {Nelson},
  {Nguyen}, {Ninan}, {N{\"o}the}, {Ogaz}, {Oh}, {Parejko}, {Parley}, {Pascual},
  {Patil}, {Patil}, {Plunkett}, {Prochaska}, {Rastogi}, {Reddy Janga},
  {Sabater}, {Sakurikar}, {Seifert}, {Sherbert}, {Sherwood-Taylor}, {Shih},
  {Sick}, {Silbiger}, {Singanamalla}, {Singer}, {Sladen}, {Sooley},
  {Sornarajah}, {Streicher}, {Teuben}, {Thomas}, {Tremblay}, {Turner},
  {Terr{\'o}n}, {van Kerkwijk}, {de la Vega}, {Watkins}, {Weaver}, {Whitmore},
  {Woillez}, {Zabalza}, \& {Astropy Contributors}}]{Astropy2018}
{Astropy Collaboration}, {Price-Whelan}, A.~M., {Sip{\H{o}}cz}, B.~M., {et~al.}
  2018, \aj, 156, 123

\bibitem[{{Benjamin} {et~al.}(2003){Benjamin}, {Churchwell}, {Babler}, {Bania},
  {Clemens}, {Cohen}, {Dickey}, {Indebetouw}, {Jackson}, {Kobulnicky},
  {Lazarian}, {Marston}, {Mathis}, {Meade}, {Seager}, {Stolovy}, {Watson},
  {Whitney}, {Wolff}, \& {Wolfire}}]{Benjamin2003}
{Benjamin}, R.~A., {Churchwell}, E., {Babler}, B.~L., {et~al.} 2003, \pasp,
  115, 953

\bibitem[{{Borlaff} {et~al.}(2021){Borlaff}, {Lopez-Rodriguez}, {Beck},
  {Stepanov}, {Ntormousi}, {Hughes}, {Tassis}, {Marcum}, {Grosset}, {Beckman},
  {Proudfit}, {Clark}, {D{\'\i}az-Santos}, {Mao}, {Reach}, {Roman-Duval},
  {Subramanian}, {Tram}, {Zweibel}, {Dale}, \& {Legacy Team}}]{Borlaff2021}
{Borlaff}, A.~S., {Lopez-Rodriguez}, E., {Beck}, R., {et~al.} 2021, \apj, 921,
  128

\bibitem[{{Cabral} \& {Leedom}(1993)}]{Cabral1993}
{Cabral}, B., \& {Leedom}, L.~C. 1993, in Proc. 20th Annual Conf. Comp. Graph.
  Interact. Tech. ({New York: ACM}), 263,
  \url{https://dl.acm.org/citation.cfm?id=166151}

\bibitem[{{Carey} {et~al.}(2009){Carey}, {Noriega-Crespo}, {Mizuno}, {Shenoy},
  {Paladini}, {Kraemer}, {Price}, {Flagey}, {Ryan}, {Ingalls}, {Kuchar},
  {Pinheiro Gon{\c c}alves}, {Indebetouw}, {Billot}, {Marleau}, {Padgett},
  {Rebull}, {Bressert}, {Ali}, {Molinari}, {Martin}, {Berriman}, {Boulanger},
  {Latter}, {Miville-Deschenes}, {Shipman}, \& {Testi}}]{Carey2009}
{Carey}, S.~J., {Noriega-Crespo}, A., {Mizuno}, D.~R., {et~al.} 2009, \pasp,
  121, 76

\bibitem[{{Clemens} {et~al.}(2020){Clemens}, {Cashman}, {Cerny}, {El-Batal},
  {Jameson}, {Marchwinski}, {Montgomery}, {Pavel}, {Pinnick}, \&
  {Taylor}}]{Clemens2020}
{Clemens}, D.~P., {Cashman}, L.~R., {Cerny}, C., {et~al.} 2020, \apjs, 249, 23

\bibitem[{{Contreras} {et~al.}(2017){Contreras}, {Rathborne}, {Guzman},
  {Jackson}, {Whitaker}, {Sanhueza}, \& {Foster}}]{Contreras2017}
{Contreras}, Y., {Rathborne}, J.~M., {Guzman}, A., {et~al.} 2017, \mnras, 466,
  340

\bibitem[{{Coud{\'e}} {et~al.}(2025){Coud{\'e}}, {Stephens}, {Myers},
  {Karnath}, {Smith}, {Guzm{\'a}n}, {Marin}, {Zucker}, {Andersson}, {Li},
  {Looney}, {Novak}, {Pillai}, {Sadavoy}, {Sanhueza}, \& {Soam}}]{Coude2025}
{Coud{\'e}}, S., {Stephens}, I.~W., {Myers}, P.~C., {et~al.} 2025, arXiv
  e-prints, arXiv:2509.25832

\bibitem[{{Crutcher} {et~al.}(2004){Crutcher}, {Nutter}, {Ward-Thompson}, \&
  {Kirk}}]{Crutcher2004}
{Crutcher}, R.~M., {Nutter}, D.~J., {Ward-Thompson}, D., \& {Kirk}, J.~M. 2004,
  \apj, 600, 279

\bibitem[{{Crutcher} {et~al.}(2010){Crutcher}, {Wandelt}, {Heiles},
  {Falgarone}, \& {Troland}}]{Crutcher2010}
{Crutcher}, R.~M., {Wandelt}, B., {Heiles}, C., {Falgarone}, E., \& {Troland},
  T.~H. 2010, \apj, 725, 466

\bibitem[{{Dobbs} {et~al.}(2011){Dobbs}, {Burkert}, \& {Pringle}}]{Dobbs2011}
{Dobbs}, C.~L., {Burkert}, A., \& {Pringle}, J.~E. 2011, \mnras, 413, 2935

\bibitem[{{Dowell} {et~al.}(2010){Dowell}, {Cook}, {Harper}, {Lin}, {Looney},
  {Novak}, {Stephens}, {Berthoud}, {Chuss}, {Crutcher}, {Dotson}, {Hildebrand},
  {Houde}, {Jones}, {Krejny}, {Lazarian}, {Moseley}, {Tassis}, {Vaillancourt},
  \& {Werner}}]{Dowell2010}
{Dowell}, C.~D., {Cook}, B.~T., {Harper}, D.~A., {et~al.} 2010, in Society of
  Photo-Optical Instrumentation Engineers (SPIE) Conference Series, Vol. 7735,
  Society of Photo-Optical Instrumentation Engineers (SPIE) Conference Series,
  6

\bibitem[{{Duarte-Cabral} \& {Dobbs}(2017)}]{DuarteCabral2017}
{Duarte-Cabral}, A., \& {Dobbs}, C.~L. 2017, \mnras, 470, 4261

\bibitem[{{Evans} {et~al.}(2021){Evans}, {Heyer}, {Miville-Desch{\^e}nes},
  {Nguyen-Luong}, \& {Merello}}]{Evans2021}
{Evans}, Neal~J., I., {Heyer}, M., {Miville-Desch{\^e}nes}, M.-A.,
  {Nguyen-Luong}, Q., \& {Merello}, M. 2021, \apj, 920, 126

\bibitem[{{Fiege} \& {Pudritz}(2000)}]{Fiege2000}
{Fiege}, J.~D., \& {Pudritz}, R.~E. 2000, \mnras, 311, 85

\bibitem[{{Friesen} {et~al.}(2017){Friesen}, {Pineda}, {co-PIs}, {Rosolowsky},
  {Alves}, {Chac{\'o}n-Tanarro}, {How-Huan Chen}, {Chun-Yuan Chen}, {Di
  Francesco}, {Keown}, {Kirk}, {Punanova}, {Seo}, {Shirley}, {Ginsburg},
  {Hall}, {Offner}, {Singh}, {Arce}, {Caselli}, {Goodman}, {Martin}, {Matzner},
  {Myers}, {Redaelli}, \& {GAS Collaboration}}]{Friesen2017}
{Friesen}, R.~K., {Pineda}, J.~E., {co-PIs}, {et~al.} 2017, \apj, 843, 63

\bibitem[{{Girart} {et~al.}(2006){Girart}, {Rao}, \& {Marrone}}]{Girart2006}
{Girart}, J.~M., {Rao}, R., \& {Marrone}, D.~P. 2006, Science, 313, 812

\bibitem[{{Goodman} {et~al.}(2014){Goodman}, {Alves}, {Beaumont}, {Benjamin},
  {Borkin}, {Burkert}, {Dame}, {Jackson}, {Kauffmann}, {Robitaille}, \&
  {Smith}}]{Goodman2014}
{Goodman}, A.~A., {Alves}, J., {Beaumont}, C.~N., {et~al.} 2014, \apj, 797, 53

\bibitem[{{Grudi{\'c}} {et~al.}(2021){Grudi{\'c}}, {Guszejnov}, {Hopkins},
  {Offner}, \& {Faucher-Gigu{\`e}re}}]{Grudic2021}
{Grudi{\'c}}, M.~Y., {Guszejnov}, D., {Hopkins}, P.~F., {Offner}, S. S.~R., \&
  {Faucher-Gigu{\`e}re}, C.-A. 2021, \mnras, 506, 2199

\bibitem[{{Grudi{\'c}} {et~al.}(2022){Grudi{\'c}}, {Guszejnov}, {Offner},
  {Rosen}, {Raju}, {Faucher-Gigu{\`e}re}, \& {Hopkins}}]{Grudic2022}
{Grudi{\'c}}, M.~Y., {Guszejnov}, D., {Offner}, S. S.~R., {et~al.} 2022,
  \mnras, 512, 216

\bibitem[{{Grudi{\'c}} {et~al.}(2019){Grudi{\'c}}, {Hopkins}, {Lee}, {Murray},
  {Faucher-Gigu{\`e}re}, \& {Johnson}}]{Grudic2019}
{Grudi{\'c}}, M.~Y., {Hopkins}, P.~F., {Lee}, E.~J., {et~al.} 2019, \mnras,
  488, 1501

\bibitem[{{Guzm{\'a}n} {et~al.}(2015){Guzm{\'a}n}, {Sanhueza}, {Contreras},
  {Smith}, {Jackson}, {Hoq}, \& {Rathborne}}]{Guzman2015}
{Guzm{\'a}n}, A.~E., {Sanhueza}, P., {Contreras}, Y., {et~al.} 2015, \apj, 815,
  130

\bibitem[{{Hacar} {et~al.}(2023){Hacar}, {Clark}, {Heitsch}, {Kainulainen},
  {Panopoulou}, {Seifried}, \& {Smith}}]{Hacar2023}
{Hacar}, A., {Clark}, S.~E., {Heitsch}, F., {et~al.} 2023, in Astronomical
  Society of the Pacific Conference Series, Vol. 534, Protostars and Planets
  VII, ed. S.~{Inutsuka}, Y.~{Aikawa}, T.~{Muto}, K.~{Tomida}, \& M.~{Tamura},
  153

\bibitem[{{Harper} {et~al.}(2018){Harper}, {Runyan}, {Dowell}, {Wirth},
  {Amato}, {Ames}, {Amiri}, {Banks}, {Bartels}, {Benford}, {Berthoud},
  {Buchanan}, {Casey}, {Chapman}, {Chuss}, {Cook}, {Derro}, {Dotson}, {Evans},
  {Fixsen}, {Gatley}, {Guerra}, {Halpern}, {Hamilton}, {Hamlin}, {Hansen},
  {Heimsath}, {Hermida}, {Hilton}, {Hirsch}, {Hollister}, {Hostetter}, {Irwin},
  {Jhabvala}, {Jhabvala}, {Kastner}, {Kov{\'a}cs}, {Lin}, {Loewenstein},
  {Looney}, {Lopez-Rodriguez}, {Maher}, {Michail}, {Miller}, {Moseley},
  {Novak}, {Pernic}, {Rennick}, {Rhody}, {Sandberg}, {Sandford}, {Santos},
  {Shafer}, {Sharp}, {Shirron}, {Siah}, {Silverberg}, {Sparr}, {Spotz},
  {Staguhn}, {Toorian}, {Towey}, {Tuttle}, {Vaillancourt}, {Voellmer},
  {Volpert}, {Wang}, \& {Wollack}}]{Harper2018}
{Harper}, D.~A., {Runyan}, M.~C., {Dowell}, C.~D., {et~al.} 2018, Journal of
  Astronomical Instrumentation, 7, 1840008

\bibitem[{{Hix} {et~al.}(2023){Hix}, {He}, \& {Ricotti}}]{Hix2023}
{Hix}, R., {He}, C.-C., \& {Ricotti}, M. 2023, \mnras, 522, 6203

\bibitem[{{Hogge} {et~al.}(2018){Hogge}, {Jackson}, {Stephens}, {Whitaker},
  {Foster}, {Camarata}, {Anish Roshi}, {Di Francesco}, {Longmore}, {Loughnane},
  {Moore}, {Rathborne}, {Sanhueza}, \& {Walsh}}]{Hogge2018}
{Hogge}, T., {Jackson}, J., {Stephens}, I., {et~al.} 2018, \apjs, 237, 27

\bibitem[{{Hull} {et~al.}(2013){Hull}, {Plambeck}, {Bolatto}, {Bower},
  {Carpenter}, {Crutcher}, {Fiege}, {Franzmann}, {Hakobian}, {Heiles}, {Houde},
  {Hughes}, {Jameson}, {Kwon}, {Lamb}, {Looney}, {Matthews}, {Mundy}, {Pillai},
  {Pound}, {Stephens}, {Tobin}, {Vaillancourt}, {Volgenau}, \&
  {Wright}}]{Hull2013}
{Hull}, C.~L.~H., {Plambeck}, R.~L., {Bolatto}, A.~D., {et~al.} 2013, \apj,
  768, 159

\bibitem[{{Hull} {et~al.}(2014){Hull}, {Plambeck}, {Kwon}, {Bower},
  {Carpenter}, {Crutcher}, {Fiege}, {Franzmann}, {Hakobian}, {Heiles}, {Houde},
  {Hughes}, {Lamb}, {Looney}, {Marrone}, {Matthews}, {Pillai}, {Pound},
  {Rahman}, {Sandell}, {Stephens}, {Tobin}, {Vaillancourt}, {Volgenau}, \&
  {Wright}}]{Hull2014}
{Hull}, C.~L.~H., {Plambeck}, R.~L., {Kwon}, W., {et~al.} 2014, \apjs, 213, 13

\bibitem[{{Jackson} {et~al.}(2010){Jackson}, {Finn}, {Chambers}, {Rathborne},
  \& {Simon}}]{Jackson2010}
{Jackson}, J.~M., {Finn}, S.~C., {Chambers}, E.~T., {Rathborne}, J.~M., \&
  {Simon}, R. 2010, \apjl, 719, L185

\bibitem[{{Jackson} {et~al.}(2006){Jackson}, {Rathborne}, {Shah}, {Simon},
  {Bania}, {Clemens}, {Chambers}, {Johnson}, {Dormody}, {Lavoie}, \&
  {Heyer}}]{Jackson2006}
{Jackson}, J.~M., {Rathborne}, J.~M., {Shah}, R.~Y., {et~al.} 2006, \apjs, 163,
  145

\bibitem[{{Jackson} {et~al.}(2013){Jackson}, {Rathborne}, {Foster}, {Whitaker},
  {Sanhueza}, {Claysmith}, {Mascoop}, {Wienen}, {Breen}, {Herpin},
  {Duarte-Cabral}, {Csengeri}, {Longmore}, {Contreras}, {Indermuehle},
  {Barnes}, {Walsh}, {Cunningham}, {Brooks}, {Britton}, {Voronkov}, {Urquhart},
  {Alves}, {Jordan}, {Hill}, {Hoq}, {Finn}, {Bains}, {Bontemps}, {Bronfman},
  {Caswell}, {Deharveng}, {Ellingsen}, {Fuller}, {Garay}, {Green}, {Hindson},
  {Jones}, {Lenfestey}, {Lo}, {Lowe}, {Mardones}, {Menten}, {Minier}, {Morgan},
  {Motte}, {Muller}, {Peretto}, {Purcell}, {Schilke}, {Bontemps}, {Schuller},
  {Titmarsh}, {Wyrowski}, \& {Zavagno}}]{Jackson2013}
{Jackson}, J.~M., {Rathborne}, J.~M., {Foster}, J.~B., {et~al.} 2013, \pasa,
  30, 57

\bibitem[{{Jadhav} {et~al.}(2025){Jadhav}, {Dewangan}, {Haj Ismail}, {Bhadari},
  {Maity}, {Kesh Yadav}, {Salouci}, {Patricio}, \& {Sharma}}]{Jadhav2025}
{Jadhav}, O.~R., {Dewangan}, L.~K., {Haj Ismail}, A., {et~al.} 2025, arXiv
  e-prints, arXiv:2504.04742

\bibitem[{{Kauffmann} {et~al.}(2008){Kauffmann}, {Bertoldi}, {Bourke}, {Evans},
  \& {Lee}}]{Kauffmann2008}
{Kauffmann}, J., {Bertoldi}, F., {Bourke}, T.~L., {Evans}, II, N.~J., \& {Lee},
  C.~W. 2008, \aap, 487, 993

\bibitem[{{Koch} \& {Rosolowsky}(2015)}]{Koch2015}
{Koch}, E.~W., \& {Rosolowsky}, E.~W. 2015, \mnras, 452, 3435

\bibitem[{{Kov{\'a}cs}(2008{\natexlab{a}})}]{Kovacs2008a}
{Kov{\'a}cs}, A. 2008{\natexlab{a}}, in Society of Photo-Optical
  Instrumentation Engineers (SPIE) Conference Series, Vol. 7020, Millimeter and
  Submillimeter Detectors and Instrumentation for Astronomy IV, ed. W.~D.
  {Duncan}, W.~S. {Holland}, S.~{Withington}, \& J.~{Zmuidzinas}, 70201S

\bibitem[{{Kov{\'a}cs}(2008{\natexlab{b}})}]{Kovacs2008b}
{Kov{\'a}cs}, A. 2008{\natexlab{b}}, in Society of Photo-Optical
  Instrumentation Engineers (SPIE) Conference Series, Vol. 7020, Millimeter and
  Submillimeter Detectors and Instrumentation for Astronomy IV, ed. W.~D.
  {Duncan}, W.~S. {Holland}, S.~{Withington}, \& J.~{Zmuidzinas}, 702007

\bibitem[{{Lee} {et~al.}(2021){Lee}, {Berthoud}, {Chen}, {Cox}, {Davidson},
  {Encalada}, {Fissel}, {Harrison}, {Kwon}, {Li}, {Li}, {Looney}, {Novak},
  {Sadavoy}, {Santos}, {Segura-Cox}, \& {Stephens}}]{LeeDennis2021}
{Lee}, D., {Berthoud}, M., {Chen}, C.-Y., {et~al.} 2021, \apj, 918, 39

\bibitem[{{Lee} {et~al.}(2015){Lee}, {Dunham}, {Myers}, {Tobin}, {Kristensen},
  {Pineda}, {Vorobyov}, {Offner}, {Arce}, {Li}, {Bourke}, {J{\o}rgensen},
  {Goodman}, {Sadavoy}, {Chandler}, {Harris}, {Kratter}, {Looney}, {Melis},
  {Perez}, \& {Segura-Cox}}]{Lee2015}
{Lee}, K.~I., {Dunham}, M.~M., {Myers}, P.~C., {et~al.} 2015, \apj, 814, 114

\bibitem[{{Li} {et~al.}(2022{\natexlab{a}}){Li}, {Immer}, {Reid}, {Sanna},
  {Rygl}, {Xu}, {Zhang}, {Brunthaler}, \& {Menten}}]{LiJJ2022}
{Li}, J.~J., {Immer}, K., {Reid}, M.~J., {et~al.} 2022{\natexlab{a}}, \apjs,
  262, 42

\bibitem[{{Li} {et~al.}(2022{\natexlab{b}}){Li}, {Lopez-Rodriguez}, {Ajeddig},
  {Andr{\'e}}, {McKee}, {Rho}, \& {Klein}}]{LiPS2022}
{Li}, P.~S., {Lopez-Rodriguez}, E., {Ajeddig}, H., {et~al.} 2022{\natexlab{b}},
  \mnras, 510, 6085

\bibitem[{{McKee} \& {Ostriker}(2007)}]{McKeeOstriker2007}
{McKee}, C.~F., \& {Ostriker}, E.~C. 2007, \araa, 45, 565

\bibitem[{{Molinari} {et~al.}(2010){Molinari}, {Swinyard}, {Bally}, {Barlow},
  {Bernard}, {Martin}, {Moore}, {Noriega-Crespo}, {Plume}, {Testi}, {Zavagno},
  {Abergel}, {Ali}, {Anderson}, {Andr{\'e}}, {Baluteau}, {Battersby},
  {Beltr{\'a}n}, {Benedettini}, {Billot}, {Blommaert}, {Bontemps}, {Boulanger},
  {Brand}, {Brunt}, {Burton}, {Calzoletti}, {Carey}, {Caselli}, {Cesaroni},
  {Cernicharo}, {Chakrabarti}, {Chrysostomou}, {Cohen}, {Compiegne}, {de
  Bernardis}, {de Gasperis}, {di Giorgio}, {Elia}, {Faustini}, {Flagey},
  {Fukui}, {Fuller}, {Ganga}, {Garcia-Lario}, {Glenn}, {Goldsmith}, {Griffin},
  {Hoare}, {Huang}, {Ikhenaode}, {Joblin}, {Joncas}, {Juvela}, {Kirk},
  {Lagache}, {Li}, {Lim}, {Lord}, {Marengo}, {Marshall}, {Masi}, {Massi},
  {Matsuura}, {Minier}, {Miville-Desch{\^e}nes}, {Montier}, {Morgan}, {Motte},
  {Mottram}, {M{\"u}ller}, {Natoli}, {Neves}, {Olmi}, {Paladini}, {Paradis},
  {Parsons}, {Peretto}, {Pestalozzi}, {Pezzuto}, {Piacentini}, {Piazzo},
  {Polychroni}, {Pomar{\`e}s}, {Popescu}, {Reach}, {Ristorcelli}, {Robitaille},
  {Robitaille}, {Rod{\'o}n}, {Roy}, {Royer}, {Russeil}, {Saraceno}, {Sauvage},
  {Schilke}, {Schisano}, {Schneider}, {Schuller}, {Schulz}, {Sibthorpe},
  {Smith}, {Smith}, {Spinoglio}, {Stamatellos}, {Strafella}, {Stringfellow},
  {Sturm}, {Taylor}, {Thompson}, {Traficante}, {Tuffs}, {Umana}, {Valenziano},
  {Vavrek}, {Veneziani}, {Viti}, {Waelkens}, {Ward-Thompson}, {White},
  {Wilcock}, {Wyrowski}, {Yorke}, \& {Zhang}}]{Molinari2010}
{Molinari}, S., {Swinyard}, B., {Bally}, J., {et~al.} 2010, \aap, 518, L100

\bibitem[{{Myers} {et~al.}(2025){Myers}, {Heyer}, {Stephens}, {Coud{\'e}},
  {Karnath}, \& {Smith}}]{Myers2025}
{Myers}, P.~C., {Heyer}, M., {Stephens}, I.~W., {et~al.} 2025, arXiv e-prints,
  arXiv:2508.05826

\bibitem[{{Myers} {et~al.}(2024){Myers}, {Stephens}, \&
  {Coud{\'e}}}]{Myers2024}
{Myers}, P.~C., {Stephens}, I.~W., \& {Coud{\'e}}, S. 2024, \apj, 962, 64

\bibitem[{{Ngoc} {et~al.}(2023){Ngoc}, {Diep}, {Hoang}, {Tram}, {Giang},
  {L{\^e}}, {Hoang}, {Phuong}, {Khang}, {Nguyen}, \& {Truong}}]{Ngoc2023}
{Ngoc}, N.~B., {Diep}, P.~N., {Hoang}, T., {et~al.} 2023, \apj, 953, 66

\bibitem[{{Ostriker} {et~al.}(2001){Ostriker}, {Stone}, \&
  {Gammie}}]{Ostriker2001}
{Ostriker}, E.~C., {Stone}, J.~M., \& {Gammie}, C.~F. 2001, \apj, 546, 980

\bibitem[{{Pattle} {et~al.}(2023){Pattle}, {Fissel}, {Tahani}, {Liu}, \&
  {Ntormousi}}]{Pattle2023}
{Pattle}, K., {Fissel}, L., {Tahani}, M., {Liu}, T., \& {Ntormousi}, E. 2023,
  in Astronomical Society of the Pacific Conference Series, Vol. 534,
  Protostars and Planets VII, ed. S.~{Inutsuka}, Y.~{Aikawa}, T.~{Muto},
  K.~{Tomida}, \& M.~{Tamura}, 193

\bibitem[{{Pattle} {et~al.}(2017){Pattle}, {Ward-Thompson}, {Berry},
  {Hatchell}, {Chen}, {Pon}, {Koch}, {Kwon}, {Kim}, {Bastien}, {Cho},
  {Coud{\'e}}, {Di Francesco}, {Fuller}, {Furuya}, {Graves}, {Johnstone},
  {Kirk}, {Kwon}, {Lee}, {Matthews}, {Mottram}, {Parsons}, {Sadavoy},
  {Shinnaga}, {Soam}, {Hasegawa}, {Lai}, {Qiu}, \& {Friberg}}]{Pattle2017}
{Pattle}, K., {Ward-Thompson}, D., {Berry}, D., {et~al.} 2017, \apj, 846, 122

\bibitem[{{Perault} {et~al.}(1996){Perault}, {Omont}, {Simon}, {Seguin},
  {Ojha}, {Blommaert}, {Felli}, {Gilmore}, {Guglielmo}, {Habing}, {Price},
  {Robin}, {de Batz}, {Cesarsky}, {Elbaz}, {Epchtein}, {Fouque}, {Guest},
  {Levine}, {Pollock}, {Prusti}, {Siebenmorgen}, {Testi}, \&
  {Tiphene}}]{Perault1996}
{Perault}, M., {Omont}, A., {Simon}, G., {et~al.} 1996, \aap, 315, L165

\bibitem[{{Planck Collaboration} {et~al.}(2016){Planck Collaboration}, {Ade},
  {Aghanim}, {Alves}, {Arnaud}, {Arzoumanian}, {Ashdown}, {Aumont},
  {Baccigalupi}, {Banday}, {Barreiro}, {Bartolo}, {Battaner}, {Benabed},
  {Beno{\^i}t}, {Benoit-L{\'e}vy}, {Bernard}, {Bersanelli}, {Bielewicz},
  {Bock}, {Bonavera}, {Bond}, {Borrill}, {Bouchet}, {Boulanger}, {Bracco},
  {Burigana}, {Calabrese}, {Cardoso}, {Catalano}, {Chiang}, {Christensen},
  {Colombo}, {Combet}, {Couchot}, {Crill}, {Curto}, {Cuttaia}, {Danese},
  {Davies}, {Davis}, {de Bernardis}, {de Rosa}, {de Zotti}, {Delabrouille},
  {Dickinson}, {Diego}, {Dole}, {Donzelli}, {Dor{\'e}}, {Douspis}, {Ducout},
  {Dupac}, {Efstathiou}, {Elsner}, {En{\ss}lin}, {Eriksen}, {Falceta-Gon{\c
  c}alves}, {Falgarone}, {Ferri{\`e}re}, {Finelli}, {Forni}, {Frailis},
  {Fraisse}, {Franceschi}, {Frejsel}, {Galeotta}, {Galli}, {Ganga}, {Ghosh},
  {Giard}, {Gjerl{\o}w}, {Gonz{\'a}lez-Nuevo}, {G{\'o}rski}, {Gregorio},
  {Gruppuso}, {Gudmundsson}, {Guillet}, {Harrison}, {Helou}, {Hennebelle},
  {Henrot-Versill{\'e}}, {Hern{\'a}ndez-Monteagudo}, {Herranz}, {Hildebrandt},
  {Hivon}, {Holmes}, {Hornstrup}, {Huffenberger}, {Hurier}, {Jaffe}, {Jaffe},
  {Jones}, {Juvela}, {Keih{\"a}nen}, {Keskitalo}, {Kisner}, {Knoche}, {Kunz},
  {Kurki-Suonio}, {Lagache}, {Lamarre}, {Lasenby}, {Lattanzi}, {Lawrence},
  {Leonardi}, {Levrier}, {Liguori}, {Lilje}, {Linden-V{\o}rnle},
  {L{\'o}pez-Caniego}, {Lubin}, {Mac{\'{\i}}as-P{\'e}rez}, {Maino},
  {Mandolesi}, {Mangilli}, {Maris}, {Martin}, {Mart{\'{\i}}nez-Gonz{\'a}lez},
  {Masi}, {Matarrese}, {Melchiorri}, {Mendes}, {Mennella}, {Migliaccio},
  {Miville-Desch{\^e}nes}, {Moneti}, {Montier}, {Morgante}, {Mortlock},
  {Munshi}, {Murphy}, {Naselsky}, {Nati}, {Netterfield}, {Noviello}, {Novikov},
  {Novikov}, {Oppermann}, {Oxborrow}, {Pagano}, {Pajot}, {Paladini},
  {Paoletti}, {Pasian}, {Perotto}, {Pettorino}, {Piacentini}, {Piat},
  {Pierpaoli}, {Pietrobon}, {Plaszczynski}, {Pointecouteau}, {Polenta},
  {Ponthieu}, {Pratt}, {Prunet}, {Puget}, {Rachen}, {Reinecke}, {Remazeilles},
  {Renault}, {Renzi}, {Ristorcelli}, {Rocha}, {Rossetti}, {Roudier},
  {Rubi{\~n}o-Mart{\'{\i}}n}, {Rusholme}, {Sandri}, {Santos}, {Savelainen},
  {Savini}, {Scott}, {Soler}, {Stolyarov}, {Sudiwala}, {Sutton}, {Suur-Uski},
  {Sygnet}, {Tauber}, {Terenzi}, {Toffolatti}, {Tomasi}, {Tristram}, {Tucci},
  {Umana}, {Valenziano}, {Valiviita}, {Van Tent}, {Vielva}, {Villa}, {Wade},
  {Wandelt}, {Wehus}, {Ysard}, {Yvon}, \& {Zonca}}]{Planck35}
{Planck Collaboration}, {Ade}, P.~A.~R., {Aghanim}, N., {et~al.} 2016, \aap,
  586, A138

\bibitem[{{Price} {et~al.}(2001){Price}, {Egan}, {Carey}, {Mizuno}, \&
  {Kuchar}}]{Price2001}
{Price}, S.~D., {Egan}, M.~P., {Carey}, S.~J., {Mizuno}, D.~R., \& {Kuchar},
  T.~A. 2001, \aj, 121, 2819

\bibitem[{{Rao} {et~al.}(2009){Rao}, {Girart}, {Marrone}, {Lai}, \&
  {Schnee}}]{Rao2009}
{Rao}, R., {Girart}, J.~M., {Marrone}, D.~P., {Lai}, S.-P., \& {Schnee}, S.
  2009, \apj, 707, 921

\bibitem[{{Reid} {et~al.}(2019){Reid}, {Menten}, {Brunthaler}, {Zheng}, {Dame},
  {Xu}, {Li}, {Sakai}, {Wu}, {Immer}, {Zhang}, {Sanna}, {Moscadelli}, {Rygl},
  {Bartkiewicz}, {Hu}, {Quiroga-Nu{\~n}ez}, \& {van Langevelde}}]{Reid2019}
{Reid}, M.~J., {Menten}, K.~M., {Brunthaler}, A., {et~al.} 2019, \apj, 885, 131

\bibitem[{{Robitaille} \& {Bressert}(2012)}]{Robitaille2012}
{Robitaille}, T., \& {Bressert}, E. 2012, {APLpy: Astronomical Plotting Library
  in Python}, Astrophysics Source Code Library, , , ascl:1208.017

\bibitem[{{Robitaille} {et~al.}(2020){Robitaille}, {Deil}, \&
  {Ginsburg}}]{Robitaille2020}
{Robitaille}, T., {Deil}, C., \& {Ginsburg}, A. 2020, {reproject: Python-based
  astronomical image reprojection}, , , ascl:2011.023

\bibitem[{{Schuller} {et~al.}(2021){Schuller}, {Urquhart}, {Csengeri},
  {Colombo}, {Duarte-Cabral}, {Mattern}, {Ginsburg}, {Pettitt}, {Wyrowski},
  {Anderson}, {Azagra}, {Barnes}, {Beltran}, {Beuther}, {Billington},
  {Bronfman}, {Cesaroni}, {Dobbs}, {Eden}, {Lee}, {Medina}, {Menten}, {Moore},
  {Montenegro-Montes}, {Ragan}, {Rigby}, {Riener}, {Russeil}, {Schisano},
  {Sanchez-Monge}, {Traficante}, {Zavagno}, {Agurto}, {Bontemps}, {Finger},
  {Giannetti}, {Gonzalez}, {Hernandez}, {Henning}, {Kainulainen}, {Kauffmann},
  {Leurini}, {Lopez}, {Mac-Auliffe}, {Mazumdar}, {Molinari}, {Motte}, {Muller},
  {Nguyen-Luong}, {Parra}, {Perez-Beaupuits}, {Schilke}, {Schneider}, {Suri},
  {Testi}, {Torstensson}, {Veena}, {Venegas}, {Wang}, \&
  {Wienen}}]{Schuller2021}
{Schuller}, F., {Urquhart}, J.~S., {Csengeri}, T., {et~al.} 2021, \mnras, 500,
  3064

\bibitem[{{Simon} {et~al.}(2006){Simon}, {Jackson}, {Rathborne}, \&
  {Chambers}}]{Simon2006}
{Simon}, R., {Jackson}, J.~M., {Rathborne}, J.~M., \& {Chambers}, E.~T. 2006,
  \apj, 639, 227

\bibitem[{{Skalidis} \& {Tassis}(2021)}]{Skalidis2021}
{Skalidis}, R., \& {Tassis}, K. 2021, \aap, 647, A186

\bibitem[{{Soler} \& {Hennebelle}(2017)}]{SolerHennebelle2017}
{Soler}, J.~D., \& {Hennebelle}, P. 2017, \aap, 607, A2

\bibitem[{{Soler} {et~al.}(2013){Soler}, {Hennebelle}, {Martin},
  {Miville-Desch{\^e}nes}, {Netterfield}, \& {Fissel}}]{Soler2013}
{Soler}, J.~D., {Hennebelle}, P., {Martin}, P.~G., {et~al.} 2013, \apj, 774,
  128

\bibitem[{{Stephens} {et~al.}(2011){Stephens}, {Looney}, {Dowell},
  {Vaillancourt}, \& {Tassis}}]{Stephens2011}
{Stephens}, I.~W., {Looney}, L.~W., {Dowell}, C.~D., {Vaillancourt}, J.~E., \&
  {Tassis}, K. 2011, \apj, 728, 99

\bibitem[{{Stephens} {et~al.}(2013){Stephens}, {Looney}, {Kwon}, {Hull},
  {Plambeck}, {Crutcher}, {Chapman}, {Novak}, {Davidson}, {Vaillancourt},
  {Shinnaga}, \& {Matthews}}]{Stephens2013}
{Stephens}, I.~W., {Looney}, L.~W., {Kwon}, W., {et~al.} 2013, \apjl, 769, L15

\bibitem[{{Stephens} {et~al.}(2017){Stephens}, {Dunham}, {Myers}, {Pokhrel},
  {Sadavoy}, {Vorobyov}, {Tobin}, {Pineda}, {Offner}, {Lee}, {Kristensen},
  {J{\o}rgensen}, {Goodman}, {Bourke}, {Arce}, \& {Plunkett}}]{Stephens2017b}
{Stephens}, I.~W., {Dunham}, M.~M., {Myers}, P.~C., {et~al.} 2017, \apj, 846,
  16

\bibitem[{{Stephens} {et~al.}(2022){Stephens}, {Myers}, {Zucker}, {Jackson},
  {Andersson}, {Smith}, {Soam}, {Battersby}, {Sanhueza}, {Hogge}, {Smith},
  {Novak}, {Sadavoy}, {Pillai}, {Li}, {Looney}, {Sugitani}, {Coud{\'e}},
  {Guzm{\'a}n}, {Goodman}, {Kusune}, {Santos}, {Zuckerman}, \&
  {Encalada}}]{Stephens2022}
{Stephens}, I.~W., {Myers}, P.~C., {Zucker}, C., {et~al.} 2022, \apjl, 926, L6

\bibitem[{{Vaillancourt}(2006)}]{Vaillancourt2006}
{Vaillancourt}, J.~E. 2006, \pasp, 118, 1340

\bibitem[{{van Hoof}(2000)}]{vanHoof2000}
{van Hoof}, P.~A.~M. 2000, \mnras, 314, 99

\bibitem[{{Wu} {et~al.}(2014){Wu}, {Sato}, {Reid}, {Moscadelli}, {Zhang}, {Xu},
  {Brunthaler}, {Menten}, {Dame}, \& {Zheng}}]{Wu2014}
{Wu}, Y.~W., {Sato}, M., {Reid}, M.~J., {et~al.} 2014, \aap, 566, A17

\bibitem[{{Zhang} {et~al.}(2019){Zhang}, {Kainulainen}, {Mattern}, {Fang}, \&
  {Henning}}]{ZhangMM2019}
{Zhang}, M., {Kainulainen}, J., {Mattern}, M., {Fang}, M., \& {Henning}, T.
  2019, \aap, 622, A52

\bibitem[{{Zhao} {et~al.}(2024){Zhao}, {Pudritz}, {Pillsworth}, {Robinson}, \&
  {Wadsley}}]{Zhao2024}
{Zhao}, B., {Pudritz}, R.~E., {Pillsworth}, R., {Robinson}, H., \& {Wadsley},
  J. 2024, \apj, 974, 240

\bibitem[{{Zucker} {et~al.}(2015){Zucker}, {Battersby}, \&
  {Goodman}}]{Zucker2015}
{Zucker}, C., {Battersby}, C., \& {Goodman}, A. 2015, \apj, 815, 23

\bibitem[{{Zucker} {et~al.}(2018){Zucker}, {Battersby}, \&
  {Goodman}}]{Zucker2018b}
---. 2018, \apj, 864, 153

\bibitem[{{Zucker} \& {Chen}(2018)}]{Zucker2018c}
{Zucker}, C., \& {Chen}, H. H.-H. 2018, \apj, 864, 152

\bibitem[{{Zucker} {et~al.}(2019){Zucker}, {Smith}, \& {Goodman}}]{Zucker2019b}
{Zucker}, C., {Smith}, R., \& {Goodman}, A. 2019, \apj, 887, 186

\end{thebibliography}

\end{document}